\documentclass[preprint]{ptephy_v1}

\preprintnumber{XXXX-XXXX} 

\usepackage{txfonts}

\usepackage[utf8]{inputenc}
\usepackage{hyperref}

\usepackage[T1]{fontenc} 
\usepackage{amsmath}
\usepackage{graphicx,epsfig,wrapfig}
\usepackage{dutchcal}
\usepackage{makeidx}
\usepackage{amsfonts}
\usepackage{amssymb}
\usepackage{mathtools}
\usepackage{amsbsy}
\usepackage[capitalize]{cleveref}
\usepackage{mathrsfs}
\usepackage{slashed}
\usepackage{bm}
\usepackage{multirow}
\usepackage{braket}

\newcommand{\bfk}{{\bf k}_{\perp}}

\newcommand{\Dp}{{\bf \Delta}_{\perp}}

\usepackage{xcolor}
\usepackage{rotating}

\begin{document}

\title{
Understanding the Valence Quark Structure of the Pion through GTMDs
}


\author{Satyajit Puhan}
\affil{Computational High Energy Physics Lab, Department of Physics, Dr. B.R. Ambedkar National Institute of Technology, Jalandhar, 144008, India \email{puhansatyajit@gmail.com}}

\author{Shubham Sharma}
\affil{Laboratory for Advanced Scientific Technologies of Mega-Science Facilities and Experiments, Moscow Institute of Physics and Technology (MIPT), 141700 Dolgoprudny, Russia \email{s.sharma.hep@gmail.com}}

\author{Narinder Kumar}
\affil{Computational Theoretical High Energy Physics Lab, Department of Physics, Doaba College, Jalandhar 144004, India\email{narinderhep@gmail.com}}

\author{Harleen Dahiya\thanks{These authors contributed equally to this work}}
\affil{Computational High Energy Physics Lab, Department of Physics, Dr. B.R. Ambedkar National Institute of Technology, Jalandhar, 144008, India \email{dahiyah@nitj.ac.in}}



%
%
%
%

\begin{abstract}%
We investigate the internal structure of the pion using generalized transverse momentum-dependent parton distributions (GTMDs) within the light-cone quark model. By solving the quark-quark correlator, we derive the twist-$2$, $3$, and $4$ quark GTMDs in terms of light-front wave functions (LFWFs). Out of the $16$ possible GTMDs, $12$ are found to be nonzero. Furthermore, we extract the valence quark transverse momentum-dependent parton distributions (TMDs) and generalized parton distributions (GPDs) from their corresponding GTMDs. Additionally, we compute the valence quark electromagnetic form factors (FFs) and parton distribution functions (PDFs) up to twist-$4$. The elastic charge radius of the pion is determined to be $0.558$ fm. Our results exhibit a qualitative agreement with predictions from other theoretical model like Nambu-Jona-Lasinio model, Light-front holographic model, and spectator model at the leading twist. This study provides a comprehensive insight into the internal structure of the pion.
\end{abstract}

\subjectindex{xxxx, xxx}

\maketitle

\section{Introduction}
Understanding the internal structure of hadrons, along with the behavior of their constituent quarks, gluons, and sea quarks, remains a long-standing challenge in particle physics \cite{Gross:2022hyw}. A complete description of hadron structure within the framework of quantum chromodynamics (QCD) \cite{Brodsky:1997de,Zhang:1997dd}, particularly in terms of hadronic matrix elements of quark-gluon field operators, continues to be an open question from both theoretical and experimental perspectives.
\par
The study of hadron structure spans both perturbative and non-perturbative regimes \cite{Magradze:1999um}. In the perturbative region (high Q$^2$), gluon contributions dominate over valence quark effects \cite{CTEQ:1993hwr}. Conversely, in the non-perturbative regime (low Q$^2$), valence quarks play a more significant role than gluons \cite{Donnachie:1993it}. To investigate valence quark distributions, non-perturbative models are employed. These distributions are then evolved to higher Q$^2$ using various evolution frameworks, such as the Dokshitzer–Gribov–Lipatov–Altarelli–Parisi (DGLAP) \cite{Markovych:2023tpa,Wang:2016sfq}, Collins-Soper-Sterman \cite{Aybat:2011ge,Aybat:2011ta}, and Balitsky-Fadin-Kuraev-Lipatov \cite{Bautista:2016xnp,Mukherjee:2023snp} formalisms.
\par
The valence quark structure of a hadron can be explored using multi-dimensional distribution functions such as parton distribution functions (PDFs) \cite{Collins:1981uw, Martin:1998sq,Gluck:1994uf,Gluck:1998xa}, generalized parton distributions (GPDs) \cite{Diehl:2003ny,Garcon:2002jb, Belitsky:2005qn,Sharma:2023ibp}, transverse momentum-dependent parton distribution functions (TMDs) \cite{Barone:2001sp, Diehl:2015uka,Puhan:2023ekt,Boussarie:2023izj,Puhan:2023hio,Sharma:2024lal}, and generalized transverse momentum-dependent parton distributions (GTMDs) \cite{Echevarria:2022ztg,Sharma:2024arf,Bhattacharya:2017bvs}, among others.  
Among these, PDFs are the simplest one-dimensional distributions, providing information about the longitudinal momentum fraction ($x$) carried by a quark within its parent hadron. PDFs can be extracted from deep inelastic scattering processes \cite{Collins:2004nx,Polchinski:2002jw}, and extensive theoretical and experimental studies have been conducted to investigate PDFs for nucleons \cite{Cheng:2023kmt,COMPASS:2010hwr,Gao:2022uhg,CMS:2012tdr}. While experimental studies on the PDFs of spin-$0$ mesons are limited \cite{Aicher:2010cb,Conway:1989fs}, upcoming facilities such as the Electron-Ion Collider \cite{AbdulKhalek:2021gbh,Accardi:2012qut} and the $12$ GeV Jefferson Lab \cite{Barry:2021osv} are expected to provide new insights into pion PDFs through the Sullivan process. To explore the spatial and transverse structure of hadrons, which cannot be accessed through PDFs alone, one must turn to three-dimensional distributions such as TMDs and GPDs. TMDs extend collinear PDFs by incorporating dependence on both the longitudinal momentum fraction ($x$) and the transverse momentum ($\bfk$) of constituent quarks. They play a crucial role in revealing spin-orbit correlations (SOCs), transverse spin effects, azimuthal asymmetries, spin densities, QCD confinement, and factorization. TMDs can be extracted from various experimental processes, including Drell-Yan scattering \cite{Tangerman:1994eh,Zhou:2009jm,Mulders:1995dh,Collins:2002kn,Bacchetta:2017gcc}, semi-inclusive deep inelastic scattering \cite{Bacchetta:2006tn,Brodsky:2002cx,Ji:2004wu,Bacchetta:2017gcc}, electron-positron annihilation, and $Z^0/W^\pm$ boson production \cite{Bacchetta:2017gcc,Catani:2015vma}. While looking into the GPDs, they carry information about the physical and mechanical properties inside the hadron. GPDs are the functions of longitudinal momentum fraction ($x$), skewness ($\xi$), and momentum transfer ($\Delta$) between the initial and final hadron. GPDs can be extracted from deeply virtual Compton scattering \cite{Ji:1996nm} and deeply virtual meson production \cite{Favart:2015umi} experimentally. For a more comprehensive understanding of hadron structure, one can study GTMDs, often regarded as the so-called mother distributions. GTMDs encapsulate nearly all the internal structural information of hadrons. However, their experimental extraction, particularly through the double Drell-Yan process, remains highly challenging. Despite these difficulties, theoretical predictions suggest promising prospects for future experimental investigations. Additionally, PDFs, TMDs, and GPDs can be derived from GTMDs through appropriate reduction relations \cite{Lorce:2011dv, Kanazawa:2014nha,Chakrabarti:2017teq,Kumar:2017xcm,Kaur:2019kpi,Sharma:2023tre,Sharma:2023qgb}.
\par
In this work, we primarily focus on the GTMDs of the spin-$0$ pion up to twist-$4$. While numerous studies have been conducted on GTMDs for nucleons, relatively little work has been dedicated to the GTMDs of the pion. As the lightest pseudoscalar meson, understanding the internal structure of the pion at both leading and higher twists is crucial. The leading-twist distribution functions contribute significantly more than subleading-twist and twist-$4$ distributions. However, higher-twist effects become relevant at relatively low $Q^2$, meaning their contributions to the scattering cross-section cannot be neglected. The $1/Q^{n-2}$ expansion of the scattering cross-section systematically introduces different twist contributions, where $Q$ represents the hard scale of the physical process and $n=2,3,4...$ denotes the twist order \cite{Bastami:2020rxn}.
\par
For spin-$0$ mesons, there exist $16$ GTMDs \cite{Meissner:2008ay}, in contrast to the $64$ GTMDs for spin-$1/2$ hadron up to twist-$4$ \cite{Meissner:2009ww}. Similarly, spin-$1/2$ nucleons possess a total of $32$ TMDs and GPDs, whereas spin-$0$ mesons have only $8$. At twist-$3$ and twist-$4$, gluon contributions also play a significant role alongside valence quarks. However, in this work, we restrict our analysis to the minimal Fock state of the meson, expressed as $|M \rangle = \sum | q \bar{q} \rangle \psi_{q \bar{q}}$. Consequently, we adopt the Wandzura-Wilczek approximation, which leads to vanishing tilde terms \cite{Anikin:2001ge}.  
\par
In this work, we have adopted the light-cone quark model (LCQM) to calculate all the distribution functions. Several studies have explored leading-twist GTMDs in different theoretical frameworks, including the light-front holographic model (LFHM) \cite{Kaur:2019kpi}, LCQM \cite{Ma:2018ysi}, and higher-twist GTMDs in the contact interaction approach \cite{Zhang:2020ecj} for the pion. Similarly, valence quark GPDs for the pion have been investigated using various models, such as LCQM \cite{Luan:2024dvc}, the AdS/QCD model \cite{Kaur:2018ewq}, and the contact interaction approach \cite{Zhang:2020ecj}. Additionally, extensive studies on valence quark TMDs of the pion have been conducted within light-front constituent models \cite{Pasquini:2014ppa}, LFHM \cite{Kaur:2019jfa, Puhan:2023ekt}, LCQM \cite{Puhan:2023ekt}, the Nambu–Jona-Lasinio (NJL) model \cite{Noguera:2015iia}, and the MIT Bag model \cite{Lu:2012hh}, among others. These theoretical models provide valuable insights into the internal structure of the pion. However, experimental data for pion GTMDs, GPDs, and TMDs remain unavailable. Some progress has been made through lattice QCD simulations, particularly for pion GPDs \cite{Chen:2019lcm} and quark TMD extractions \cite{Cerutti:2022lmb}. To achieve a comprehensive understanding of the pion structure, we have systematically studied GTMDs, GPDs, TMDs, PDFs, and electromagnetic form factors (EMFFs) up to twist-$4$, along with their interrelations, within the LCQM framework. LCQM is a non-perturbative approach that is inherently gauge-invariant and relativistic. It primarily focuses on valence quarks, which are the fundamental constituents governing the overall structure and intrinsic properties of hadrons.
\par
This work primarily focuses on the GTMDs of spin-$0$ mesons up to twist-$4$, addressing a critical gap in understanding the pion structure. Using the LCQM, we compute and analyze multidimensional distributions and their interrelations, shedding light on valence quark contributions and the significance of higher-twist effects at low $Q^2$. To achieve this, we solve the quark-quark correlator for valence-quark GTMDs at all twist orders in the zero skewness limit ($\xi=0$). At $\xi=0$, the twist-2 GTMD $H^k_1$ vanishes, leaving only three nonzero GTMDs at the leading twist. Additionally, we identify six GTMDs at twist-$3$ and three at twist-$4$ that remain nonzero. We present these distributions through three-dimensional visualizations. By integrating GTMDs over transverse momentum $\bfk$, we obtain the corresponding GPDs. In the $\xi=0$ limit, the twist-$3$ GPDs $H_2(x,0,-\Delta^2_\perp)$ and $F_2(x,0,-\Delta^2_\perp)$ vanish. Similarly, in the $\Delta_\perp \to 0$ limit, we extract TMDs from GTMDs. Our analysis reveals that the T-odd TMDs vanish, leaving four nonzero T-even TMDs up to twist-$4$. Furthermore, we compute PDFs and form factors (FFs) from the extracted TMDs and GPDs. Finally, we discuss the positivity constraints and sum rules governing all the distribution functions.  
\section{Light-Cone Quark Model \label{secmodel}}
LCQM plays a crucial role in describing the hadronic system in terms of its constituent quarks and gluons. It provides a useful framework for relativistically modeling hadrons using quark and gluon degrees of freedom. By employing the light-cone (LC) Fock-state expansion, the mesonic wave function can be expressed as \cite{Lepage:1980fj,Brodsky:1997de,Pasquini:2023aaf}.
 \begin{eqnarray}
|M\rangle &=& \sum |q\bar{q}\rangle \psi_{q\bar{q}}
        + \sum
        |q\bar{q}g\rangle \psi_{q\bar{q}g} + \sum
        |q\bar{q}gg\rangle \psi_{q\bar{q}g g} + \cdots  \, ,
\end{eqnarray}
where $|M\rangle$ represents the meson eigenstate. In this work, we restrict our calculations to the lowest-order meson state. However, we aim to incorporate higher Fock-state contributions in future studies. The mesonic wave function, formulated within the framework of LC quantization of QCD using a multi-particle Fock-state expansion, can be expressed as \cite{Puhan:2023ekt, Qian:2008px,Brodsky:2000xy}.
\begin{eqnarray}\label{fockstate}
|M (P^+, \mathbf{P}_\perp, S_z) \rangle
   &=&\sum_{n,\lambda_i}\int\prod_{i=1}^n \frac{\mathrm{d} x_i \mathrm{d}^2
        \mathbf{k}_{\perp i}}{\sqrt{x_i}~16\pi^3}
 16\pi^3  \delta\Big(1-\sum_{i=1}^n x_i\Big)\nonumber\\
 &&\delta^{(2)}\Big(\sum_{i=1}^n \mathbf{k}_{\perp i}\Big) \psi_{n/M}(x_i,\mathbf{k}_{\perp i},\lambda_i)   | n ; x_i P^+, x_i \mathbf{P}_\perp+\mathbf{k}_{\perp i},
        \lambda_i \rangle .
\end{eqnarray}
Here, $|M (P, S_z) \rangle$ represents the meson eigenstate, where $P=(P^+,P^-,P_{\perp})$ denotes the meson's total momentum. The quantities $\lambda_i$ and $S_z$ correspond to the helicity of the $i$-th constituent and the longitudinal spin projection of the target, respectively. For a spin-$0$ meson, we have $S_z=0$. The momentum of the $i$-th constituent is given by $\mathbf{k_i}=(\mathbf{k}^+_i,\mathbf{k}^-_i,\mathbf{k}_{i \perp})$. The longitudinal momentum fraction carried by an active quark is defined as $x=\frac{\mathbf{k}^+}{P^+}$. Since we are working within the lower Fock-state approximation, we consider the minimal state description of Eq.~(\ref{fockstate}) in the form of a quark-antiquark pair, which can be expressed as
\begin{eqnarray}
|M(P, S_Z=0)\rangle &=& \sum_{\lambda_i,\lambda_j}\int
\frac{\mathrm{d} x \mathrm{d}^2
        \mathbf{k}_{\perp}}{\sqrt{x_1x_2}16\pi^3}
           \Psi(x,\mathbf{k}_{\perp},\lambda_i, 
          \lambda_j)|x, x  
 \mathbf{P}_\perp+\mathbf{k}_{\perp},
        \lambda_i,\lambda_j \rangle
        .
        \label{meson}
\end{eqnarray}
Here, $x_1=x$ and $x_2=(1-x)$ represent the longitudinal momentum fractions of the quark and antiquark, respectively.  
The four-momentum vectors of the meson ($P$), the active quark  ($k_1$), and the antiquark ($k_2$) in the LC frame are defined as
\begin{eqnarray}
P&\equiv&\bigg(P^+,\frac{{M_{\pi}}^2}{P^+},\textbf{0}_\perp \bigg)\label{n1},\\
k_1&\equiv&\bigg(x_1 P^+, \frac{\textbf{k}_\perp^2+m_u^2}{x_1 P^+},\textbf{k}_\perp \bigg),\\
k_2&\equiv&\bigg(x_2 P^+, \frac{\textbf{k}_\perp^2+m_{\bar d}^2}{x_2 P^+},-\textbf{k}_\perp \bigg),
\label{n3}
\end{eqnarray}
with ${M_{\pi}}$ being the  invariant mass of the composite pion system, defined in terms of the quark mass $m_u$ and the anti-quark mass $m_{\bar d}$ as
\begin{eqnarray}
    {M_{\pi}}^2=\frac{\bfk^2+m^2_u}{x_1} +\frac{\bfk^2+m^2_{\bar d}}{x_2}\, .
\end{eqnarray}
The LC meson wave function \( \Psi(x, \mathbf{k}_{\perp}, \lambda_i, \lambda_j) \) in Eq. (\ref{meson}) describes the meson state with different spin and helicity projections of the constituent quarks. It can be expressed as  
\begin{eqnarray}
\Psi_{\pi}(x,\textbf{k}_\perp, \lambda_i, \lambda_j)= \chi(x,\textbf{k}_\perp, \lambda_i, \lambda_j) \psi_{\pi}(x, \textbf{k}_\perp).\
\label{space}
\end{eqnarray}
Here, $\chi(x,\textbf{k}_\perp, \lambda_i, \lambda_j)$ and $\psi_{\pi}(x, \textbf{k}_\perp)$ represent the spin and momentum space wave functions of the pion, respectively. The momentum space wave function can be expressed using the Brodsky-Huang-Lepage framework \cite{Qian:2008px,Xiao:2002iv} for the quark-antiquark sector as  
\begin{eqnarray}
\psi_{\pi}(x,\textbf{k}_\perp)= A \ {\rm exp} \Bigg[-\frac{\frac{\textbf{k}^2_\perp+m_u^2}{x_1}+\frac{\textbf{k}^2_\perp+m^2_{\bar d}}{x_2}}{8 \beta^2}
-\frac{(m_u^2-m_{\bar d}^2)^2}{8 \beta^2 \bigg(\frac{\textbf{k}^2_\perp+m_u^2}{x_1}+\frac{\textbf{k}^2_\perp+m_{\bar d}^2}{x_2}\bigg)} + \frac{m^2_u +m^2_{\bar d}}{4 \beta^2}\Bigg]\, ,
\label{bhl-k}
\end{eqnarray}
%
%
where $A$ and $\beta$ are the meson's normalization constant and harmonic scale parameters, respectively. The normalization constant $A$ can be determined using the following normalization condition as
\begin{eqnarray}
    \int  \frac{d^2 \bfk}{16 \pi^3}|\psi_{\pi}(x,\textbf{k}_\perp)|^2=1.
\end{eqnarray}
The function $\chi(x,\textbf{k}_\perp, \lambda_i, \lambda_j)$ in Eq. (\ref{space}) is the front-form spin wave function, which can be obtained from the instant-form representation via the Melosh-Wigner rotation \cite{Qian:2008px,Xiao:2002iv} or derived from the quark-meson vertex \cite{Choi:1996mq,Qian:2008px}. It is well known that understanding the Melosh-Wigner rotation is crucial for addressing the ``proton spin puzzle,'' as this effect is inherently relativistic and arises due to the transverse motion of quarks inside the hadron \cite{Qian:2008px,Xiao:2002iv}. The transformation between the instant-form state $\Phi(intstant)$ and the front-form state $\Phi(F)$ is given by
\begin{eqnarray}
\Phi_i^\uparrow(instant)&=&-\frac{[\textbf{k}_i^R \Phi_i^\downarrow(F)-(\textbf{k}_i^+ +m_{u(\bar d)})\Phi_i^\uparrow(F)]}{\omega_i},\label{instant-front1}\\
\Phi_i^\downarrow(instant)&=&\frac{[\textbf{k}_i^L\Phi_i^\uparrow(F)+(\textbf{k}_i^+ +m_{u(\bar d)})\Phi_i^\downarrow(F)]}{\omega_i}.
\label{instant-front}
\end{eqnarray}
Here, $\Phi(F)$ is a two-component Dirac spinor, and $\textbf{k}_i^{R(L)}=\textbf{k}_i^1 \pm \iota \textbf{k}_i^2$. The parameter $\omega_i$ is defined as $\omega_i=1/ \sqrt{2 \textbf{k}^+_i (\textbf{k}^0+m_{u(\bar d)})}$. By applying the different momentum forms from Eqs. (\ref{n1})-(\ref{n3}) in the Melosh-Wigner rotation, the front-form spin wave function is obtained in terms of the coefficient $\kappa_{0}^F(x,\textbf{k}_\perp, \lambda_1, \lambda_2)$ as
\begin{eqnarray}
\chi(x,\textbf{k}_\perp, \lambda_i, \lambda_j)=\sum_{\lambda_1, \lambda_2}\kappa_{0}^F(x,\textbf{k}_\perp, \lambda_1, \lambda_2) \Phi_1^{\lambda_1}(F) \Phi_2^{\lambda_2}(F).
\end{eqnarray}
These spin-wave function coefficients with different helicities satisfy the normalization relation
\begin{eqnarray}
\sum_{\lambda_1,\lambda_2} \kappa_0^{F*}(x, \textbf{k}_\perp, \lambda_1, \lambda_2)\kappa_0^F(x, \textbf{k}_\perp, \lambda_1, \lambda_2)=1.
\end{eqnarray}
Similarly, the spin-wave function can also be obtained using an appropriately chosen quark-meson vertex \cite{Choi:1996mq,Qian:2008px}, expressed as
\begin{eqnarray}
    \chi(x,\textbf{k}_\perp, \lambda_i, \lambda_j) = \bar u (k_1,\lambda_i) \frac{\gamma_5}{\sqrt{2}\sqrt{{M_{\pi}}^2-(m^2_u-m^2_{\bar d}})} v(k_2,\lambda_j) \, .
\end{eqnarray}
Here, $u$ and $v$ denote the Dirac spinors. Both approaches lead to the same spin wave function. The spin wave function for different helicities is given by \cite{Qian:2008px}
\begin{equation}
\left\{
  \begin{array}{lll}
    \chi(x,\mathbf{k}_\perp,\uparrow,\uparrow)&=&\frac{1}{\sqrt{2}}\omega^{-1}(-\textbf{k}^L),\\
    \chi(x,\mathbf{k}_\perp,\uparrow,\downarrow)&=&\frac{1}{\sqrt{2}}\omega^{-1}(x_2m_u+x_1 m_{\bar d}),\\
    \chi(x,\mathbf{k}_\perp,\downarrow,\uparrow)&=&\frac{1}{\sqrt{2}}\omega^{-1}(-x_2m_u-x_1 m_{\bar d}),\\
    \chi(x,\mathbf{k}_\perp,\downarrow,\downarrow)&=&\frac{1}{\sqrt{2}}\omega^{-1}(-\textbf{k}^{R}),
  \end{array}
\right.
\end{equation}
where $\omega=\sqrt{x_1x_2[M_\pi^2-(m_u-m_{\bar d})^2]}$. The two-particle Fock-state in Eq.~(\ref{meson}) can be expressed in terms of LC wave functions (LCWFs), incorporating all possible helicity configurations of its constituent quark and antiquark, as
\begin{eqnarray}
\ket{M (P^+,\textbf{P}_\perp,S_z=0)}&=&\int \frac{{ {\rm d}x  \rm d}^2\textbf{k}_\perp}{2 (2 \pi)^3 \sqrt{x_1 x_2}}\big[\Psi_\pi(x,\textbf{k}_\perp, \uparrow, \uparrow)\ket{x P^+, x_1 P_\perp+\textbf{k}_\perp, \uparrow, \uparrow} \nonumber\\
&&+\Psi_{\pi}(x,\textbf{k}_\perp, \downarrow, \downarrow)\ket{x P^+,x_1 P_\perp+ \textbf{k}_\perp, \downarrow, \downarrow}+\Psi_{\pi}(x,\textbf{k}_\perp, \downarrow, \uparrow)\nonumber\\
&&\ket{x P^+,x_1 P_\perp+ \textbf{k}_\perp, \downarrow, \uparrow}+\Psi_{\pi}(x,\textbf{k}_\perp, \uparrow, \downarrow)\ket{x P^+,x_1 P_\perp+ \textbf{k}_\perp, \uparrow, \downarrow}\big].
\label{overlap}
\end{eqnarray}
As we are calculating the GTMDs of the pion, we have taken the quark mass as $m_u = m_{\bar{d}} = m = 0.2$ GeV. The harmonic scale parameter  $\beta$ is taken to be $0.410$ GeV \cite{Puhan:2023ekt}. These input parameters have been fitted with pion mass through the variational principle method. 
\section{Generalized Transverse Momentum-Dependent Parton Distributions (GTMDs)}
For a spin-$0$ pseudoscalar meson, the valence quark GTMDs can be expressed in terms of the quark-quark correlator up to twist-$4$ as \cite{Meissner:2008ay}
\begin{eqnarray}
\label{coor}
      \Phi_q^{[\Gamma]}&=&\frac{1}{2}\int\frac{\mathrm{d}z^-\mathrm{d}^2\vec{z}_\perp}{2(2\pi)^3}e^{i \textbf{k}\cdot z} \nonumber\\
      && \langle M (P^{*+},\textbf{P}^{*}_\perp,S_z=0)|\bar\psi(-z/2)\Gamma\mathcal{W}(-z/2,z/2)\psi(z/2)|M (P^+,\textbf{P}_\perp,S_z=0) \rangle|_{z^+=0} ,
\end{eqnarray}
where $\Phi_q^{[\Gamma]}=\Phi_q^{[\Gamma]}(x, k_{\perp}, \Delta_{\perp},\bfk\cdot\Dp)$ represents different GTMDs at zero skewness, defined as the longitudinal momentum fraction difference between the initial and final hadron, is given by $\xi=-\frac{(P^+-P^{*+})}{(P^++P^{*+})}=-\frac{\Delta^+}{(P^++P^{*+})}$. The states  $M (P^{*+},\textbf{P}^{*}_\perp,S_z=0)$ and $M (P^+,\textbf{P}_\perp,S_z=0)$ correspond to the final and initial hadronic states, respectively. The momentum transfer between the initial and final states is given by $\Delta=P^*-P$. However, in this work, we take $\xi=0$. The four-vector $z=(z^+,z^-,z_\perp)$ represents the position, and $\psi(z/2)$ is the quark field operator. The Wilson line $\mathcal{W}(-z/2,z/2)$ ensures gauge invariance of the bilocal quark field operators in the correlation functions \cite{Bacchetta:2020vty} and determines the path of the quark field. In this work, we set $\mathcal{W}(0,z) = 1$. In addition, we have chosen the planes $\bfk$ and $\Dp$ as parallel. The matrix $\Gamma$ represents the gamma matrices, whose specific form depends on the choice of GTMDs. The structure of $\Phi_q^{[\Gamma]}$ at different orders of twist and the hard scale $Q$ can be written as 
\begin{eqnarray}
    \Phi_q^{[\Gamma]}= \Phi_{q(twist-2)}^{[\Gamma]}+\frac{1}{Q}\Phi_{q(twist-3)}^{[\Gamma]}+\frac{1}{Q^2}\Phi_{q(twist-4)}^{[\Gamma]}+.....
\end{eqnarray}
Substituting the meson wave function from Eq.~(\ref{overlap}) into the above equation, we obtain
\begin{align}\label{gtmd}
      \Phi_q^{[\Gamma]}
      &= \frac{1}{2 (2\pi)^{3}} \sum_{\lambda_{i}} \sum_{\lambda_{j}}\left[\Psi^{*}_{\pi}\left(x, \boldsymbol{k}_{\perp}^{\prime\prime},\lambda_{j},\uparrow\right) \Psi_{\pi}\left(x, \boldsymbol{k}_{\perp}^{\prime},\lambda_{i},\uparrow\right)
    	\right.\left.+\Psi^{*}_{\pi}\left(x, \boldsymbol{k}_{\perp}^{\prime\prime},\lambda_{f},\downarrow\right) \Psi_{\pi}\left(x, \boldsymbol{k}_{\perp}^{\prime},\lambda_{i},\downarrow\right) \right] \nonumber
    	\\&\times \frac{\bar u_{\lambda_{j}}\left(x P^{+}, \boldsymbol{k}_{\perp}+\frac{\boldsymbol{\Delta}_{\perp}}{2}\right)  \Gamma u_{\lambda_{i}}\left(x P^{+}, \boldsymbol{k}_{\perp}-\frac{\boldsymbol{\Delta}_{\perp}}{2}\right)}{2 x P^{+}}.
\end{align}
Here, $\lambda_i$ and $\lambda_j$ denote the helicities of the initial and final quarks, respectively. The quantity $u_\lambda$ represents the Dirac spinor. The transverse momenta of the active quark in the final and initial states are denoted by $\bfk^{\prime\prime}$ and $\bfk^{\prime}$, respectively, and can be defined as
\begin{eqnarray}
    \bfk^{\prime\prime}= \bfk +(1-x)\frac{\Dp}{2} 
    \nonumber\\
     \bfk^{\prime}= \bfk -(1-x)\frac{\Dp}{2}.
\end{eqnarray}
Now, the choice of $\Gamma$ determines the twist of the GTMDs. For twist-$2$ GTMDs, we have  
$\Gamma = (\gamma^+, \gamma^+\gamma_5, \iota \sigma^{j+}\gamma_5)$.  
For twist-$3$ GTMDs, $\Gamma$ takes the forms  
$\Gamma = (1, \gamma_5, \gamma^j, \gamma^j\gamma_5, \iota \sigma^{ij}\gamma_5, \iota \sigma^{+-}\gamma_5)$.  
Similarly, for twist-$4$ GTMDs, the choice  
$\Gamma = (\gamma^-, \gamma^-\gamma_5, \iota \sigma^{j-}\gamma_5)$  
yields results analogous to those at twist-$2$. The explicit results and forms of GTMDs for different twists are discussed below.
%
%
%
\subsection{Twist-2 GTMDs}
\par
\begin{figure}[ht]
		\centering
		\begin{minipage}[c]{1\textwidth}\begin{center}				(a)\includegraphics[width=.45\textwidth]{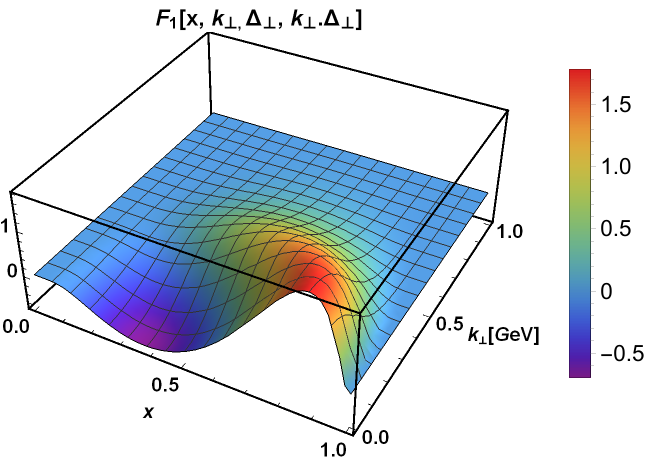}
				(b)\includegraphics[width=.45\textwidth]{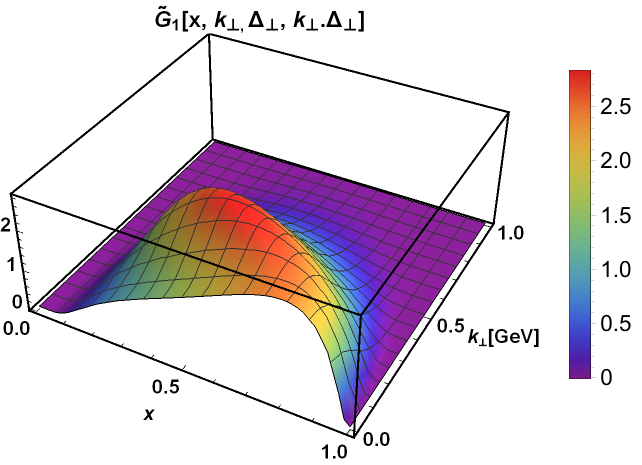}
			\end{center}
		\end{minipage}
		\begin{minipage}[c]{1\textwidth}\begin{center}
				(c)\includegraphics[width=.45\textwidth]{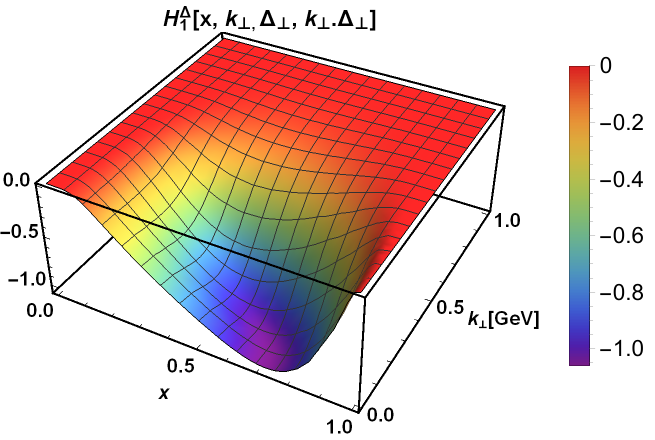}
			\end{center}
		\end{minipage}
		\caption{(Color online) Twist-2 quark  GTMDs plotted as a function of longitudinal momentum fraction $x$ and transverse momentum $\bfk$ at a fixed value of $\Delta_\perp=1$ GeV.}
		\label{realtmds1}
	\end{figure}
\begin{figure}[ht]
		\centering
		\begin{minipage}[c]{1\textwidth}\begin{center}
				(a)\includegraphics[width=.45\textwidth]{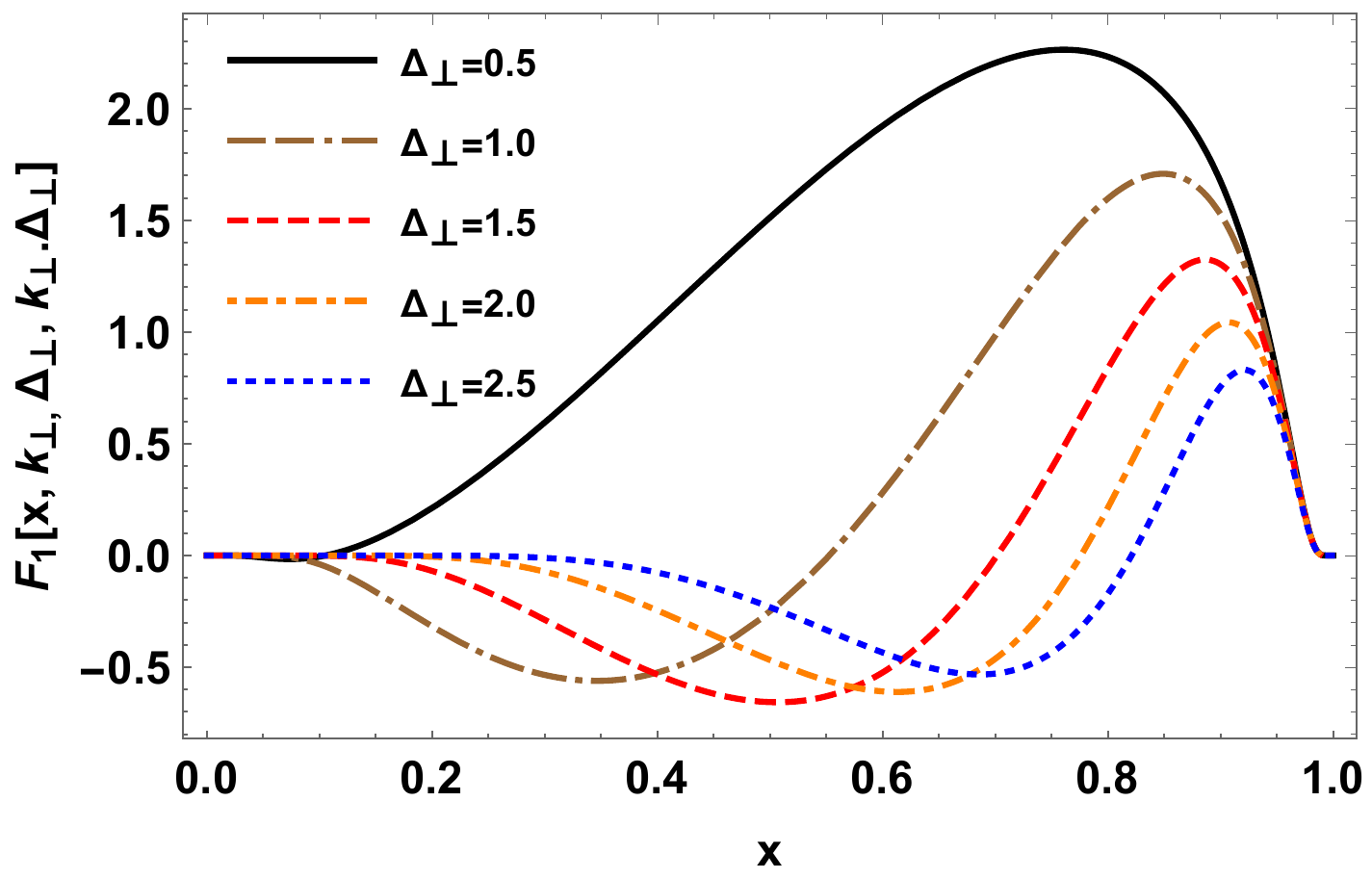}
				(b)\includegraphics[width=.45\textwidth]{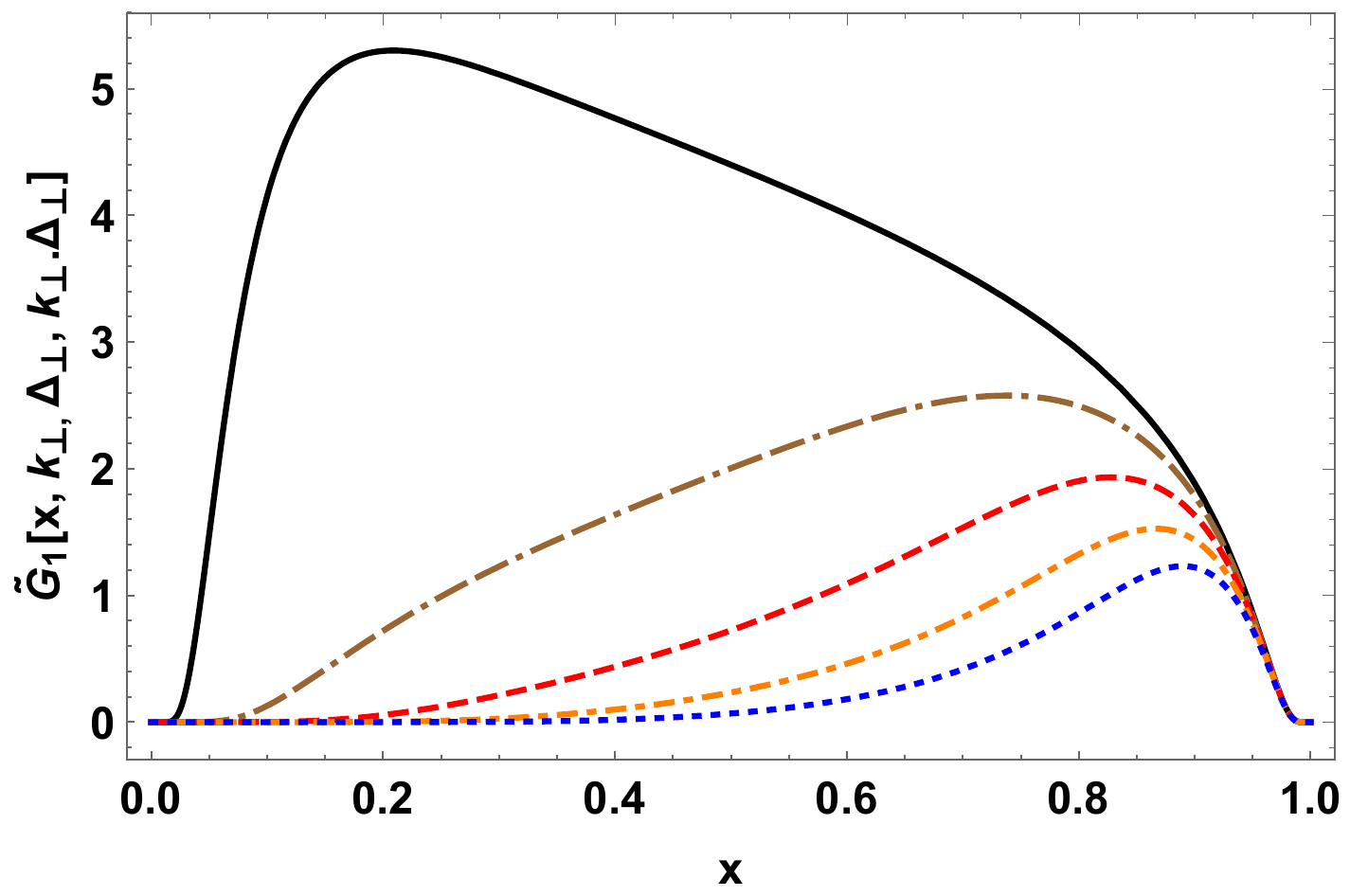}
			\end{center}
		\end{minipage}
		\begin{minipage}[c]{1\textwidth}\begin{center}
				(c)\includegraphics[width=.45\textwidth]{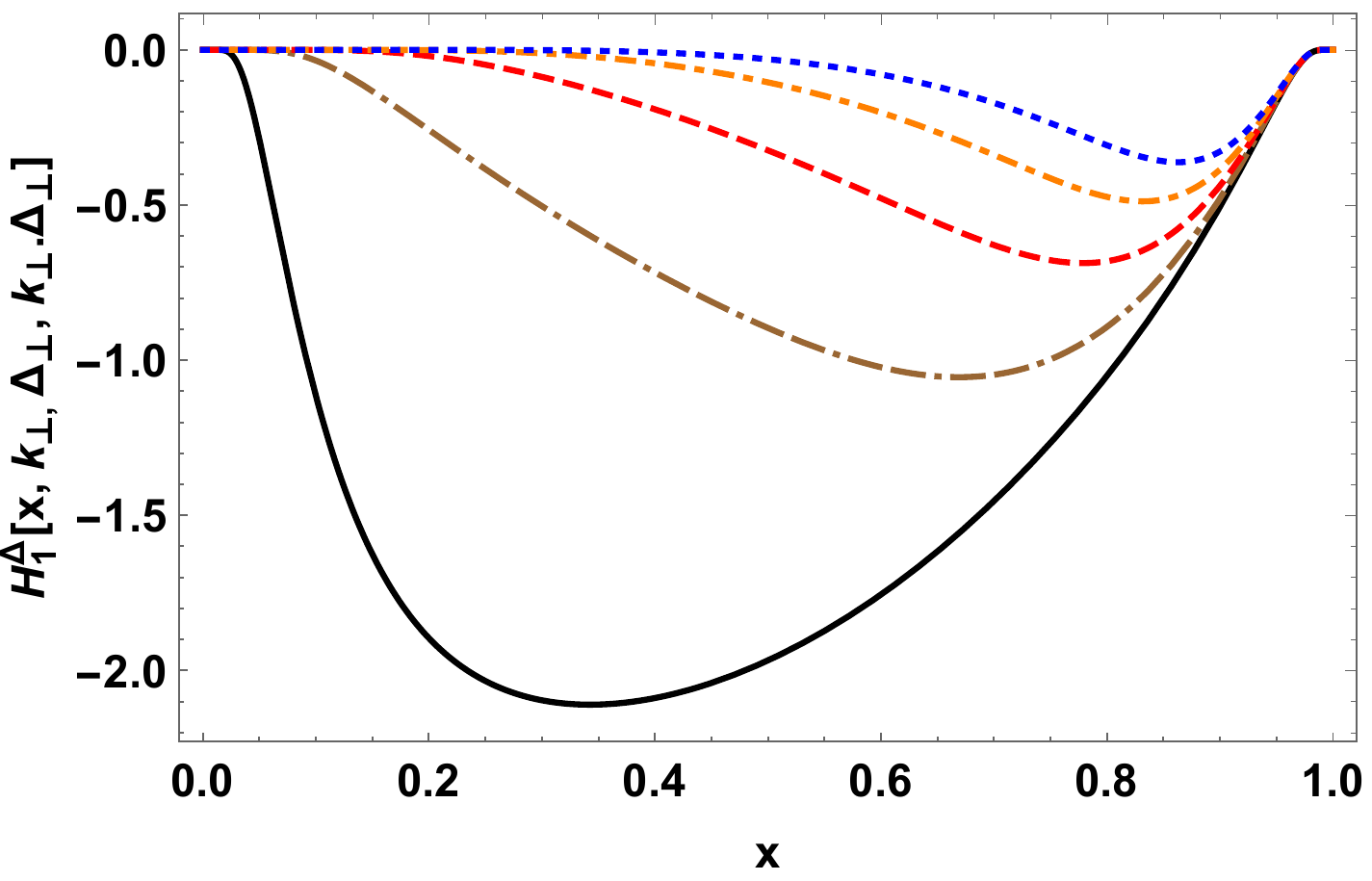}
			\end{center}
		\end{minipage}
		\caption{(Color online) Twist-2 valence quark GTMDs plotted as a function of longitudinal momentum fraction $x$ at a fixed transverse momentum $\mathbf{k}_\perp=0.1$ GeV and $\Delta_\perp=0.5, 1, 1.5, 2,$ and $2.5$ GeV.}
		\label{2drealtmds1}
	\end{figure}
         \begin{figure}[ht]
		\centering				\includegraphics[width=.45\textwidth]{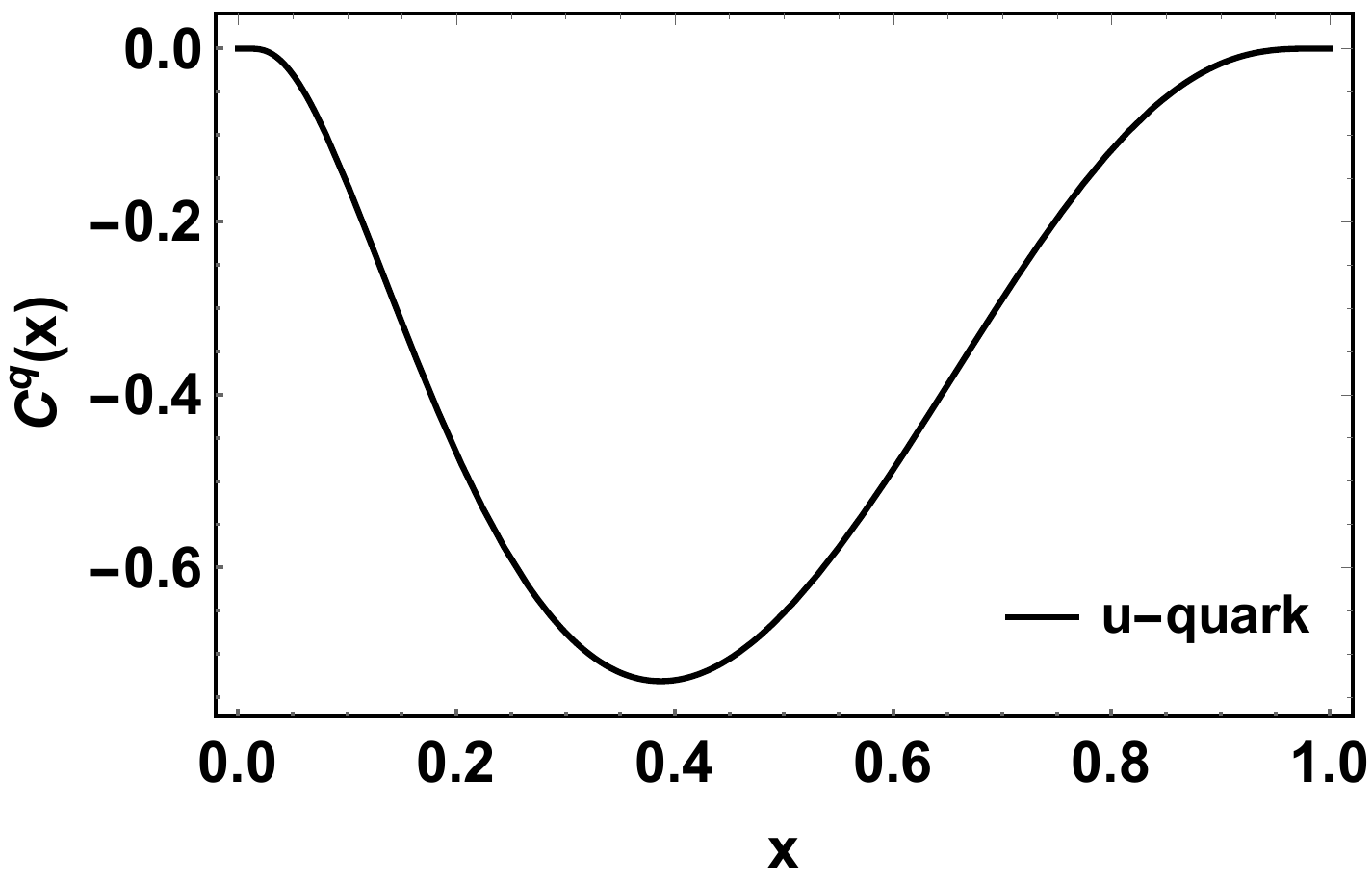}
				
		\caption{(Color online) Spin-orbit correlation of the $u$-quark in the pion.}
		\label{soc}
	\end{figure}
At leading-order twist, there are a total of $4$ GTMDs for spin--$0$ pseudoscalar mesons, each corresponding to the various forms of the Dirac matrix gamma structure as \cite{Meissner:2008ay}
\begin{eqnarray}
\Phi_q^{[\gamma^+]}       
& = & F_1 (x, k_{\perp}, \Delta_{\perp},\bfk\cdot\Dp)\,, \vphantom{\frac{1}{1}} 
\label{e:gtmd_1a} \\
\Phi_q^{[\gamma^+ \gamma_5]}
& = & 
 \frac{i\varepsilon_\perp^{ij} k_\perp^i \Delta_\perp^j}{M_\pi^2} \, \tilde{G}_1 (x, k_{\perp}, \Delta_{\perp},\bfk\cdot\Dp)\,, \\
\Phi_q^{[i\sigma^{j+}\gamma_5]}
& = & 
 \frac{i\varepsilon_\perp^{ij} k_\perp^i}{M_\pi} \, H^k_1(x, k_{\perp}, \Delta_{\perp},\bfk\cdot\Dp) 
+\frac{i\varepsilon_\perp^{ij} \Delta_\perp^i}{M_\pi} \, H^\Delta_1(x, k_{\perp}, \Delta_{\perp},\bfk\cdot\Dp) \,. 
\label{e:gtmd_3a}
\end{eqnarray}
With the antisymmetric tensor defined as \( \varepsilon_\perp^{ij} = \varepsilon^{-+ij} \) and \( \varepsilon^{0123} = 1 \), and the sigma matrix given by \( \sigma^{ij} = \frac{i}{2} [\gamma^i, \gamma^j] \), the GTMDs can now be expressed in terms of LFWFs using Eq.~\ref{gtmd} as
\begin{eqnarray}
  \Phi_q^{[\gamma^+]}
  &=& \frac{1}{2 (2\pi)^{3}} \Bigg[\Psi^{*}_{\pi}\left(x, \boldsymbol{k}_{\perp}^{\prime\prime},\uparrow,\uparrow\right) \Psi_{\pi}\left(x, \boldsymbol{k}_{\perp}^{\prime},\uparrow,\uparrow\right) 
  + \Psi^{*}_{\pi}\left(x, \boldsymbol{k}_{\perp}^{\prime\prime},\downarrow,\uparrow\right) \Psi_{\pi}\left(x, \boldsymbol{k}_{\perp}^{\prime},\downarrow,\uparrow\right) \nonumber \\
  &+& \Psi^*_{\pi}\left(x, \boldsymbol{k}_{\perp}^{\prime\prime},\uparrow,\downarrow\right) \Psi_{\pi}\left(x, \boldsymbol{k}_{\perp}^{\prime},\uparrow,\downarrow\right) 
  + \Psi^{*}_{\pi}\left(x, \boldsymbol{k}_{\perp}^{\prime\prime},\downarrow,\downarrow\right) \Psi_{\pi}\left(x, \boldsymbol{k}_{\perp}^{\prime},\downarrow,\downarrow\right) \Bigg]\, ,
\end{eqnarray}
  \begin{eqnarray}
  \Phi_q^{[\gamma^+\gamma_5]}
  &=& \frac{1}{2 (2\pi)^{3}} \Bigg[\Psi^{*}_{\pi}\left(x, \boldsymbol{k}_{\perp}^{\prime\prime},\uparrow,\uparrow\right) \Psi_{\pi}\left(x, \boldsymbol{k}_{\perp}^{\prime},\uparrow,\uparrow\right) 
  - \Psi^{*}_{\pi}\left(x, \boldsymbol{k}_{\perp}^{\prime\prime},\downarrow,\uparrow\right) \Psi_{\pi}\left(x, \boldsymbol{k}_{\perp}^{\prime},\downarrow,\uparrow\right) \nonumber \\
  &+& \Psi^{*}_{\pi}\left(x, \boldsymbol{k}_{\perp}^{\prime\prime},\uparrow,\downarrow\right) \Psi_{\pi}\left(x, \boldsymbol{k}_{\perp}^{\prime},\uparrow,\downarrow\right) 
  - \Psi^{*}_{\pi}\left(x, \boldsymbol{k}_{\perp}^{\prime\prime},\downarrow,\downarrow\right) \Psi_{\pi}\left(x, \boldsymbol{k}_{\perp}^{\prime},\downarrow,\downarrow\right) \Bigg]\, ,
\\
  \Phi_q^{[i\sigma^{2+}\gamma_5]}
  &=& \frac{1}{2 (2\pi)^{3}} \Bigg[ (i) \Psi^{*}_{\pi}\left(x, \boldsymbol{k}_{\perp}^{\prime\prime},\downarrow,\uparrow\right) \Psi_{\pi}\left(x, \boldsymbol{k}_{\perp}^{\prime},\uparrow,\uparrow\right) -(i)\Psi^{*}_{\pi}\left(x, \boldsymbol{k}_{\perp}^{\prime\prime},\uparrow,\uparrow\right) \Psi_{\pi}\left(x, \boldsymbol{k}_{\perp}^{\prime},\downarrow,\uparrow\right) \nonumber \\
  &-& (i)\Psi^{*}_{\pi}\left(x, \boldsymbol{k}_{\perp}^{\prime\prime},\uparrow,\downarrow\right) \Psi_{\pi}\left(x, \boldsymbol{k}_{\perp}^{\prime},\downarrow,\downarrow\right) + (i)\Psi^{*}_{\pi}\left(x, \boldsymbol{k}_{\perp}^{\prime\prime},\downarrow,\downarrow\right) \Psi_{\pi}\left(x, \boldsymbol{k}_{\perp}^{\prime},\uparrow,\downarrow\right) \Bigg]\, .
\end{eqnarray}
Now, the explicit form of twist-$2$ GTMDs at zero skewness is found to be  
\begin{eqnarray}
    F_1 & = & \frac{1}{2 (2 \pi)^3}\Bigg[ (m_u^2+\bfk^2-\frac{(1-x)^2\Delta_\perp^2}{4})\Bigg]\frac{\psi^*(x,\bfk^{\prime\prime})\psi(x,\bfk^{\prime})}{\omega^+\omega^-}\, ,
    \\
    \tilde{G}_1 & =& \frac{M^2_\pi}{2 (2 \pi)^3} (1-x)\frac{\psi^*(x,\bfk^{\prime\prime})\psi(x,\bfk^{\prime})}{\omega^+\omega^-}\, ,
\\
   H^\Delta_1& =& -\frac{M_\pi}{2 (2 \pi)^3} m_u(1-x)\frac{\psi^*(x,\bfk^{\prime\prime})\psi(x,\bfk^{\prime})}{\omega^+\omega^-}\, ,
       \end{eqnarray}
  \begin{eqnarray}
   H^k_1&=&0\, ,
\end{eqnarray}
where
\begin{eqnarray}
    \omega^+=\sqrt{x_1x_2[\frac{(\bfk^{\prime \prime2}+m^2_u)}{x_1x_2}-(m_u-m_{\bar d})^2]}\, , \ \ \ \  \omega^-=\sqrt{x_1x_2[\frac{(\bfk^{\prime2}+m^2_u)}{x_1x_2}-(m_u-m_{\bar d})^2]}\, .
\end{eqnarray}
At the leading twist, there are a total of $4$ GTMDs for the pion, in contrast to the $16$ GTMDs for spin-$1/2$ nucleons. However, at $\xi=0$, only $3$ quark GTMDs remain non-zero. These quark GTMDs are plotted as functions of the longitudinal momentum fraction $x$ and transverse momentum $\bfk$ (in GeV) at a fixed value of $\Delta_\perp=1.0$ GeV in Fig.~\ref{realtmds1}. The unpolarized GTMD $F_1(x, k_{\perp}, \Delta_{\perp},\bfk\cdot\Dp)$ exhibits a positive distribution at $\Delta_\perp=0$, whereas with increase in $\Delta_\perp$, the distribution develops in both positive and negative regions, as shown in Fig.~\ref{2drealtmds1} (a). Additionally, the distribution shifts toward larger values of $x$ while its peak decreases with increasing $\Delta_\perp$. This behavior indicates that as the momentum transfer between the initial and final hadron increases, the momentum fraction carried by the quark decreases. The GTMD $\tilde{G}_1 (x, k_{\perp}, \Delta_{\perp},\bfk\cdot\Dp)$ exhibits positive distributions across the entire range of $x$, $\bfk$ (in GeV), and $\Delta_\perp$ (in GeV). This GTMD carries information about the SOC of the hadrons. The SOC can be calculated as \cite{Kaur:2019jow}
\begin{eqnarray}
C^q=\int dx d^2{\bf k}_\perp \frac{{\bf k}^2_\perp}{M_\pi^2} \tilde{G}_1(x, k_{\perp}, 0,0).
\end{eqnarray} 
The coefficient $C^q$ is found to be $-0.36$, indicating that the quark spin and orbital angular momentum (OAM) are anti-aligned. Similar results were obtained in the Nambu–Jona-Lasinio (NJL) model ($C^q = -0.491$) and the Dyson-Schwinger equation (DSE) model ($C^q = -0.374$) for the pion case \cite{Zhang:2024adr}. The distribution of $C^q$ as a function of $x$ is shown in Fig.~\ref{soc}, where it exhibits a negative trend, as expected for a pseudoscalar meson \cite{Acharyya:2024enp,Lorce:2025ayr}. Since the skewness parameter is $\xi=0$ in our case, the GTMD $H^k_1(x, k_{\perp}, \Delta_{\perp},\bfk\cdot\Dp)$ comes out to be zero. The GTMD $H^\Delta_1(x, k_{\perp}, \Delta_{\perp},\bfk\cdot\Dp)$ exhibits a decreasing trend with increasing $\Delta_\perp$, while the distribution shifts towards higher values of $x$. This behavior suggests that at higher momentum transfer $\Delta_\perp$, the quark distributions become more concentrated near $x=1$. Similar GTMD behavior has also been observed in Refs.~\cite{Ma:2018ysi, Zhang:2020ecj, Meissner:2008ay}. Furthermore, we have analyzed the average transverse momentum $\langle \bfk \rangle$ carried by the valence quark for each GTMD as a function of $\Delta_\perp$, as shown in Fig.~\ref{average}. The average transverse momentum $\langle \bfk \rangle$ is given by
\begin{eqnarray}
    \langle \bfk (\Delta_\perp) \rangle = \frac{\int d x d^2\bfk \bfk Y(x, k_{\perp}, \Delta_{\perp},\bfk\cdot\Dp)}{\int d x d^2\bfk  Y(x, k_{\perp}, \Delta_{\perp},\bfk\cdot\Dp)},
\end{eqnarray}
where $Y$ represents different GTMDs of the pion. In Fig.~\ref{average}~(a), we observe that up to a certain value of $\Delta_\perp$, the average transverse momentum $\langle \bfk \rangle$ increases and then begins to decrease, except for the GTMD $F_1(x, k_{\perp}, \Delta_{\perp},\bfk\cdot\Dp)$. The quark GTMDs $H^k_1(x, k_{\perp}, \Delta_{\perp},\bfk\cdot\Dp)$ and $\tilde{G}_1(x, k_{\perp}, \Delta_{\perp},\bfk\cdot\Dp)$ exhibit nearly similar kind of trends for average transverse momentum, whereas $\langle \bfk \rangle$ for $F_1(x, k_{\perp}, \Delta_{\perp},\bfk\cdot\Dp)$ approaches zero beyond $\Delta_\perp = 4$~GeV. The $F_1(x, k_{\perp}, \Delta_{\perp},\bfk\cdot\Dp)$ GTMD carries the highest average transverse momentum compared to other GTMDs at $\Delta_\perp=0$.
\subsection{Twist-3 GTMDs}
At the subleading twist, or twist-$3$, GTMDs are expressed using various structures of the Dirac matrix as \cite{Meissner:2008ay}
\begin{eqnarray}
\Phi_q^{[1]} & = & \frac{M_\pi}{P^+}
 \bigg[ E_2 (x, k_{\perp}, \Delta_{\perp},\bfk\cdot\Dp)\bigg] \,,
\\
\Phi_q^{[\gamma_5]} & = & \frac{M_\pi}{P^+}
 \bigg[ \frac{i\varepsilon_\perp^{ij} k_\perp^i \Delta_\perp^j}{M_\pi^2} \, \tilde{E}_2(x, k_{\perp}, \Delta_{\perp},\bfk\cdot\Dp) \bigg] \,,
\\
\Phi_q^{[\gamma^j]} & = & \frac{M_\pi}{P^+}
 \bigg[ \frac{k_\perp^j}{M_\pi} \, F^k_2 (x, k_{\perp}, \Delta_{\perp},\bfk\cdot\Dp) + \frac{\Delta_\perp^j}{M_\pi} \, F^\Delta_2 (x, k_{\perp}, \Delta_{\perp},\bfk\cdot\Dp)\bigg] \,,
\\
\Phi_q^{[\gamma^j\gamma_5]} & = & \frac{M_\pi}{P^+}
 \bigg[ \frac{i\varepsilon_\perp^{ij} k_\perp^i}{M_\pi} \, G^k_2(x, k_{\perp}, \Delta_{\perp},\bfk\cdot\Dp)
       +\frac{i\varepsilon_\perp^{ij} \Delta_\perp^i}{M_\pi} \, G^\Delta_2(x, k_{\perp}, \Delta_{\perp},\bfk\cdot\Dp) \bigg] \,,
\\
\Phi_q^{[i\sigma^{ij}\gamma_5]} & = & \frac{M_\pi}{P^+}
 \bigg[ i\varepsilon_\perp^{ij} H_2(x, k_{\perp}, \Delta_{\perp},\bfk\cdot\Dp) \bigg] \,,
\\
\Phi_q^{[i\sigma^{+-}\gamma_5]} & = & \frac{M_\pi}{P^+}
 \bigg[ \frac{i\varepsilon_\perp^{ij} k_\perp^i \Delta_\perp^j}{M_\pi^2} \, \tilde{H}_2 (x, k_{\perp}, \Delta_{\perp},\bfk\cdot\Dp)\bigg] \,.
\end{eqnarray}
The overlap form of twist-$3$ GTMDs can be expressed as  
\begin{eqnarray}
    \Phi_q^{[1]}
    &=& \frac{1}{2 (2\pi)^{3}} \Bigg[\frac{m_u}{x P^+}\Bigg(\Psi^{*}_{\pi}\left(x, \boldsymbol{k}_{\perp}^{\prime\prime},\uparrow,\uparrow\right) \Psi_{\pi}\left(x, \boldsymbol{k}_{\perp}^{\prime},\uparrow,\uparrow\right) 
  + \Psi^{*}_{\pi}\left(x, \boldsymbol{k}_{\perp}^{\prime\prime},\downarrow,\uparrow\right) \Psi_{\pi}\left(x, \boldsymbol{k}_{\perp}^{\prime},\downarrow,\uparrow\right) \nonumber \\
  &+& \Psi^*_{\pi}\left(x, \boldsymbol{k}_{\perp}^{\prime\prime},\uparrow,\downarrow\right) \Psi_{\pi}\left(x, \boldsymbol{k}_{\perp}^{\prime},\uparrow,\downarrow\right) 
  + \Psi^{*}_{\pi}\left(x, \boldsymbol{k}_{\perp}^{\prime\prime},\downarrow,\downarrow\right) \Psi_{\pi}\left(x, \boldsymbol{k}_{\perp}^{\prime},\downarrow,\downarrow\right)\Bigg)\nonumber\\ &-& 
  \frac{(\Delta_1+ i \Delta_2)}{2 x P^+}\Bigg( \Psi^{*}_{\pi}\left(x, \boldsymbol{k}_{\perp}^{\prime\prime},\downarrow,\uparrow\right) \Psi_{\pi}\left(x, \boldsymbol{k}_{\perp}^{\prime},\uparrow,\uparrow\right)+\Psi^{*}_{\pi}\left(x, \boldsymbol{k}_{\perp}^{\prime\prime},\downarrow,\downarrow\right) \Psi_{\pi}\left(x, \boldsymbol{k}_{\perp}^{\prime},\uparrow,\downarrow\right)\Bigg)\nonumber\\ &+& 
  \frac{(\Delta_1- i \Delta_2)}{2 x P^+}\Bigg( \Psi^{*}_{\pi}\left(x, \boldsymbol{k}_{\perp}^{\prime\prime},\uparrow,\uparrow\right) \Psi_{\pi}\left(x, \boldsymbol{k}_{\perp}^{\prime},\downarrow,\uparrow\right)+\Psi^{*}_{\pi}\left(x, \boldsymbol{k}_{\perp}^{\prime\prime},\uparrow,\downarrow\right) \Psi_{\pi}\left(x, \boldsymbol{k}_{\perp}^{\prime},\downarrow,\downarrow\right)\Bigg)\Bigg]\, ,
  \\
  \Phi_q^{[\gamma_5]}
  & = & \frac{-1}{2 (2\pi)^{3}} \Bigg[
  \frac{(\Delta_1+ i \Delta_2)}{2 x P^+}\Bigg( \Psi^{*}_{\pi}\left(x, \boldsymbol{k}_{\perp}^{\prime\prime},\downarrow,\uparrow\right) \Psi_{\pi}\left(x, \boldsymbol{k}_{\perp}^{\prime},\uparrow,\uparrow\right)+\Psi^{*}_{\pi}\left(x, \boldsymbol{k}_{\perp}^{\prime\prime},\downarrow,\downarrow\right) \Psi_{\pi}\left(x, \boldsymbol{k}_{\perp}^{\prime},\uparrow,\downarrow\right)\Bigg)\nonumber\\ &+& 
  \frac{(\Delta_1- i \Delta_2)}{2 x P^+}\Bigg( \Psi^{*}_{\pi}\left(x, \boldsymbol{k}_{\perp}^{\prime\prime},\uparrow,\uparrow\right) \Psi_{\pi}\left(x, \boldsymbol{k}_{\perp}^{\prime},\downarrow,\uparrow\right)+\Psi^{*}_{\pi}\left(x, \boldsymbol{k}_{\perp}^{\prime\prime},\uparrow,\downarrow\right) \Psi_{\pi}\left(x, \boldsymbol{k}_{\perp}^{\prime},\downarrow,\downarrow\right)\Bigg)\Bigg]\, ,
  \\
  \Phi_q^{[\gamma^2]}
  & = & \frac{1}{2 (2\pi)^{3}} \Bigg[
  \frac{(2 k_2+i \Delta_1)}{2 x P^+}\Bigg( \Psi^{*}_{\pi}\left(x, \boldsymbol{k}_{\perp}^{\prime\prime},\uparrow,\uparrow\right) \Psi_{\pi}\left(x, \boldsymbol{k}_{\perp}^{\prime},\uparrow,\uparrow\right)+\Psi^{*}_{\pi}\left(x, \boldsymbol{k}_{\perp}^{\prime\prime},\uparrow,\downarrow\right) \Psi_{\pi}\left(x, \boldsymbol{k}_{\perp}^{\prime},\uparrow,\downarrow\right)\Bigg)\nonumber\\ &+& 
  \frac{(2 k_2-i \Delta_1)}{2 x P^+}\Bigg( \Psi^{*}_{\pi}\left(x, \boldsymbol{k}_{\perp}^{\prime\prime},\downarrow,\uparrow\right) \Psi_{\pi}\left(x, \boldsymbol{k}_{\perp}^{\prime},\downarrow,\uparrow\right)+\Psi^{*}_{\pi}\left(x, \boldsymbol{k}_{\perp}^{\prime\prime},\downarrow,\downarrow\right) \Psi_{\pi}\left(x, \boldsymbol{k}_{\perp}^{\prime},\downarrow,\downarrow\right)\Bigg)\Bigg]\, ,
      \end{eqnarray}
  \begin{eqnarray}
  \Phi_q^{[\gamma^1\gamma_5]}
  &=& \frac{1}{2 (2\pi)^{3}} \Bigg[\frac{m_u}{x P^+}\Bigg(\Psi^{*}_{\pi}\left(x, \boldsymbol{k}_{\perp}^{\prime\prime},\uparrow,\uparrow\right) \Psi_{\pi}\left(x, \boldsymbol{k}_{\perp}^{\prime},\downarrow,\uparrow\right) 
  + \Psi^{*}_{\pi}\left(x, \boldsymbol{k}_{\perp}^{\prime\prime},\downarrow,\uparrow\right) \Psi_{\pi}\left(x, \boldsymbol{k}_{\perp}^{\prime},\uparrow,\uparrow\right) \nonumber \\
  &+& \Psi^*_{\pi}\left(x, \boldsymbol{k}_{\perp}^{\prime\prime},\uparrow,\downarrow\right) \Psi_{\pi}\left(x, \boldsymbol{k}_{\perp}^{\prime},\downarrow,\downarrow\right) 
  + \Psi^{*}_{\pi}\left(x, \boldsymbol{k}_{\perp}^{\prime\prime},\downarrow,\downarrow\right) \Psi_{\pi}\left(x, \boldsymbol{k}_{\perp}^{\prime},\uparrow,\downarrow\right)\Bigg)\nonumber\\ &+& 
  \frac{(2 k_1-i \Delta_2)}{2 x P^+}\Bigg( \Psi^{*}_{\pi}\left(x, \boldsymbol{k}_{\perp}^{\prime\prime},\uparrow,\uparrow\right) \Psi_{\pi}\left(x, \boldsymbol{k}_{\perp}^{\prime},\uparrow,\uparrow\right)+\Psi^{*}_{\pi}\left(x, \boldsymbol{k}_{\perp}^{\prime\prime},\uparrow,\downarrow\right) \Psi_{\pi}\left(x, \boldsymbol{k}_{\perp}^{\prime},\uparrow,\downarrow\right)\Bigg)\nonumber\\ &-& 
  \frac{(2 k_1+i\Delta_2)}{2 x P^+}\Bigg( \Psi^{*}_{\pi}\left(x, \boldsymbol{k}_{\perp}^{\prime\prime},\downarrow,\uparrow\right) \Psi_{\pi}\left(x, \boldsymbol{k}_{\perp}^{\prime},\downarrow,\uparrow\right)+\Psi^{*}_{\pi}\left(x, \boldsymbol{k}_{\perp}^{\prime\prime},\downarrow,\downarrow\right) \Psi_{\pi}\left(x, \boldsymbol{k}_{\perp}^{\prime},\downarrow,\downarrow\right)\Bigg)\Bigg]\, ,
  \\
 \Phi_q^{[i\sigma^{12}\gamma_5]} 
 & = & \frac{i}{2 (2\pi)^{3}} \Bigg[
  \frac{(k_1-ik_2)}{x P^+}\Bigg( \Psi^{*}_{\pi}\left(x, \boldsymbol{k}_{\perp}^{\prime\prime},\uparrow,\uparrow\right) \Psi_{\pi}\left(x, \boldsymbol{k}_{\perp}^{\prime},\downarrow,\uparrow\right)+\Psi^{*}_{\pi}\left(x, \boldsymbol{k}_{\perp}^{\prime\prime},\uparrow,\downarrow\right) \Psi_{\pi}\left(x, \boldsymbol{k}_{\perp}^{\prime},\downarrow,\downarrow\right)\Bigg)\nonumber\\ &-& 
  \frac{(k_1+ ik_2)}{ x P^+}\Bigg( \Psi^{*}_{\pi}\left(x, \boldsymbol{k}_{\perp}^{\prime\prime},\downarrow,\uparrow\right) \Psi_{\pi}\left(x, \boldsymbol{k}_{\perp}^{\prime},\uparrow,\uparrow\right)+\Psi^{*}_{\pi}\left(x, \boldsymbol{k}_{\perp}^{\prime\prime},\downarrow,\downarrow\right) \Psi_{\pi}\left(x, \boldsymbol{k}_{\perp}^{\prime},\uparrow,\downarrow\right)\Bigg)\Bigg]\, ,
  \\
   \Phi_q^{[i\sigma^{+-}\gamma_5]}
   &=& \frac{1}{2 (2\pi)^{3}} \Bigg[\frac{2 m_u}{x P^+}\Bigg(\Psi^{*}_{\pi}\left(x, \boldsymbol{k}_{\perp}^{\prime\prime},\uparrow,\uparrow\right) \Psi_{\pi}\left(x, \boldsymbol{k}_{\perp}^{\prime},\uparrow,\uparrow\right) 
  - \Psi^{*}_{\pi}\left(x, \boldsymbol{k}_{\perp}^{\prime\prime},\downarrow,\uparrow\right) \Psi_{\pi}\left(x, \boldsymbol{k}_{\perp}^{\prime},\downarrow,\uparrow\right) \nonumber \\
  &+& \Psi^*_{\pi}\left(x, \boldsymbol{k}_{\perp}^{\prime\prime},\uparrow,\downarrow\right) \Psi_{\pi}\left(x, \boldsymbol{k}_{\perp}^{\prime},\uparrow,\downarrow\right) 
  - \Psi^{*}_{\pi}\left(x, \boldsymbol{k}_{\perp}^{\prime\prime},\downarrow,\downarrow\right) \Psi_{\pi}\left(x, \boldsymbol{k}_{\perp}^{\prime},\downarrow,\downarrow\right)\Bigg)\nonumber\\ &-& 
  \frac{2(k_1- i k_2)}{x P^+}\Bigg( \Psi^{*}_{\pi}\left(x, \boldsymbol{k}_{\perp}^{\prime\prime},\uparrow,\uparrow\right) \Psi_{\pi}\left(x, \boldsymbol{k}_{\perp}^{\prime},\downarrow,\uparrow\right)+\Psi^{*}_{\pi}\left(x, \boldsymbol{k}_{\perp}^{\prime\prime},\uparrow,\downarrow\right) \Psi_{\pi}\left(x, \boldsymbol{k}_{\perp}^{\prime},\downarrow,\downarrow\right)\Bigg)\nonumber\\ &-& 
  \frac{2(k_1+ i k_2)}{x P^+}\Bigg( \Psi^{*}_{\pi}\left(x, \boldsymbol{k}_{\perp}^{\prime\prime},\downarrow,\uparrow\right) \Psi_{\pi}\left(x, \boldsymbol{k}_{\perp}^{\prime},\uparrow,\uparrow\right)+\Psi^{*}_{\pi}\left(x, \boldsymbol{k}_{\perp}^{\prime\prime},\downarrow,\downarrow\right) \Psi_{\pi}\left(x, \boldsymbol{k}_{\perp}^{\prime},\uparrow,\downarrow\right)\Bigg)\Bigg]\, .
\end{eqnarray}
The explicit form of twist-$3$ GTMDs is found to be  
\begin{eqnarray}
     E_2 & = & \frac{1}{2 (2 \pi)^3}\frac{m_u}{x M_\pi}\Bigg[ (m_u^2+\bfk^2-\frac{(1-x)^2\Delta_\perp^2}{4}+(1-x)\frac{\Delta^2_\perp}{2})\Bigg]\frac{\psi^*(x,\bfk^{\prime\prime})\psi(x,\bfk^{\prime})}{\omega^+\omega^-}\, ,
    \\
    \tilde{E}_2&=&0 \, ,
    \\
    F^k_2 &=& \frac{1}{2 (2 \pi)^3}\frac{1}{4 x}\Bigg[ 4(m_u^2+\bfk^2-\frac{(1-x)^2\Delta_\perp^2}{4})+2 (1-x)\Delta^2_\perp)\Bigg]\frac{\psi^*(x,\bfk^{\prime\prime})\psi(x,\bfk^{\prime})}{\omega^+\omega^-}\, ,
    \\
   F^\Delta_2 &=&-\frac{1}{2 (2 \pi)^3} \frac{1}{2 x}\Bigg[ (1-x)\bfk.\Delta_\perp\Bigg]\frac{\psi^*(x,\bfk^{\prime\prime})\psi(x,\bfk^{\prime})}{\omega^+\omega^-}\ \, ,
   \\
  G^k_2 &=&\frac{1}{2 (2 \pi)^3} \frac{1}{ x}\Bigg[ (1-x)\bfk.\Delta_\perp\Bigg]\frac{\psi^*(x,\bfk^{\prime\prime})\psi(x,\bfk^{\prime})}{\omega^+\omega^-}\ \, ,
   \\
   G^\Delta_2 &=&  \frac{-1}{2 (2 \pi)^3}\frac{1}{ x}\Bigg[-(m_u^2+\bfk^2-\frac{(1-x)^2\Delta_\perp^2}{4})+2 (1-x)m_u^2+2 (1-x)\bfk^2\Bigg]\frac{\psi^*(x,\bfk^{\prime\prime})\psi(x,\bfk^{\prime})}{\omega^+\omega^-}\, , \nonumber\\
   \\ 
   H_2&=& \frac{1}{2 (2 \pi)^3}\frac{m_u}{ x M_\pi}\Bigg[ (1-x)\bfk.\Delta_\perp\Bigg]\frac{\psi^*(x,\bfk^{\prime\prime})\psi(x,\bfk^{\prime})}{\omega^+\omega^-}\, ,
          \end{eqnarray}
  \begin{eqnarray}
  \tilde{H}_2 &=& 0 .  
\end{eqnarray}
\begin{figure}[ht]
		\centering
		\begin{minipage}[c]{1\textwidth}\begin{center}
				(a)\includegraphics[width=.45\textwidth]{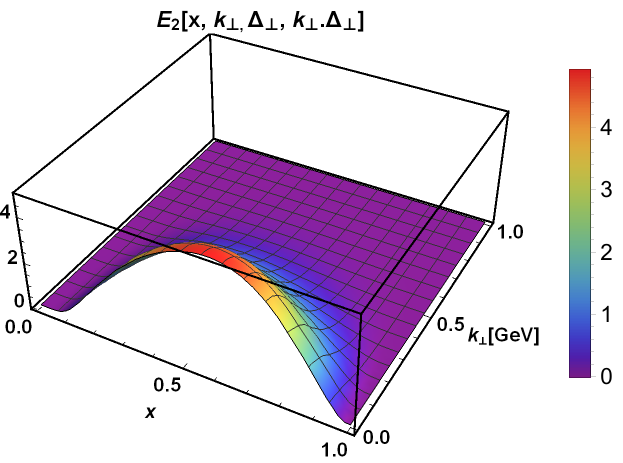}
				(b)\includegraphics[width=.45\textwidth]{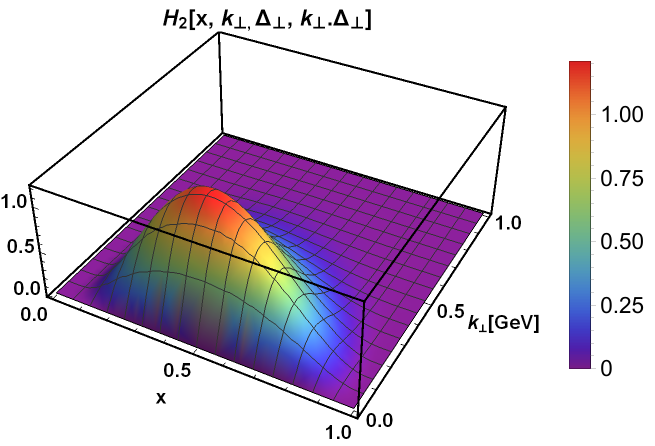}
			\end{center}
		\end{minipage}
		\begin{minipage}[c]{1\textwidth}\begin{center}
				(c)\includegraphics[width=.45\textwidth]{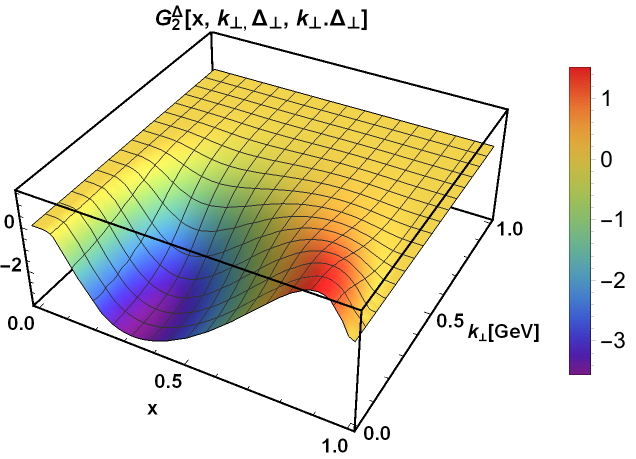}
				(d)\includegraphics[width=.45\textwidth]{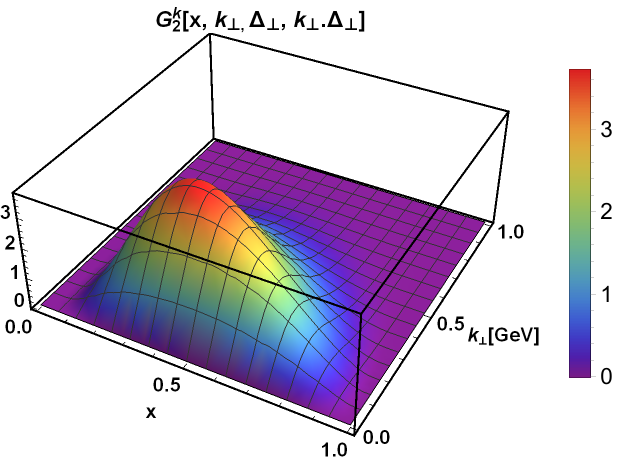}
                \end{center}
		\end{minipage}
        \begin{minipage}[c]{1\textwidth}\begin{center}
				(e)\includegraphics[width=.45\textwidth]{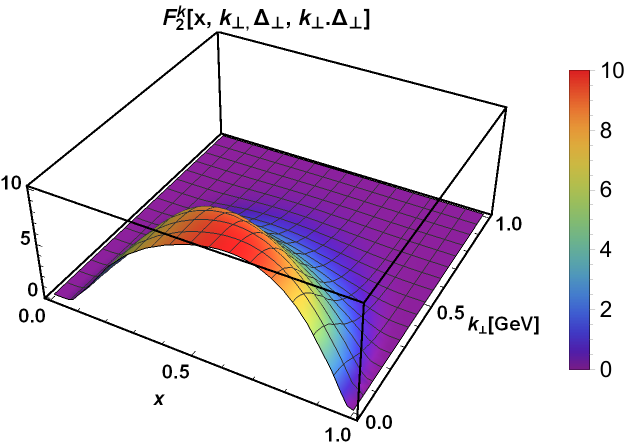}
                (f)\includegraphics[width=.45\textwidth]{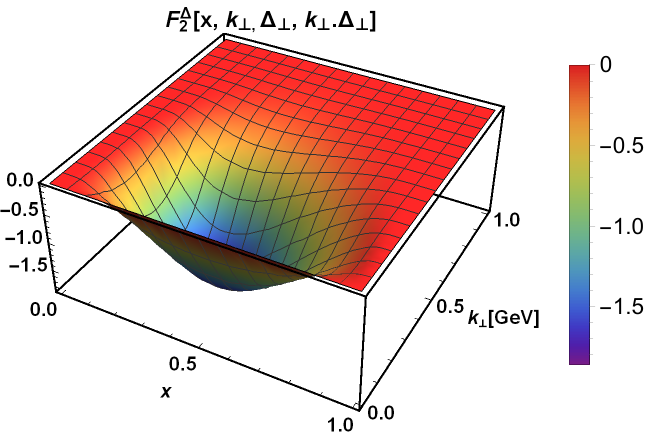}
			\end{center}
		\end{minipage}
		\caption{(Color online) Twist-$3$ quark GTMDs plotted as a function of $x$ and $\bfk$ at fixed values of $\Delta_\perp=1$ and $2$ GeV.}
		\label{t3realtmds2}
	\end{figure}
    \begin{figure}[ht]
		\centering
		\begin{minipage}[c]{1\textwidth}\begin{center}
				(a)\includegraphics[width=.45\textwidth]{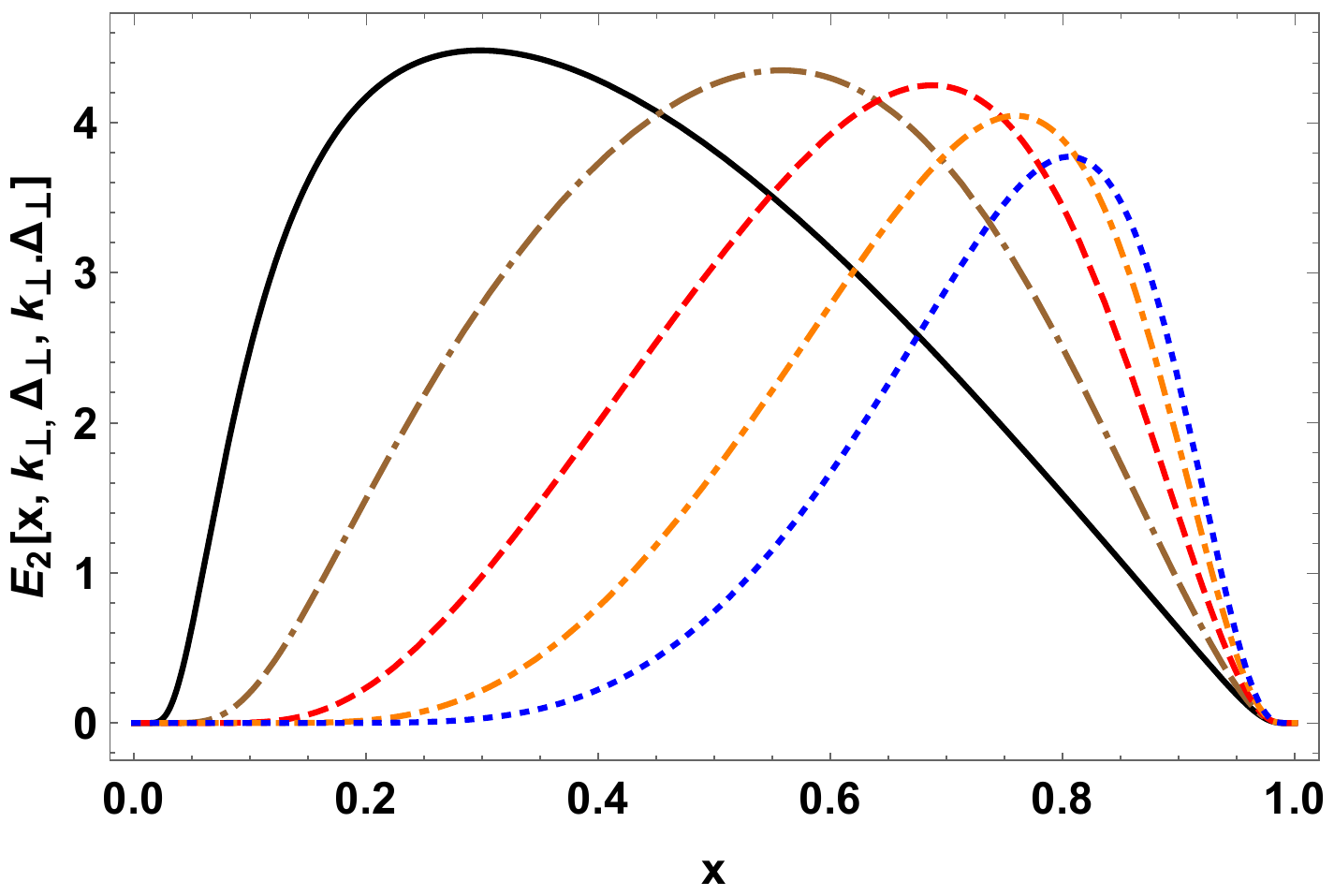}
				(b)\includegraphics[width=.45\textwidth]{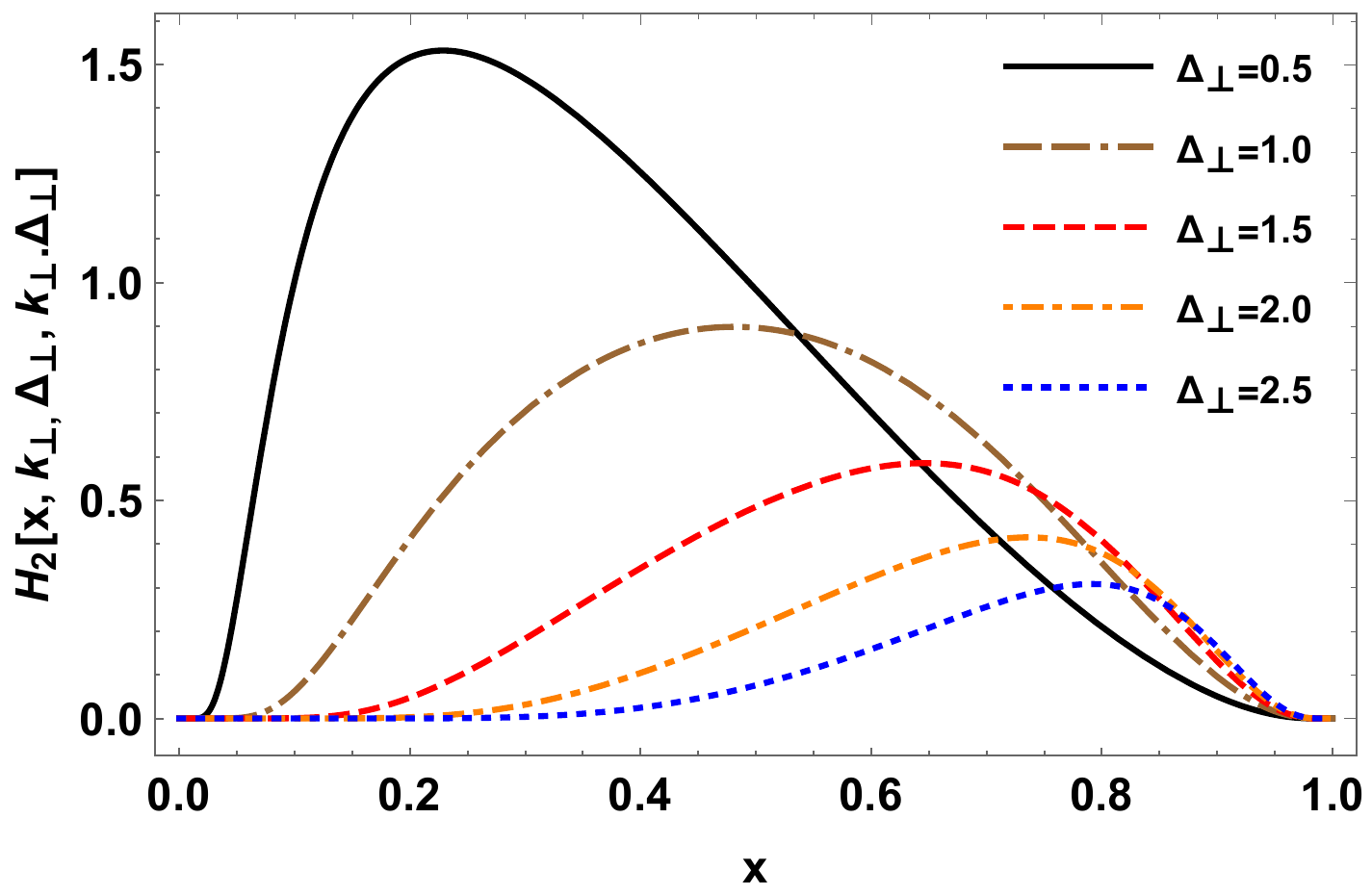}
			\end{center}
		\end{minipage}
		\begin{minipage}[c]{1\textwidth}\begin{center}
				(c)\includegraphics[width=.45\textwidth]{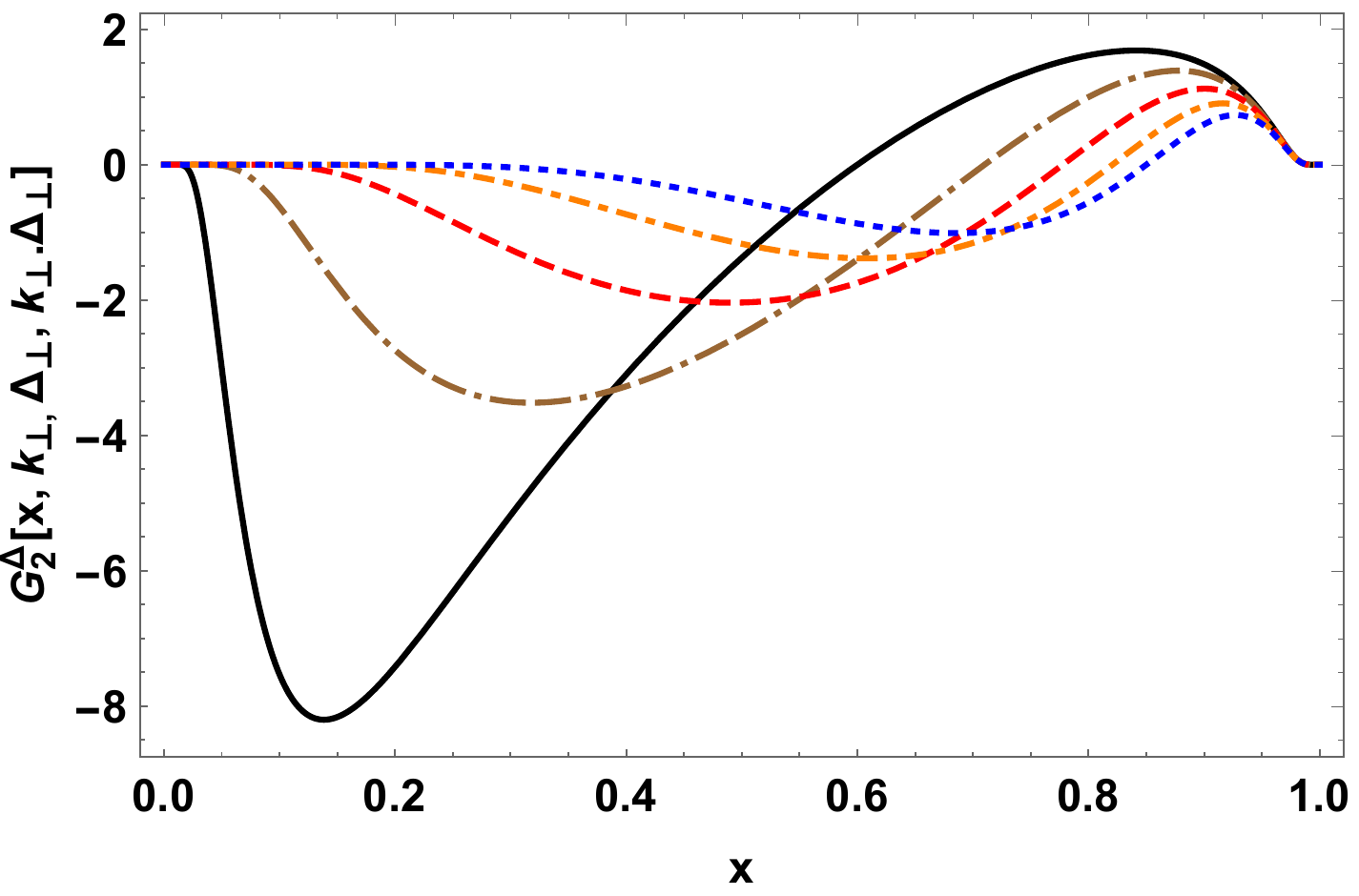}
                (c)\includegraphics[width=.45\textwidth]{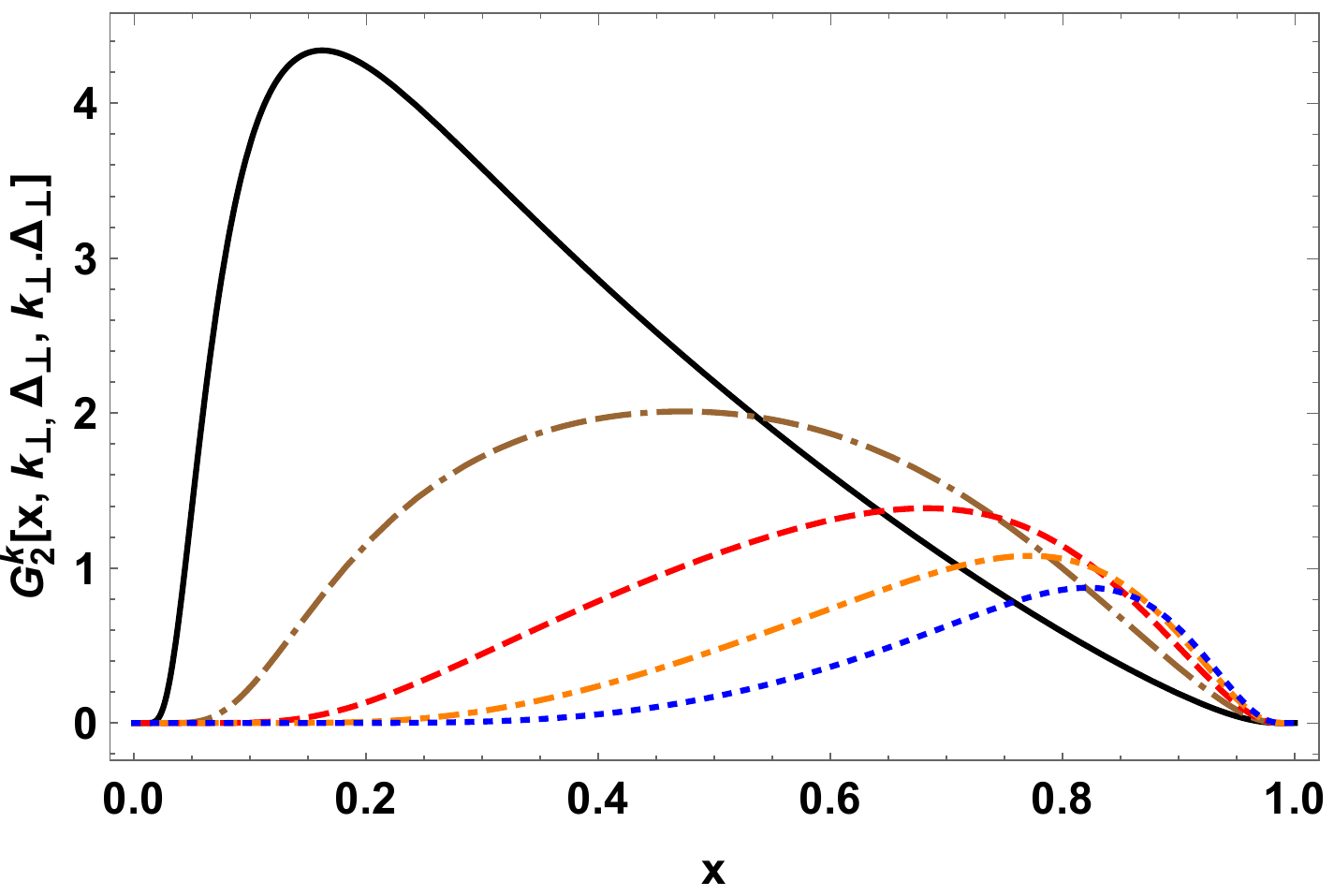}
                \end{center}
		\end{minipage}
        \begin{minipage}[c]{1\textwidth}\begin{center}
				(e)\includegraphics[width=.45\textwidth]{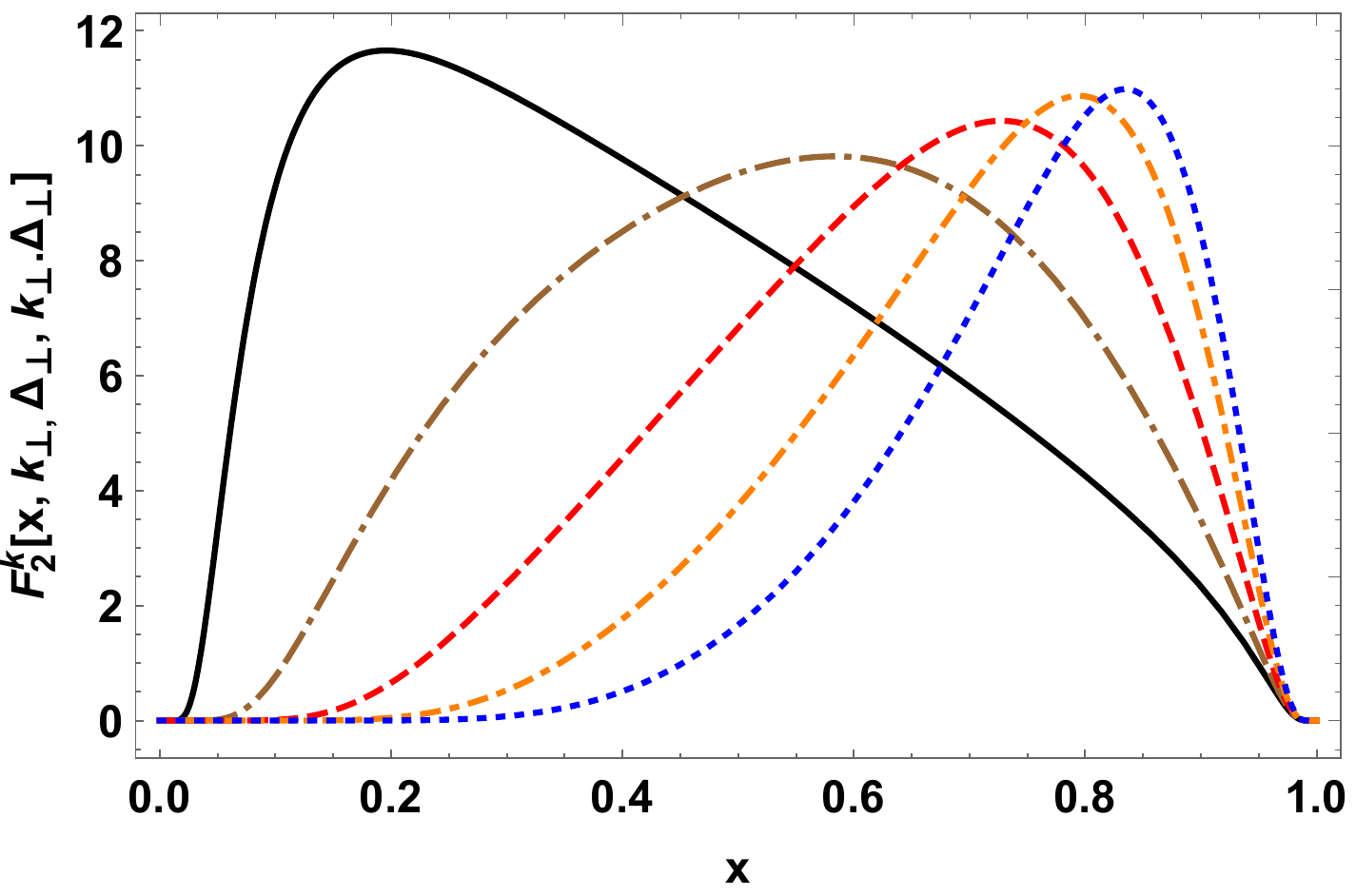}
				(f)\includegraphics[width=.45\textwidth]{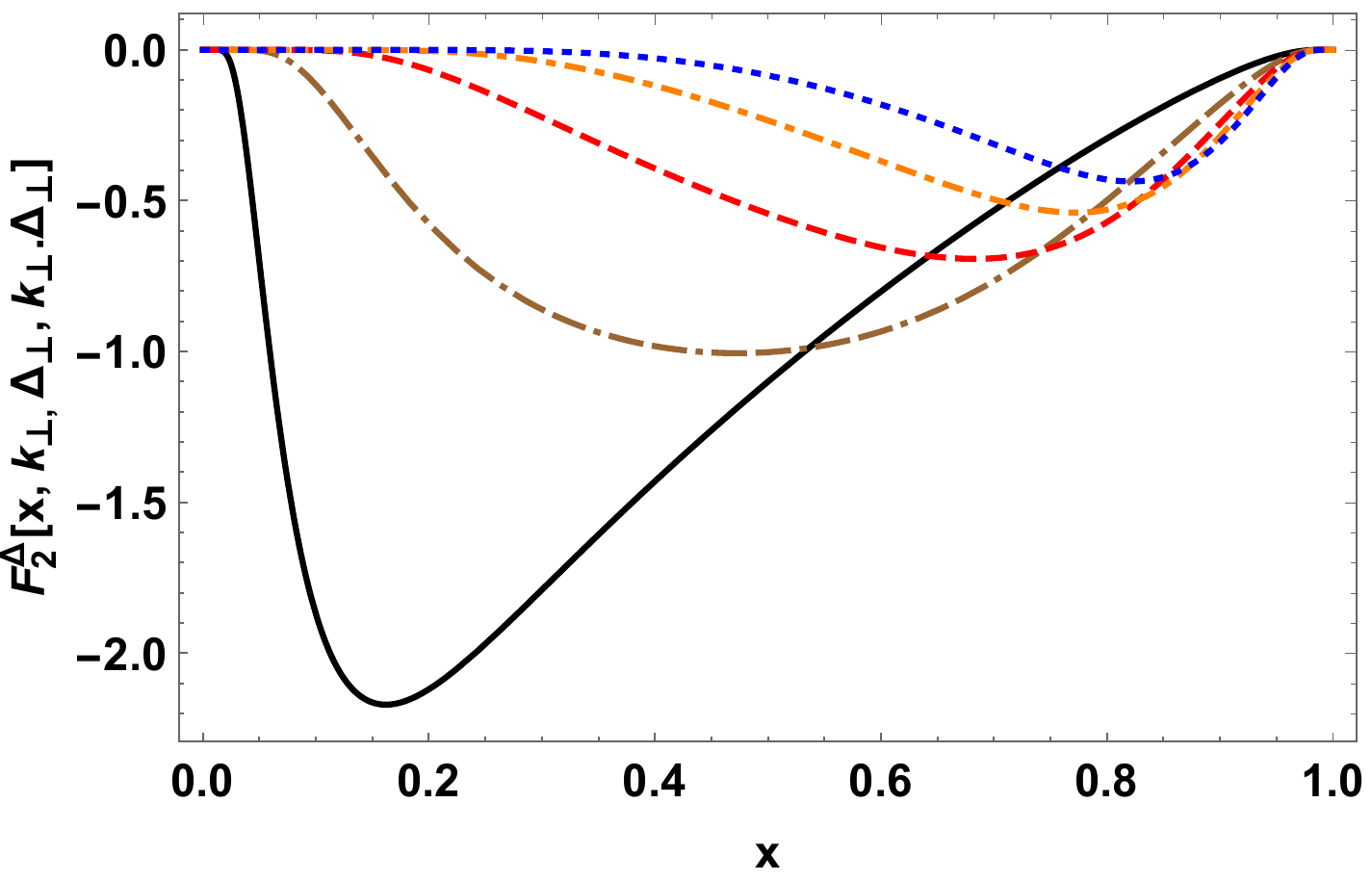}
			\end{center}
		\end{minipage}
		\caption{(Color online) 
        Twist-$3$ quark GTMDs plotted as a function of $x$ at a fixed transverse momentum of $\mathbf{k}_\perp=0.1$ GeV for different values of $\Delta_\perp=0.5, 1, 1.5, 2,$ and $2.5$ GeV.}
		\label{12drealtmds2}
	\end{figure}
For the case of twist-$3$, there are a total of $8$ GTMDs for spin-$0$ pseudoscalar mesons, compared to $32$ for the case of nucleons \cite{Meissner:2009ww, Meissner:2008ay}. In our analysis, the GTMDs $\tilde{E}$ and $\tilde{H}_2$ are found to be zero. All the GTMDs satisfy the hermiticity property as \cite{Meissner:2008ay}
\begin{eqnarray}
    Y^*(x,k_{\perp}, \Delta_{\perp},\bfk\cdot\Dp:\eta)&=&\pm Y(x,k_{\perp}, \Delta_{\perp},-\bfk\cdot\Dp:\eta),
\end{eqnarray}
The GTMDs $F_1$, $\tilde{G}_1$, $h_1^\Delta$, $E_2$, $F_2^k$, $G^\Delta_2$, $H_2$, $\tilde{H}_2$, $F_3$, $\tilde{G}_3$, and $H^\Delta_3$ correspond to a positive hermiticity sign and are T-even in nature. In contrast, the remaining GTMDs exhibit negative hermiticity and are T-odd. The twist-$3$ GTMDs are plotted as functions of $x$ and $\bfk$ at a fixed momentum transfer of $\Delta_\perp=1$ GeV in Fig.~\ref{t3realtmds2}. The GTMDs $E_2$, $G_2^k$, $H_2$ and $F_2^k$ exhibit positive distributions, whereas $F^\Delta_2$ show negative distributions throughout. The GTMD $G_2^\Delta$ exhibits both positive and negative distributions. The unpolarized GTMD $E_2$ carries crucial information about the quark PDF $e(x)$ and the scalar form factor $F_2(Q^2)$ \cite{Efremov:2002qh}. In Fig.~\ref{12drealtmds2} (a), we observe that the distribution of $E_2$ decreases smoothly and shifts toward higher $x$ as the momentum transfer $\Delta_\perp$ increases, at a fixed transverse momentum of $\bfk=0.1$ GeV. A similar behavior is observed for other GTMDs. Meanwhile, for $F^k_2$, the distribution exhibits a slight increase in its peak value as $\Delta_\perp$ increases from $2$ to $2.5$ GeV as shown in Fig.~\ref{12drealtmds2} (e). Additionally, we note that all GTMDs exhibit significant quark distributions in the small-$x$ region when the transverse momentum transfer $\Delta_\perp$ is low. This suggests that as the final-state pion acquires greater momentum transfer, the quark GTMDs shift toward higher $x$, with a decrease in overall distribution. In Fig.~\ref{average} (b), we present the variation of the average transverse momentum $\langle \bfk \rangle$ for twist-$3$ GTMDs as a function of $\Delta_\perp$. There is no theoretical prediction for higher twist GTMDs in other models, however the authors in Ref. \cite{Meissner:2008ay} have provided the explicit form of higher twist GTMDs in the spectator model for pion. Both the $\tilde{E}$ and $\tilde{H}_2$ GTMDs are also zero in the spectator model, like Furthermoreours. Furthermore, we observe that the average transverse momentum initially increases with increasing $\Delta_\perp$ up to $\Delta_\perp=3$ GeV, beyond which it starts to decrease. All the GTMDs reach the maximum value of average transverse momenta in between $\Delta_\perp=2$ and $3$ GeV. Notably, the GTMD $G_2^k$ and $F_2^\Delta$ carry the same average transverse momenta throughout all values of $\Delta_\perp$ and carry the maximum value of $\langle\bfk\rangle$ at $\Delta_\perp=0$ i.e, $0.29$ GeV compared to other twist-$3$ GTMDs.
\subsection{Twist-4 GTMDs}
\begin{figure}[ht]
		\centering
		\begin{minipage}[c]{1\textwidth}\begin{center}
				(a)\includegraphics[width=.45\textwidth]{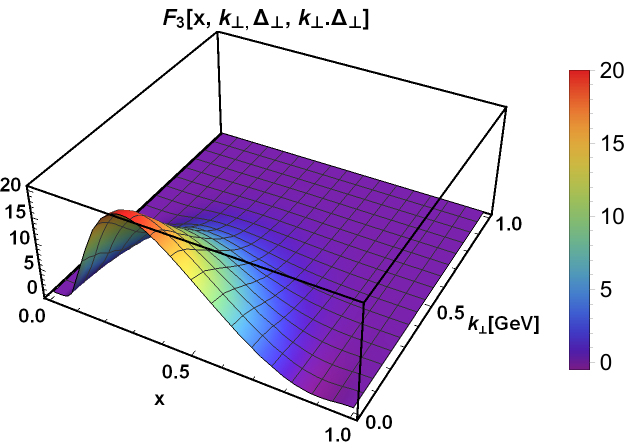}
				(b)\includegraphics[width=.45\textwidth]{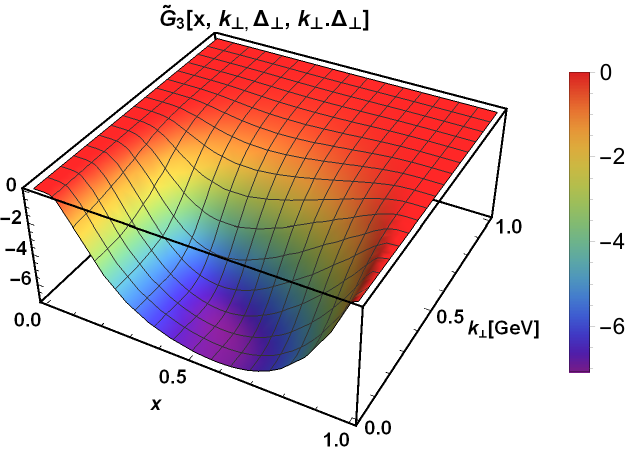}
			\end{center}
		\end{minipage}
		\begin{minipage}[c]{1\textwidth}\begin{center}
				(c)\includegraphics[width=.45\textwidth]{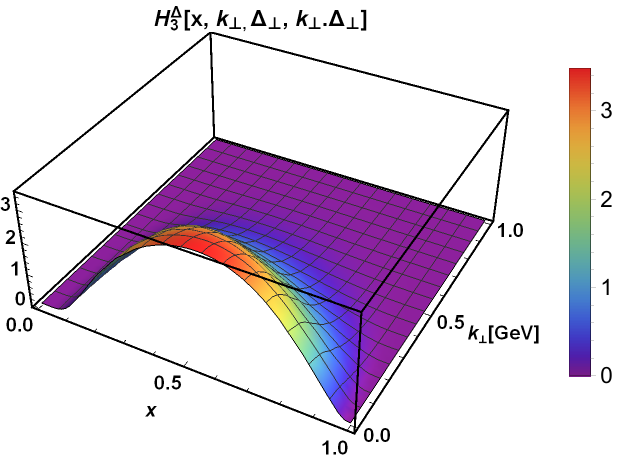}
			\end{center}
		\end{minipage}
		\caption{(Color online) Twist-4 valence quark GTMDs plotted as a function of $x$ and $\bfk$ at a fixed value of $\Delta_\perp=1.0$ GeV and $\theta=0$.}
		\label{21realtmds3}
	\end{figure}

\begin{figure}[ht]
		\centering
		\begin{minipage}[c]{1\textwidth}\begin{center}
				(a)\includegraphics[width=.45\textwidth]{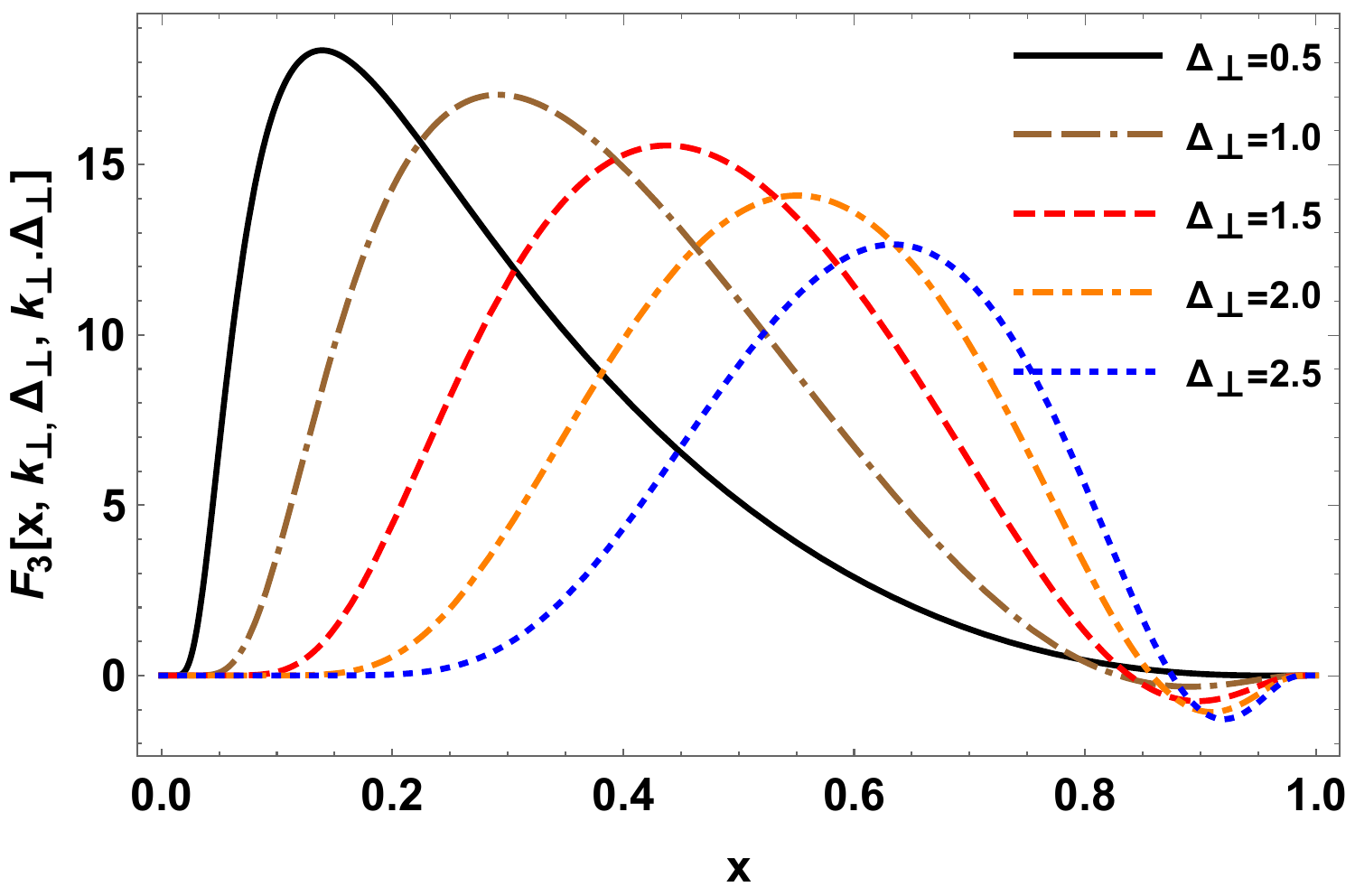}
				(b)\includegraphics[width=.45\textwidth]{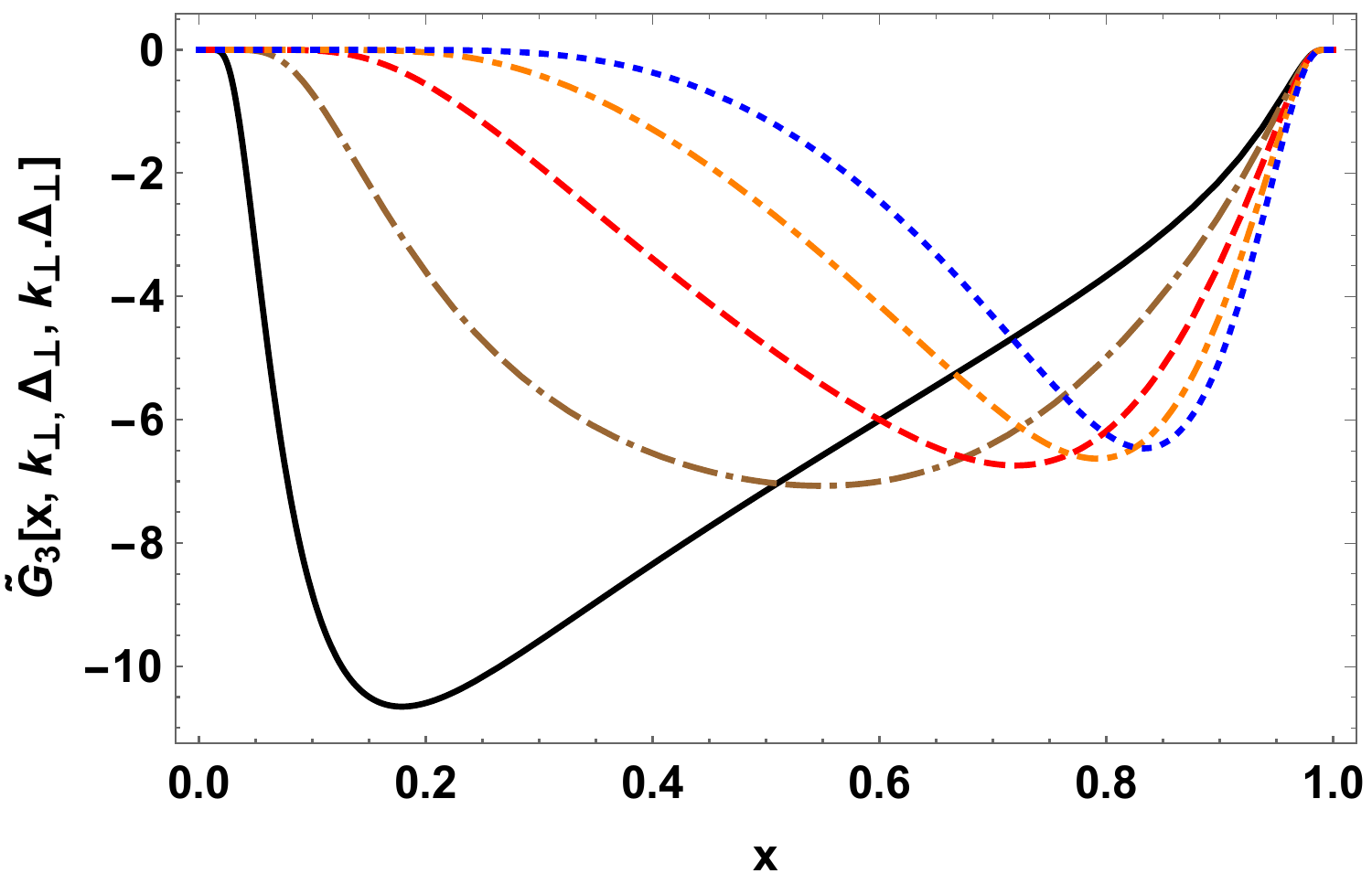}
			\end{center}
		\end{minipage}
		\begin{minipage}[c]{1\textwidth}\begin{center}
				(c)\includegraphics[width=.45\textwidth]{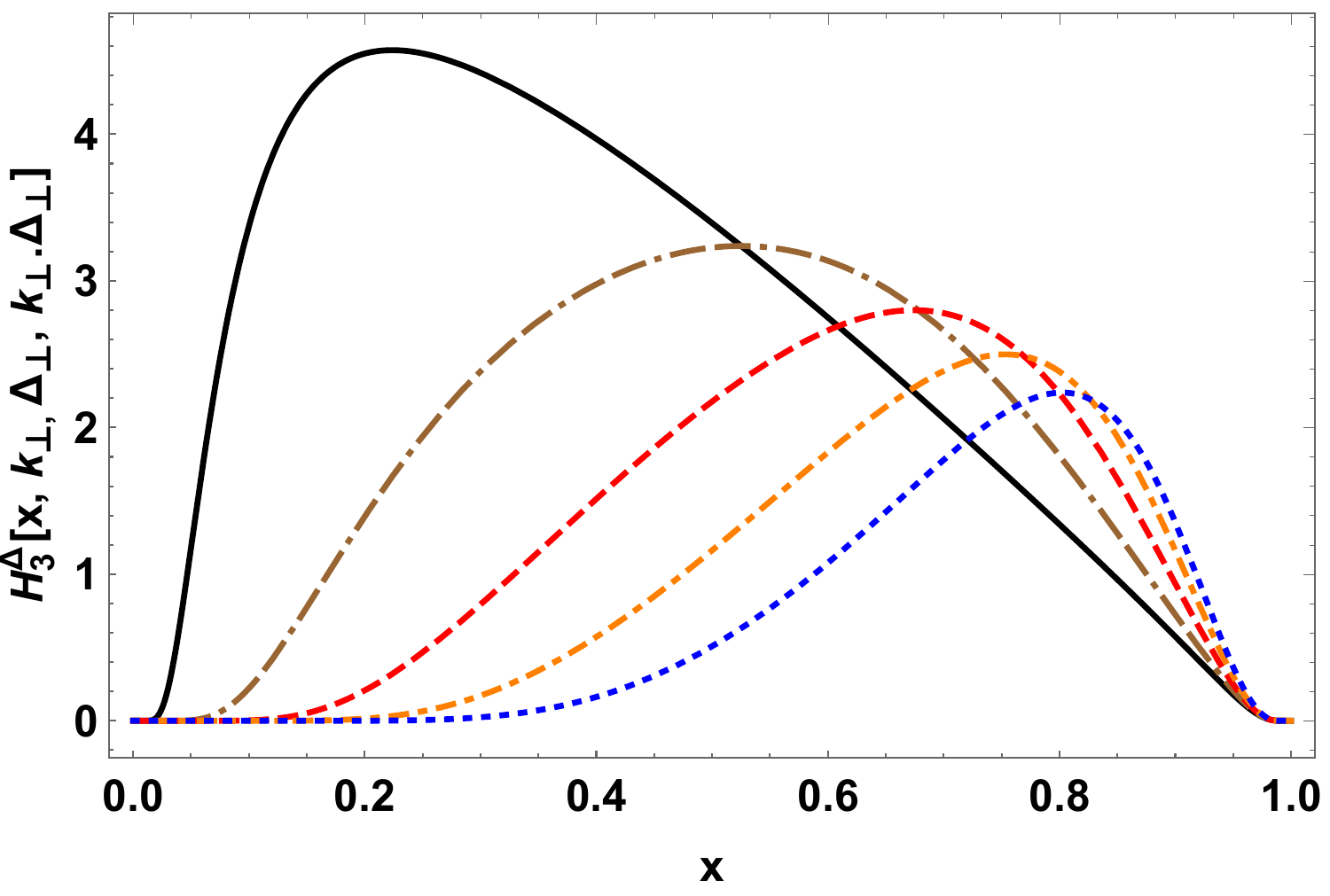}
			\end{center}
		\end{minipage}
		\caption{(Color online) Twist-4 quark GTMDs plotted as a function of $x$ at a fixed transverse momentum of $\mathbf{k}_\perp=0.1$ GeV for different values of $\Delta_\perp=0.5, 1, 1.5, 2,$ and $2.5$ GeV.}
		\label{32realtmds3}
	\end{figure}

\begin{figure}[ht]
		\centering
		\begin{minipage}[c]{1\textwidth}\begin{center}				(a)\includegraphics[width=.45\textwidth]{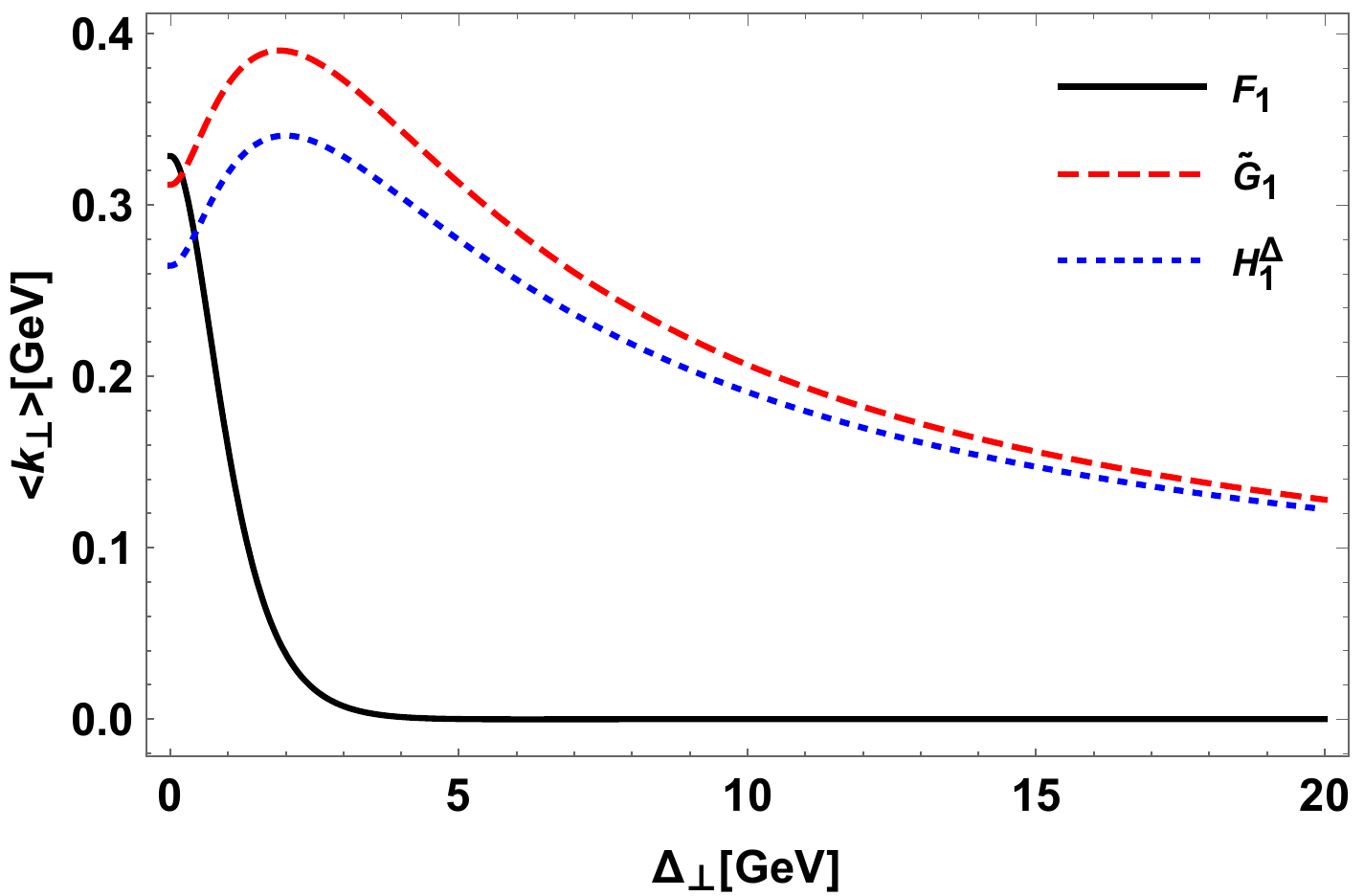}
				(b)\includegraphics[width=.45\textwidth]{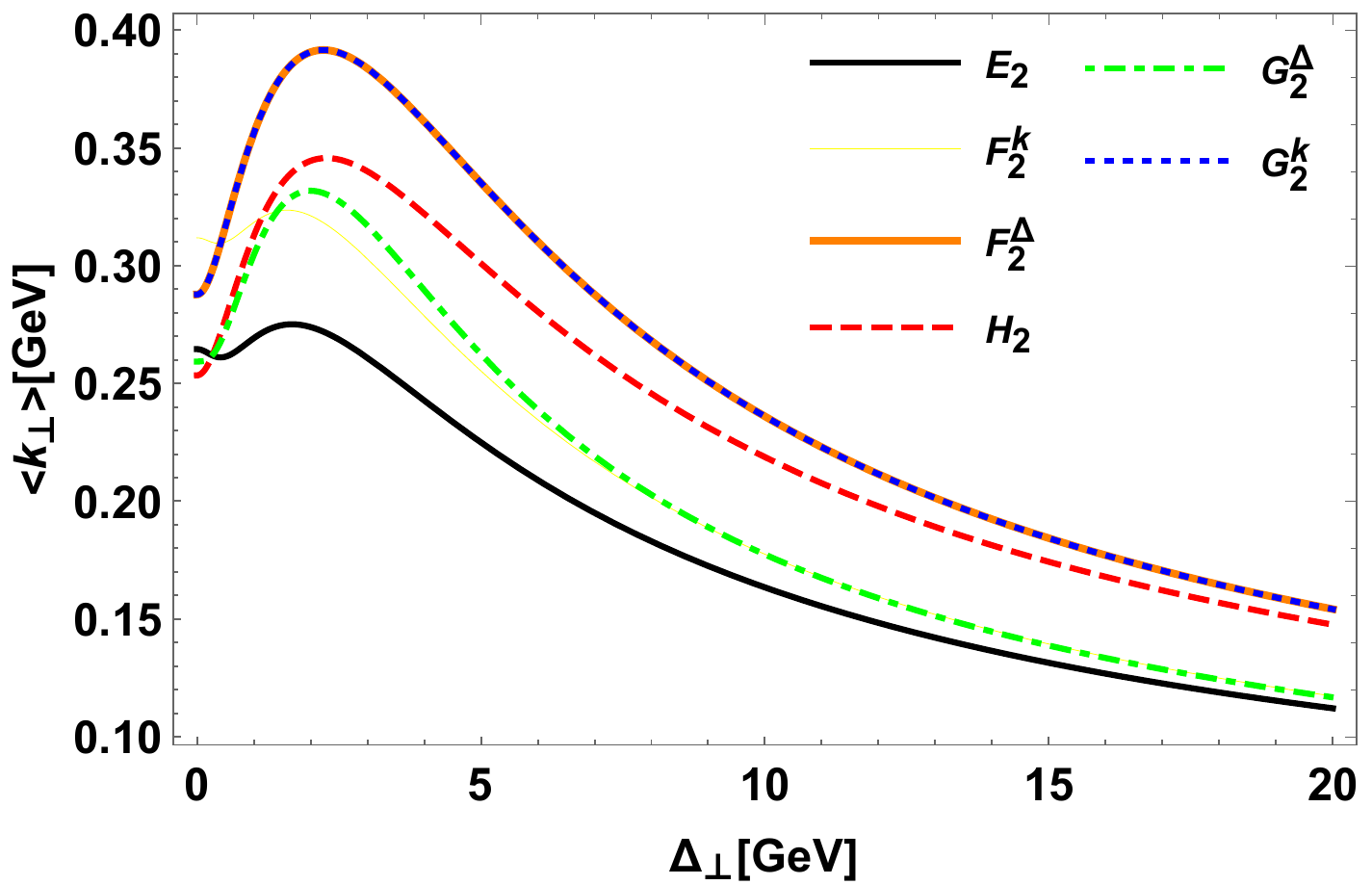}
			\end{center}
		\end{minipage}
		\begin{minipage}[c]{1\textwidth}\begin{center}
				(c)\includegraphics[width=.45\textwidth]{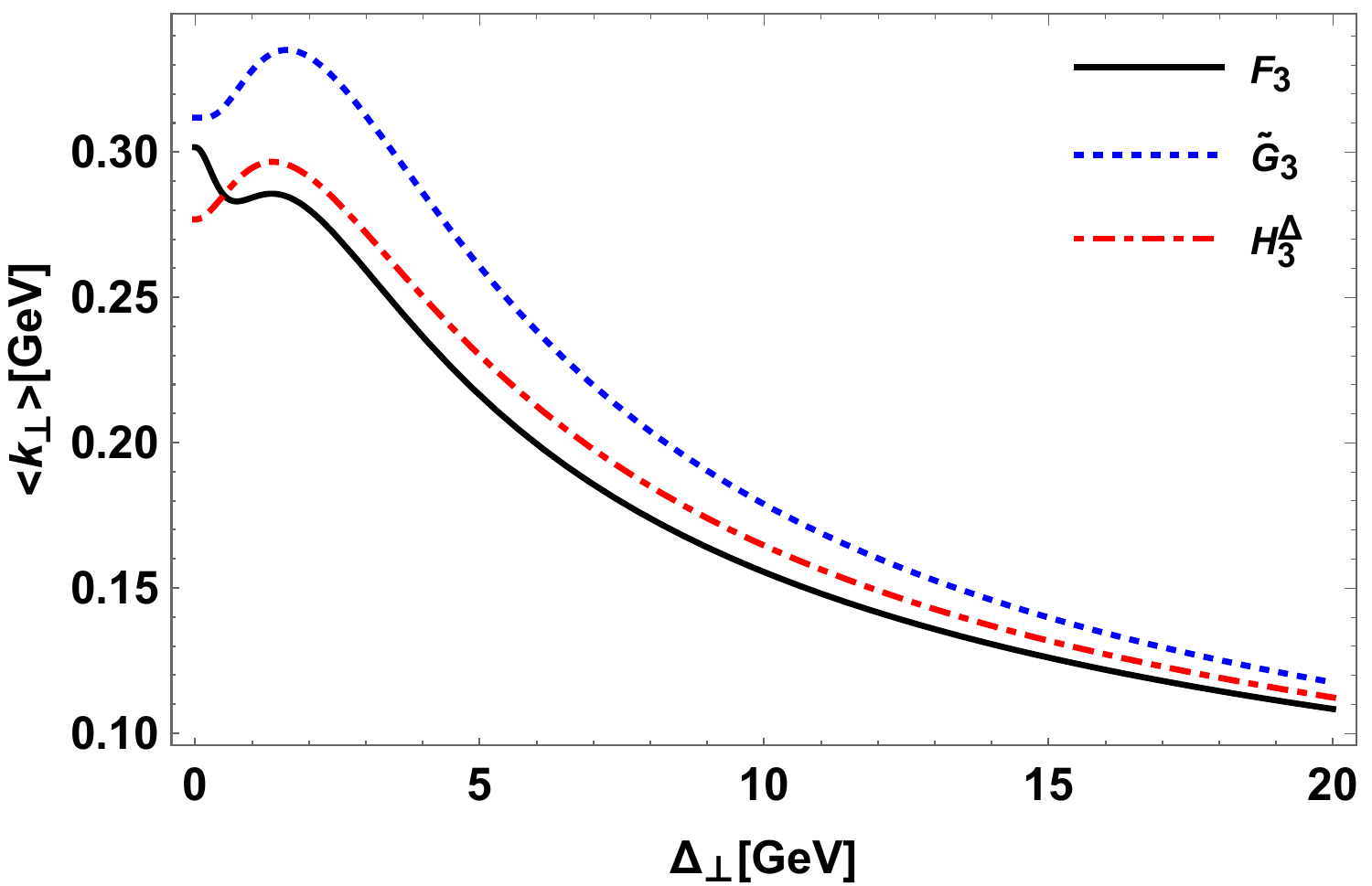}
			\end{center}
		\end{minipage}
		\caption{(Color online) The average transverse momenta of the GTMDs plotted as a function of $\Delta_\perp$ for (a) twist-$2$, (b) twist-$3$, and (c) twist-$4$.}
		\label{average}
	\end{figure}

The twist-$4$ GTMDs for the case of the pion can be expressed in terms of different Dirac matrices as \cite{Meissner:2008ay}
\begin{eqnarray}
\Phi_q^{[\gamma^-]} & = &\frac{M^2_\pi}{(P^+)^2} \bigg[F_3 (x, k_{\perp}, \Delta_{\perp},\bfk\cdot\Dp)\,, \vphantom{\frac{1}{1}}\bigg] 
\label{e:gtmd_1b} \\
\Phi_q^{[\gamma^- \gamma_5]} & = & \frac{M^2_\pi}{(P^+)^2}
 \bigg[\frac{i\varepsilon_\perp^{ij} k_\perp^i \Delta_\perp^j}{M_\pi^2} \, \tilde{G}_3 (x, k_{\perp}, \Delta_{\perp},\bfk\cdot\Dp)\bigg], \\
\Phi_q^{[i\sigma^{j-}\gamma_5]} & = & \frac{M^2_\pi}{(P^+)^2}\bigg[
 \frac{i\varepsilon_\perp^{ij} k_\perp^i}{M_\pi} \, H^k_3(x, k_{\perp}, \Delta_{\perp},\bfk\cdot\Dp) 
+\frac{i\varepsilon_\perp^{ij} \Delta_\perp^i}{M_\pi} \, H^\Delta_3(x, k_{\perp}, \Delta_{\perp},\bfk\cdot\Dp) \bigg]. 
\label{e:gtmd_3b}
\end{eqnarray}
The LFWFs overlap representation for the twist-4 GTMDs is given by
\begin{eqnarray}
    \Phi_q^{[\gamma^-]}
    &=& \frac{1}{2 (2\pi)^{3}}\frac{1}{4 x^2 (P^+)^2} \Bigg[(4 m^2_u+ (2k_2+i \Delta_1)^2)+ (2 k_1-i \Delta_2)^2)\Bigg(\Psi^{*}_{\pi}\left(x, \boldsymbol{k}_{\perp}^{\prime\prime},\uparrow,\uparrow\right) \Psi_{\pi}\left(x, \boldsymbol{k}_{\perp}^{\prime},\uparrow,\uparrow\right) 
  \nonumber\\ &+& \Psi^{*}_{\pi}\left(x, \boldsymbol{k}_{\perp}^{\prime\prime},\uparrow,\downarrow\right) \Psi_{\pi}\left(x, \boldsymbol{k}_{\perp}^{\prime},\uparrow,\downarrow\right) \Bigg)+(4 m^2_u+ (2k_2-i \Delta_1)^2)+ (2 k_1+i \Delta_2)^2) \nonumber\\ &\times&\Bigg(\Psi^*_{\pi}\left(x, \boldsymbol{k}_{\perp}^{\prime\prime},\downarrow,\uparrow\right) \Psi_{\pi}\left(x, \boldsymbol{k}_{\perp}^{\prime},\downarrow,\uparrow\right) 
  + \Psi^{*}_{\pi}\left(x, \boldsymbol{k}_{\perp}^{\prime\prime},\downarrow,\downarrow\right) \Psi_{\pi}\left(x, \boldsymbol{k}_{\perp}^{\prime},\downarrow,\downarrow\right)\Bigg)\nonumber\\ &+& 
  4 m_u(\Delta_1- i \Delta_2)\Bigg( \Psi^{*}_{\pi}\left(x, \boldsymbol{k}_{\perp}^{\prime\prime},\uparrow,\uparrow\right) \Psi_{\pi}\left(x, \boldsymbol{k}_{\perp}^{\prime},\downarrow,\uparrow\right)+\Psi^{*}_{\pi}\left(x, \boldsymbol{k}_{\perp}^{\prime\prime},\uparrow,\downarrow\right) \Psi_{\pi}\left(x, \boldsymbol{k}_{\perp}^{\prime},\downarrow,\downarrow\right)\Bigg)\nonumber
      \end{eqnarray}
  \begin{eqnarray}
  &-& 
  4 m_u(\Delta_1+ i \Delta_2)\Bigg( \Psi^{*}_{\pi}\left(x, \boldsymbol{k}_{\perp}^{\prime\prime},\downarrow,\uparrow\right) \Psi_{\pi}\left(x, \boldsymbol{k}_{\perp}^{\prime},\uparrow,\uparrow\right)+\Psi^{*}_{\pi}\left(x, \boldsymbol{k}_{\perp}^{\prime\prime},\downarrow,\downarrow\right) \Psi_{\pi}\left(x, \boldsymbol{k}_{\perp}^{\prime},\uparrow,\downarrow\right)\Bigg)\Bigg]\, ,
 \\
  \Phi_q^{[\gamma^-\gamma_5]}
  &=& \frac{1}{2 (2\pi)^{3}}\frac{1}{4 x^2 (P^+)^2} \Bigg[(-4 m^2_u+ (2k_2+i \Delta_1)^2)+ (2 k_1-i \Delta_2)^2)\Bigg(\Psi^{*}_{\pi}\left(x, \boldsymbol{k}_{\perp}^{\prime\prime},\uparrow,\uparrow\right) \Psi_{\pi}\left(x, \boldsymbol{k}_{\perp}^{\prime},\uparrow,\uparrow\right) 
  \nonumber\\ &+& \Psi^{*}_{\pi}\left(x, \boldsymbol{k}_{\perp}^{\prime\prime},\uparrow,\downarrow\right) \Psi_{\pi}\left(x, \boldsymbol{k}_{\perp}^{\prime},\uparrow,\downarrow\right) \Bigg)+(4 m^2_u-(2k_2-i \Delta_1)^2)- (2 k_1+i \Delta_2)^2) \nonumber\\ &\times&\Bigg(\Psi^*_{\pi}\left(x, \boldsymbol{k}_{\perp}^{\prime\prime},\downarrow,\uparrow\right) \Psi_{\pi}\left(x, \boldsymbol{k}_{\perp}^{\prime},\downarrow,\uparrow\right) 
  + \Psi^{*}_{\pi}\left(x, \boldsymbol{k}_{\perp}^{\prime\prime},\downarrow,\downarrow\right) \Psi_{\pi}\left(x, \boldsymbol{k}_{\perp}^{\prime},\downarrow,\downarrow\right)\Bigg)\nonumber\\ &+& 
  8 m_u(k_1- i k_2)\Bigg( \Psi^{*}_{\pi}\left(x, \boldsymbol{k}_{\perp}^{\prime\prime},\uparrow,\uparrow\right) \Psi_{\pi}\left(x, \boldsymbol{k}_{\perp}^{\prime},\downarrow,\uparrow\right)+\Psi^{*}_{\pi}\left(x, \boldsymbol{k}_{\perp}^{\prime\prime},\uparrow,\downarrow\right) \Psi_{\pi}\left(x, \boldsymbol{k}_{\perp}^{\prime},\downarrow,\downarrow\right)\Bigg)\nonumber\\ &+& 
  8 m_u(k_1+ i k_2)\Bigg( \Psi^{*}_{\pi}\left(x, \boldsymbol{k}_{\perp}^{\prime\prime},\downarrow,\uparrow\right) \Psi_{\pi}\left(x, \boldsymbol{k}_{\perp}^{\prime},\uparrow,\uparrow\right)+\Psi^{*}_{\pi}\left(x, \boldsymbol{k}_{\perp}^{\prime\prime},\downarrow,\downarrow\right) \Psi_{\pi}\left(x, \boldsymbol{k}_{\perp}^{\prime},\uparrow,\downarrow\right)\Bigg)\Bigg]\, ,
\\
   \Phi_q^{[i\sigma^{1-}\gamma_5]}
   &=& \frac{1}{2 (2\pi)^{3}}\frac{1}{4 x^2 (P^+)^2} \Bigg[(4 m_u (2 k_1-i \Delta_2))\Bigg(\Psi^{*}_{\pi}\left(x, \boldsymbol{k}_{\perp}^{\prime\prime},\uparrow,\uparrow\right) \Psi_{\pi}\left(x, \boldsymbol{k}_{\perp}^{\prime},\uparrow,\uparrow\right) 
  \nonumber\\ &+& \Psi^{*}_{\pi}\left(x, \boldsymbol{k}_{\perp}^{\prime\prime},\uparrow,\downarrow\right) \Psi_{\pi}\left(x, \boldsymbol{k}_{\perp}^{\prime},\uparrow,\downarrow\right) \Bigg)+(4 m_u (-2k_1-i \Delta_2)) \nonumber\\ &\times&\Bigg(\Psi^*_{\pi}\left(x, \boldsymbol{k}_{\perp}^{\prime\prime},\downarrow,\uparrow\right) \Psi_{\pi}\left(x, \boldsymbol{k}_{\perp}^{\prime},\downarrow,\uparrow\right) 
  + \Psi^{*}_{\pi}\left(x, \boldsymbol{k}_{\perp}^{\prime\prime},\downarrow,\downarrow\right) \Psi_{\pi}\left(x, \boldsymbol{k}_{\perp}^{\prime},\downarrow,\downarrow\right)\Bigg)\nonumber\\ &+& 
  4\Bigg(m^2_u-(k_1-i k_2)^2+\frac{(\Delta_1-i \Delta_2)^2}{4}\Bigg)\Bigg( \Psi^{*}_{\pi}\left(x, \boldsymbol{k}_{\perp}^{\prime\prime},\uparrow,\uparrow\right) \Psi_{\pi}\left(x, \boldsymbol{k}_{\perp}^{\prime},\downarrow,\uparrow\right)\nonumber\\
  &+&\Psi^{*}_{\pi}\left(x, \boldsymbol{k}_{\perp}^{\prime\prime},\uparrow,\downarrow\right) \Psi_{\pi}\left(x, \boldsymbol{k}_{\perp}^{\prime},\downarrow,\downarrow\right)\Bigg)+
  4\Bigg(m^2_u-(k_1+i k_2)^2+\frac{(\Delta_1+i \Delta_2)^2}{4}\Bigg)\nonumber\\
  &\times&\Bigg( \Psi^{*}_{\pi}\left(x, \boldsymbol{k}_{\perp}^{\prime\prime},\downarrow,\uparrow\right) \Psi_{\pi}\left(x, \boldsymbol{k}_{\perp}^{\prime},\uparrow,\uparrow\right)+\Psi^{*}_{\pi}\left(x, \boldsymbol{k}_{\perp}^{\prime\prime},\downarrow,\downarrow\right) \Psi_{\pi}\left(x, \boldsymbol{k}_{\perp}^{\prime},\uparrow,\downarrow\right)\Bigg)\Bigg]\, .
\end{eqnarray}
The explicit form of twist-4 GTMDs is found to be
\begin{eqnarray}
    F_3 &=& \frac{1}{2 (2 \pi)^3}\frac{1}{4 x^2 M^2_\pi}\Bigg[ (m_u^2+\bfk^2-\frac{\Delta_\perp^2}{4})(4 m_u^2+4 \bfk^2-(1-x)^2\Delta_\perp^2) \nonumber\\
    &&+4 m^2_u (1-x)\Delta^2_\perp
    +4 (1-x)((\bfk\Delta_\perp)^2-(\bfk.\Delta_\perp)^2)\Bigg]\frac{\psi^*(x,\bfk^{\prime\prime})\psi(x,\bfk^{\prime})}{\omega^+\omega^-}\, ,
    \\
    \tilde{G}_3&=& \frac{1}{2 (2 \pi)^3}\frac{1}{4 x^2}\Bigg[-4 (m_u^2+\bfk^2-\frac{(1-x)^2\Delta_\perp^2}{4})\nonumber\\
    &&-(1-x)(-4 m_u^2-4 \bfk^2+\Delta_\perp^2)
    \Bigg]\frac{\psi^*(x,\bfk^{\prime\prime})\psi(x,\bfk^{\prime})}{\omega^+\omega^-}\, ,
    \\
    H^k_3&=& 0\, ,
        \end{eqnarray}
  \begin{eqnarray}
    H^\Delta_3&=& \frac{1}{2 (2 \pi)^3}\frac{1}{4 x^2 M_\pi}\Bigg[4 m_u( m_u^2+\bfk^2-\frac{(1-x)^2\Delta_\perp^2}{4}) \nonumber\\
    &&-(1-x)m_u(4 m_u^2+4 \bfk^2-\Delta^2_\perp)\Bigg]\frac{\psi^*(x,\bfk^{\prime\prime})\psi(x,\bfk^{\prime})}{\omega^+\omega^-}\, ,
\end{eqnarray}
The twist-$4$ GTMDs are essentially a replication of the twist-$2$ GTMDs and have been studied for completeness in the GTMD framework. At present, studying these GTMDs experimentally or through lattice simulations remains a challenging task. However, this analysis provides valuable insight into the complete internal structure of the pion. At the twist-$4$ level, there are four GTMDs for the pion. Among them, $F_3$, $\tilde{G}_3$, and $H^\Delta_3$ are T-even, whereas $H^k_3$ is T-odd in nature and coming out to be zero in our case. A similar kind of observation was predicted for $H^\Delta_3$ in Ref. \cite{Meissner:2008ay}. In Fig.~\ref{21realtmds3}, we present these GTMDs as $x$ and $\bfk$ functions at a fixed value of $\Delta_\perp=1.0$ GeV and $\theta=0$. Due to the factor of $1/x^2$, these GTMDs exhibit higher peak values compared to twist-$2$ and twist-$3$ GTMDs. At $\Delta_\perp=1.0$ GeV. $\tilde{G}_3$ and $H_3^\Delta$ have peak distributions around $x=0.5$, whereas $F_3$ is primarily distributed around $x\leq 0.5$. Similar to twist-$3$ quark distributions, these GTMDs also exhibit concentration in the low-$x$ region at small $\Delta_\perp$. The $F_3$ quark GTMD displays negative distributions in the range $0.7\leq x\leq 1$ for $\Delta_\perp\geq0.5$ GeV. We found that $\tilde{G}_3$ and $H^k_3$ GTMDs show  opposite behavior with incresae in $\Delta_\perp$ as shown in Fig. \ref{32realtmds3}. $H_3^\Delta$ shows a negative behavior for $\Delta_\perp\le 1$ GeV in the region $0\le x\le 0.1$ as shown in Fig. \ref{32realtmds3} (c). Compared to twist-$2$ and twist-$3$ GTMDs, twist-$4$ GTMDs become more concentrated near $x=0.8$ at larger values of $\Delta_\perp$, as illustrated in Fig.~\ref{32realtmds3}. The behavior of the average transverse momentum of quarks follows a similar trend as observed for other twist GTMDs and is depicted in Fig.~\ref{average} (c). However, in the case of the $F_3$ GTMDs, the average momenta initially decrease to $\Delta_\perp = 0.65$ GeV, then abruptly increase until $\Delta_\perp = 1.65$ GeV. After that, they exhibit gradually decreasing. This behavior is unique to this GTMD.
%
    
    \begin{figure}[ht]
		\centering
		\begin{minipage}[c]{1\textwidth}\begin{center}
				(a)\includegraphics[width=.45\textwidth]{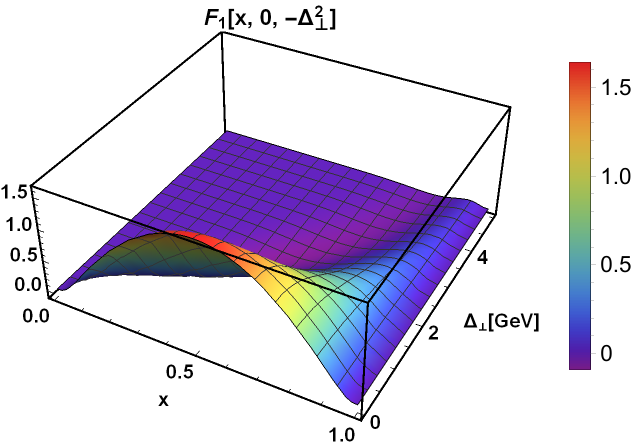}
				(b)\includegraphics[width=.45\textwidth]{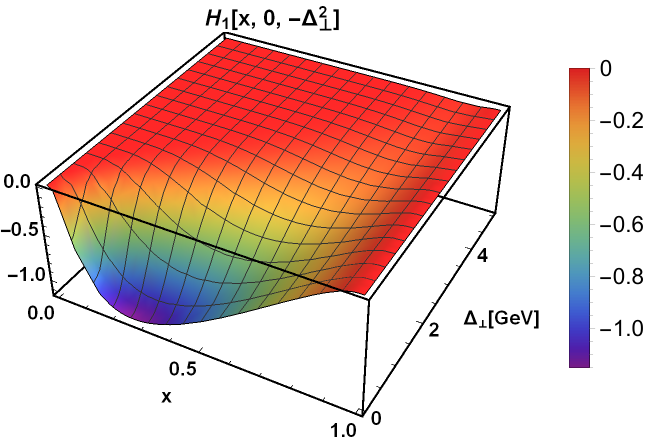}
			\end{center}
		\end{minipage}
		\begin{minipage}[c]{1\textwidth}\begin{center}
				(c)\includegraphics[width=.45\textwidth]{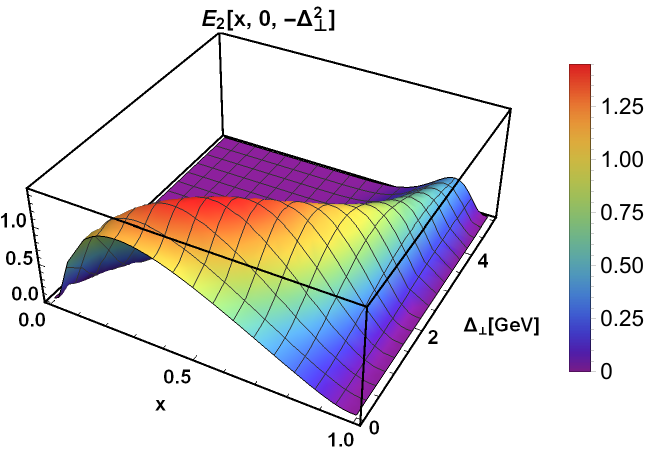}
				(d)\includegraphics[width=.45\textwidth]{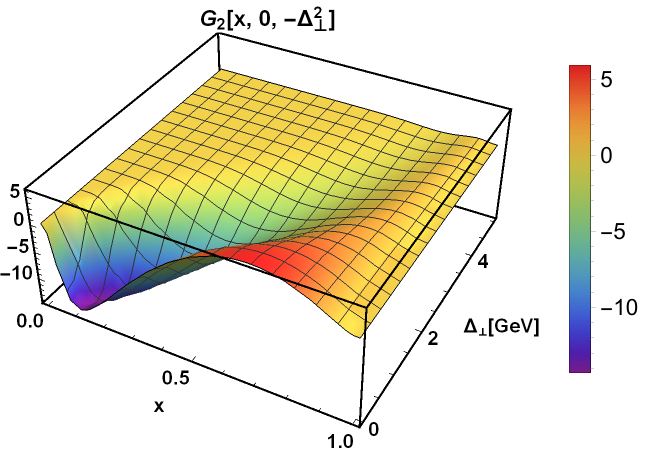}
                \end{center}
		\end{minipage}
        \begin{minipage}[c]{1\textwidth}\begin{center}
				(e)\includegraphics[width=.45\textwidth]{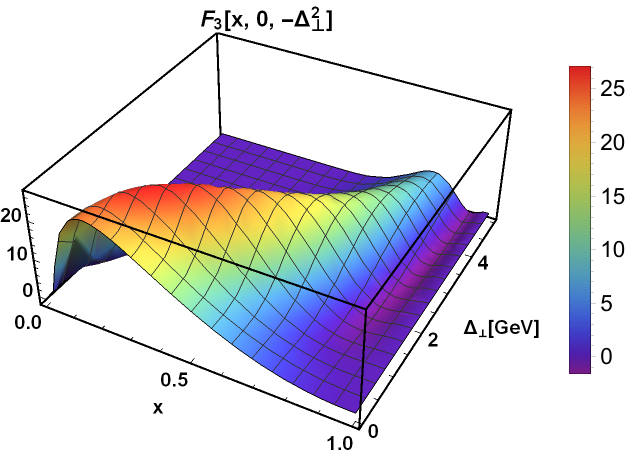}
				(f)\includegraphics[width=.45\textwidth]{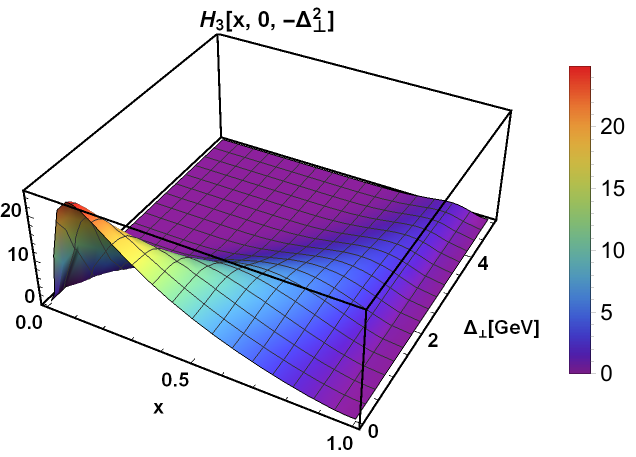}
			\end{center}
		\end{minipage}
		\caption{(Color online)
        Quark 
        GPDs plotted as a function of longitudinal momentum fraction $x$ and $\Delta_\perp$. The upper panel corresponds to twist-$2$, the middle panel to twist-$3$, and the lower panel to twist-$4$.}
		\label{realtmds4}
	\end{figure}
        \begin{figure}[ht]
		\centering
		\begin{minipage}[c]{1\textwidth}\begin{center}
				(a)\includegraphics[width=.45\textwidth]{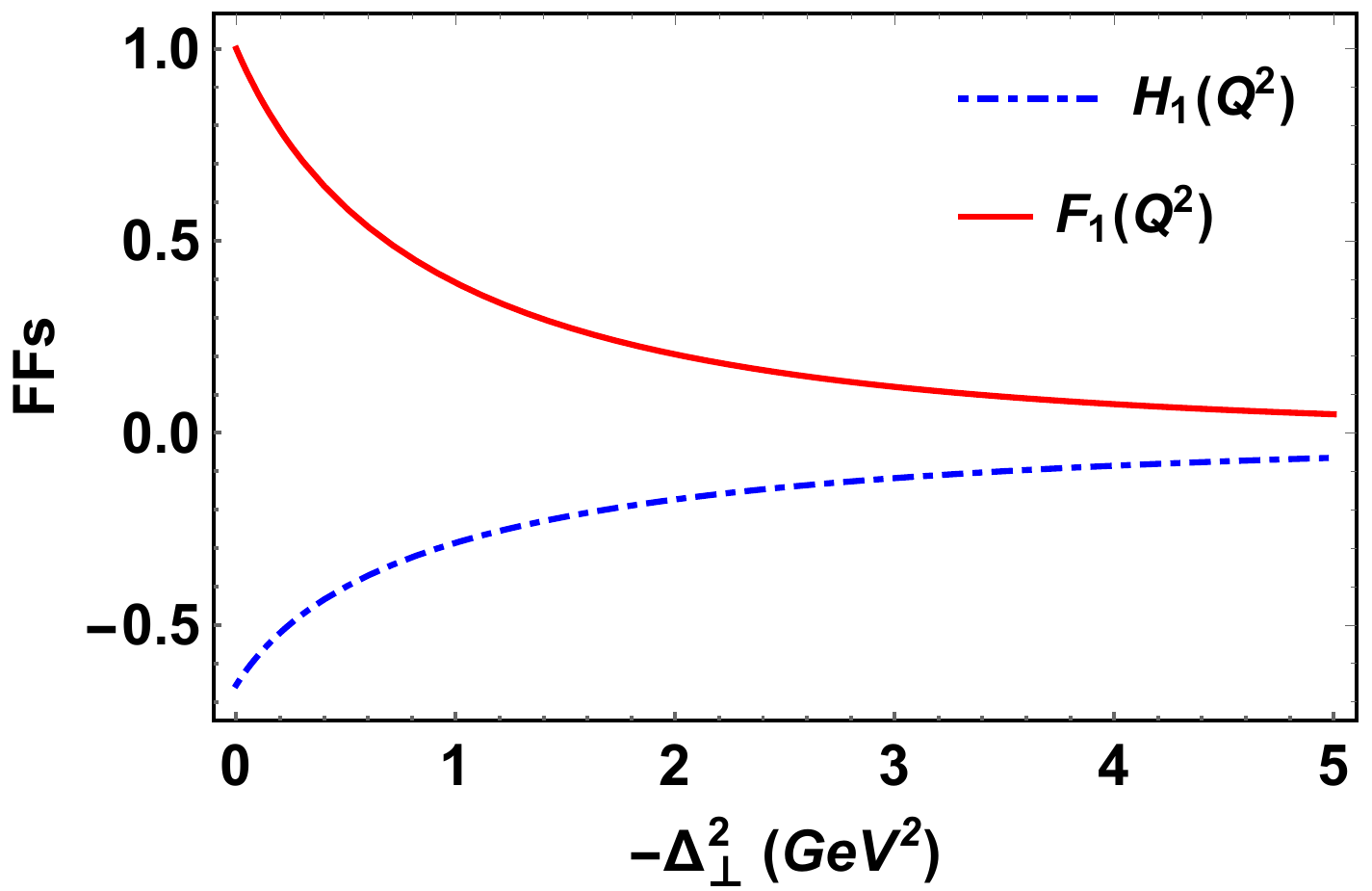}
				(b)\includegraphics[width=.45\textwidth]{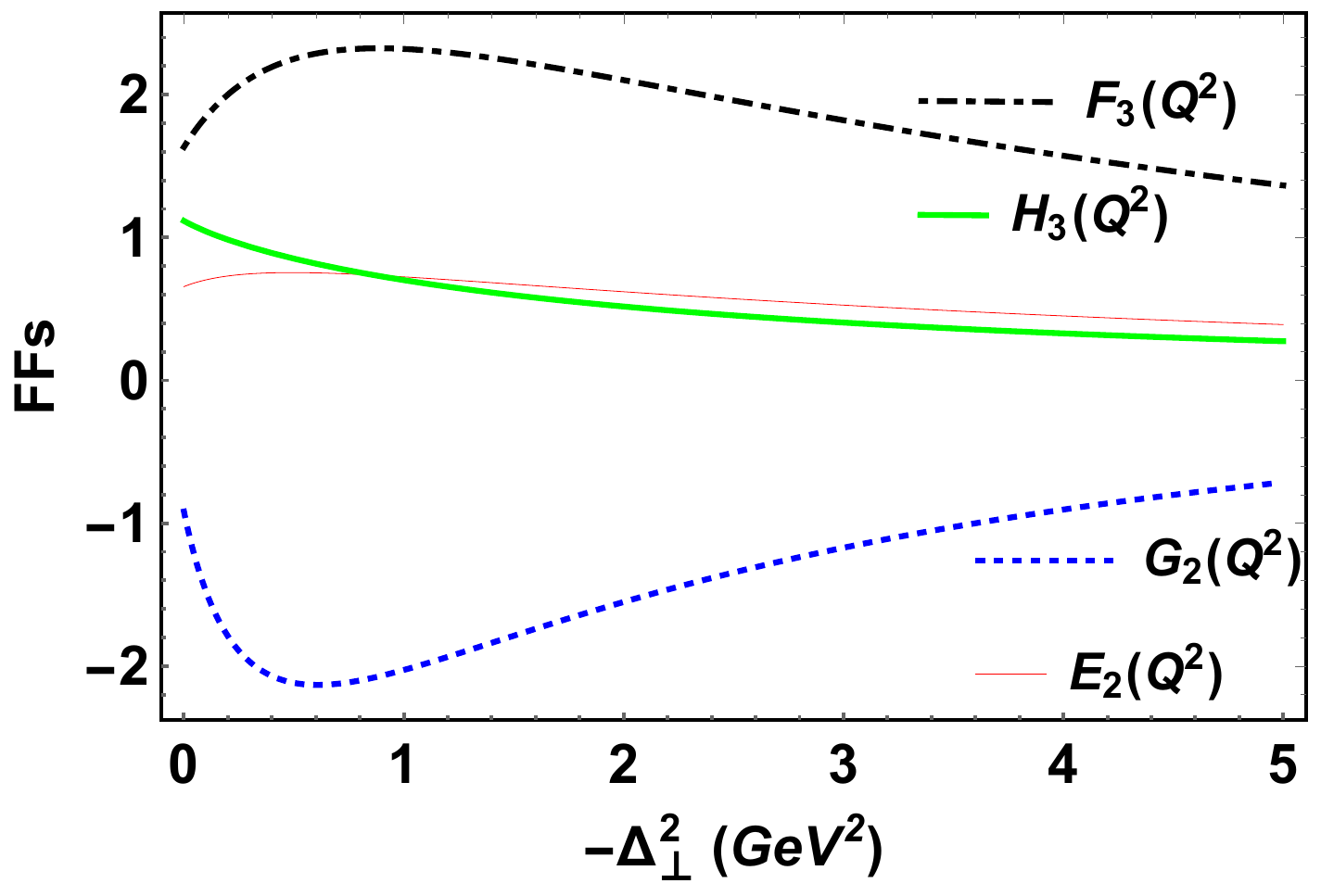}
			\end{center}
		\end{minipage}
		\caption{(Color online) The elastic EMFFs corresponding to different GPDs plotted as a function of $-\Delta^2_\perp$ GeV$^2$ for (a) twist-$2$, (b) twist-$3$, and twist-$4$.}
		\label{realtmds6}
	\end{figure}

    \begin{figure}[ht]
		\centering
		\begin{minipage}[c]{1\textwidth}\begin{center}
				(a)\includegraphics[width=.45\textwidth]{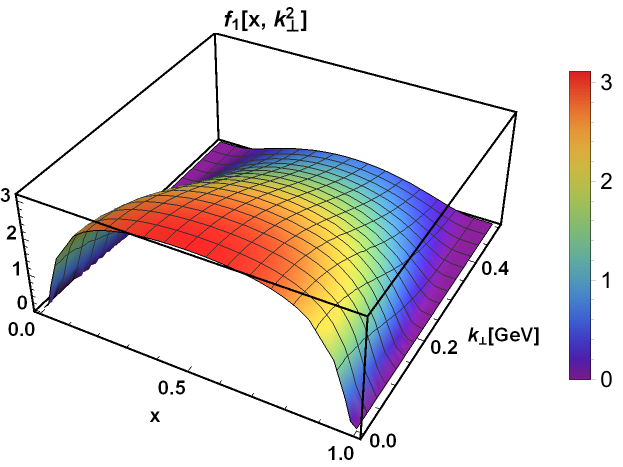}
				(b)\includegraphics[width=.45\textwidth]{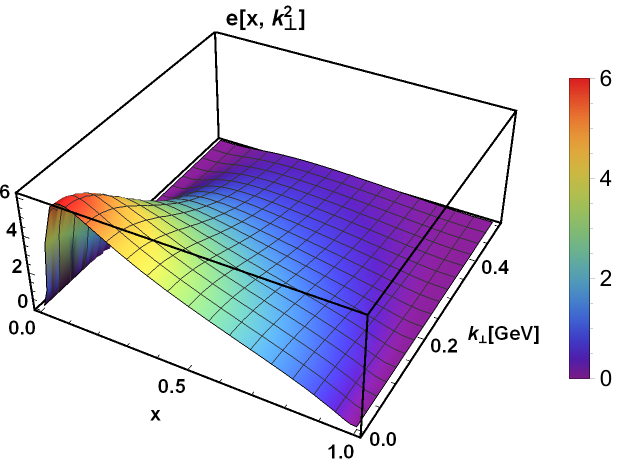}
			\end{center}
		\end{minipage}
		\begin{minipage}[c]{1\textwidth}\begin{center}
				(c)\includegraphics[width=.45\textwidth]{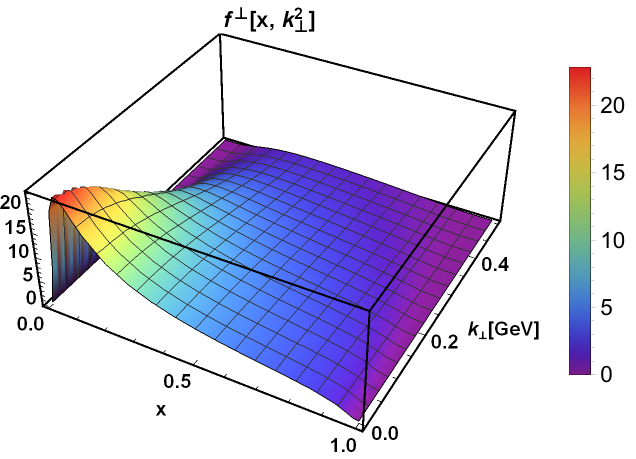}
				(d)\includegraphics[width=.45\textwidth]{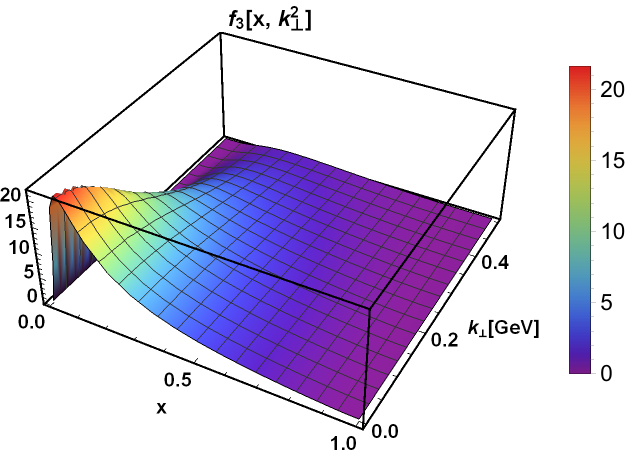}
                \end{center}
		\end{minipage}
		\caption{(Color online) 
        The unpolarized quark transverse momentum 
        TMDs of the pion plotted as a function of $x$ and $\bfk$ up to twist-$4$. (a) Leading twist unpolarized quark $f_1(x,\bfk^2)$ TMD, (b) twist-3 $e(x,\bfk^2)$ TMD, (c) twist-3 $f^\perp(x,\bfk^2)$ TMD and (d) twist-4 $f_3(x,\bfk^2)$ TMD.}
		\label{realtmds5}
	\end{figure}


     \begin{figure}[ht]
		\centering
\includegraphics[width=.45\textwidth]{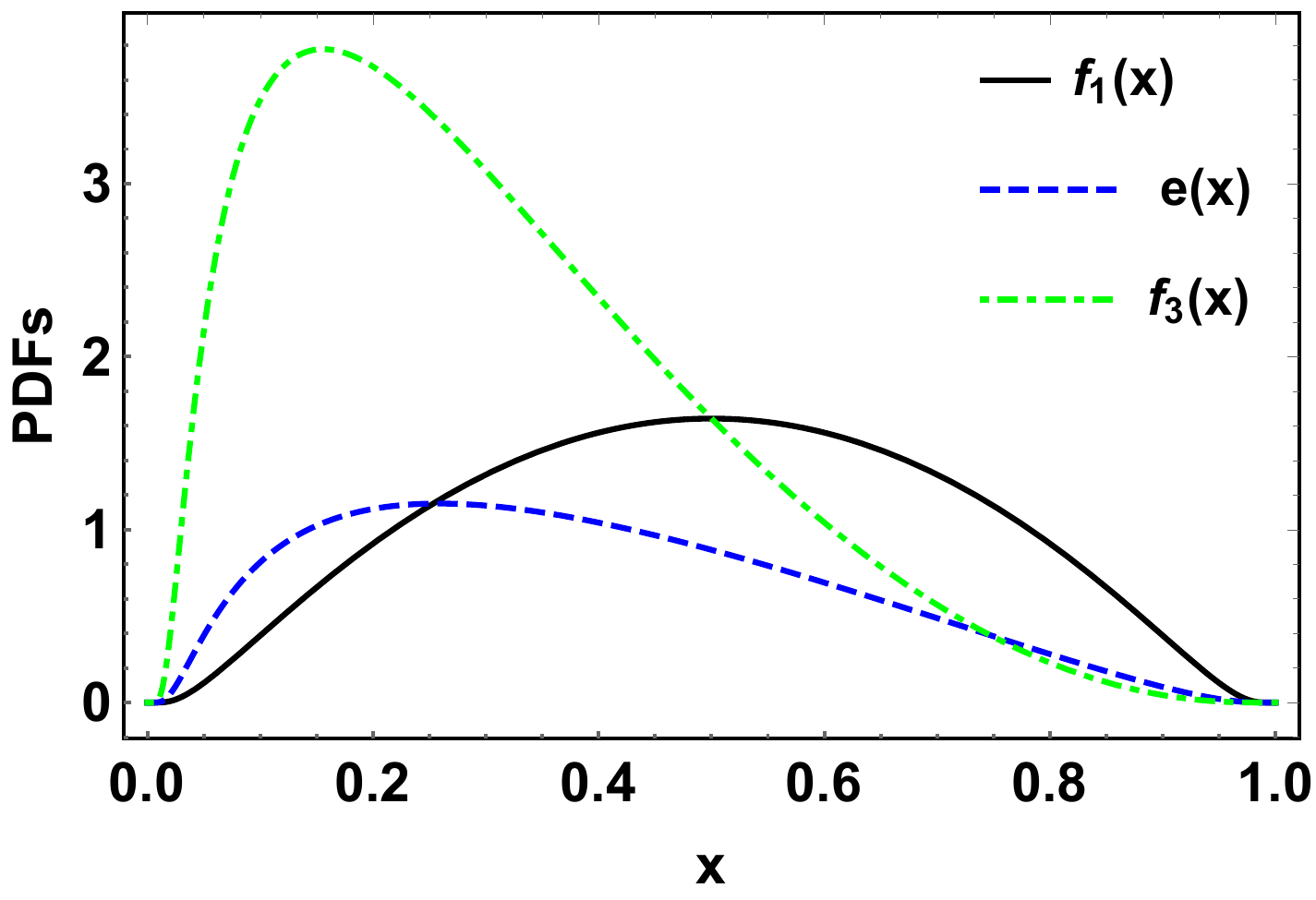}
\caption{(Color online) The unpolarized valence quark
PDFs plotted as a function of longitudinal momentum fraction $x$ up to twist-4.}
		\label{realtmds7}
	\end{figure}
%
 %
%
\section{GTMDs to GPDs}
There are a total of $8$ GPDs for the pion up to twist-$4$. These GPDs are $F_1$ (twist-2), $H_1$ (twist-$2$), $E_2$ (twist-3), $F_2$ (twist-3), $G_2$ (twist-3), $H_2$ (twist-3), $F_3$ (twist-4), and $H_3$ (twist-4). Among these, $F_1$, $F_2$, $G_2$, and $F_3$ are chiral-even, while the remaining GPDs are chiral-odd. These GPDs can be computed either by solving the GPD correlator \cite{Meissner:2008ay,Luan:2024dvc} or by integrating the corresponding GTMDs over the transverse momentum of the quark \cite{Meissner:2008ay}. Previously, higher-twist GPDs have been computed using the GPD correlator approach, as discussed in Ref.~\cite{Luan:2024dvc}. However, in this work, we calculate all GPDs from GTMDs at zero skewness, where its variables are $(x,\xi=0,-\Delta^2_\perp)$. At this limit, both $F_2(x,\xi=0,-\Delta^2_\perp)$ and $H_2(x,\xi=0,-\Delta^2_\perp)$ vanish, as noted in Ref.~\cite{Meissner:2008ay}. The remaining GPDs can be obtained from different GTMDs as \cite{Meissner:2008ay}
\begin{eqnarray}
     F_1(x,\xi=0,-\Delta^2_\perp)&=& \int d^2\bfk F_1, \\
   H_1(x,\xi=0,-\Delta^2_\perp)&=& \int d^2\bfk \Bigg[\frac{\bfk.
   \Dp}{\Dp^2}H^k_1+H^\Delta_1\Bigg],\\
   E_2(x,\xi=0,-\Delta^2_\perp)&=& \int d^2\bfk E_2, \\
    G_2(x,\xi=0,-\Delta^2_\perp)&=& \int d^2\bfk \Bigg[\frac{\bfk.
   \Dp}{\Dp^2}G^k_2+G^\Delta_2\Bigg], \\
   F_3(x,\xi=0,-\Delta^2_\perp)&=& \int d^2\bfk F_3, \\
    H_3(x,\xi=0,-\Delta^2_\perp)&=& \int d^2\bfk \Bigg[\frac{\bfk.
   \Dp}{\Dp^2}H^k_3+H^\Delta_3\Bigg].
\end{eqnarray}
In our case, the $H_1^k$  and $H_3^k$ GTMDs are zero, which simplifies our leading twist and twist-4 calculations. The twist-2 GPDs $F_1(x,0,-\Delta^2_\perp)$ and $H_1(x,0,-\Delta^2_\perp)$ are referred to as the vector (no spin-flip) and tensor (spin-flip) GPDs, respectively. The GPD $F_1(x,0,-\Delta^2_\perp)$ is directly related to the elastic electromagnetic form factor, gravitational FFs, and the unpolarized quark PDF $f_1(x)$, whereas $H_1(x,0,-\Delta^2_\perp)$ provides insight into the dependence of the pion’s quark distributions on their polarization perpendicular to the pion’s direction of motion (transversity) \cite{QCDSF:2007ifr}. At sub-leading twist, $E_2(x,0,-\Delta^2_\perp)$ and $G_2(x,0,-\Delta^2_\perp)$ correspond to the scalar and axial vector GPDs, respectively. The twist-4 GPDs have been studied primarily for academic purposes. All these GPDs are plotted as functions of $x$ and $\Delta_\perp$ in Fig.~\ref{realtmds4}. We observe that most GPDs exhibit quark distributions concentrated in the small-$x$ region for low values of $-\Delta_\perp^2$ (in GeV$^2$), except for $F_1(x,0,-\Delta^2_\perp)$. Additionally, the distribution shifts towards the higher-$x$ region with increasing $-\Delta_\perp^2$, indicating that quark distributions are more prominent when the momentum transfer to the final-state pion is low and vice versa. Only $F_1(x,0,-\Delta^2_\perp)$ exhibits symmetry about $x=0.5$ at $\Delta_\perp=0$ GeV. The twist-4 GPD $F_3(x,0,-\Delta^2_\perp)$ attains the highest peak compared to other GTMDs. The SOC of the pion can also be extracted from the leading-twist GPDs as \cite{Acharyya:2024enp,Zhang:2024adr}
\begin{eqnarray}
    C^q= \int d x~\Big(\frac{m_u}{M_\pi}H_1(x,0,0)-F_1(x,0,0)\Big).
\end{eqnarray}
The value of $C^q$ is found to be the same as obtained from the GTMDs result of ours.  
The FFs corresponding to each GPD can be calculated as  
\begin{eqnarray}
    FF(Q^2)=\int d x ~\text{GPD}(x,0, -\Delta^2_\perp),
\end{eqnarray}
with $Q^2 = \Delta^2 = -\Delta_\perp^2$.  
The vector $F_1(Q^2)$ and tensor $H_1(Q^2)$ EMFFs correspond to the twist-$2$ GPDs and have been plotted as a function of $-\Delta^2_\perp$ (in GeV$^2$) in Fig.~\ref{realtmds6} (a). The twist-$3$ and twist-$4$ FFs are presented in Fig.~\ref{realtmds6} (b). Both the leading twist FFs exhibit a smooth decreasing behavior with increasing momentum transfer. However, the higher twist FFs show an increasing distribution and then decrease with an increase in $-\Delta^2_\perp$ (in GeV$^2$). The twist-$3$ $G_2(Q^2)$ and twist-$2$ $H_1(Q^2)$ FFs show negative distributions, whereas the other FFs display positive distributions. There have been numerous theoretical and experimental studies on leading-twist FFs \cite{QCDSF:2007ifr,Nematollahi:2024wrj,Broniowski:2024oyk,Zhang:2021shm,Arrington:2021biu,Roberts:2021nhw,Dorokhov:2011ew,Amendolia:1984nz,Choi:2024ptc,Allen:1998hb}, while very few studies have reported results for higher-twist FFs in the case of the pion \cite{Alexandrou:2021ztx, Ropertz:2018stk, Wang:2022mrh, Born:1991dp}. The vector FF $F_1(Q^2)$ was compared with experimental and lattice simulation data in our previous study \cite{Puhan:2024jaw}. However, no experimental data is available for the other FFs. Lattice simulation data exists for the tensor $H_1(Q^2)$ FF in Ref. \cite{QCDSF:2007ifr} and the scalar $E_2(Q^2)$ FF in Ref. \cite{Alexandrou:2021ztx}. The scalar FF exhibits a peak value above $1$ in Ref. \cite{Alexandrou:2021ztx}, whereas in our case, it is found to be lower. The sum rule for EMFFs is satisfied with $F_1(Q^2=0) = 1$, while $H_1(Q^2=0)$ is found to be $0.658$. The elastic charge radius can be accessed through $F_1(Q^2)$ as  
\begin{eqnarray}
    \langle r^2_\pi \rangle= -6 \frac{\partial F_1(Q^2)}{\partial Q^2}\Big|_{Q^2\rightarrow0}.
\end{eqnarray}
The charge radius of the pion is found to be $\langle r^2_\pi \rangle^{1/2} = 0.558$ fm, which is smaller compared to the NA7 experimental result of $0.660$ fm \cite{Amendolia:1984nz}.
\section{GTMDs to TMDs}
For the case of spin-$0$ pseudoscalar mesons, there are a total of $8$ TMDs up to twist-$4$ \cite{Meissner:2008ay}, compared to $32$ TMDs for spin-$1/2$ hadrons \cite{Meissner:2009ww} and $72$ TMDs for spin-$1$ hadrons \cite{Kumano:2020ijt}. Out of these $8$ TMDs, $4$ are T-even, while the remaining $4$ are T-odd in nature. These TMDs can be obtained from GTMDs by taking $\Delta = 0$ as \cite{Meissner:2008ay}
\begin{eqnarray}
    f_1(x,\bfk^2)&=& F_1(x,k_\perp,\Delta_\perp=0,\bfk\cdot\Dp=0),\\
    h_1^\perp(x,\bfk^2)&=& H_1^k(x,k_\perp,\Delta_\perp=0,\bfk\cdot\Dp=0),\\
    e(x,\bfk^2)&=& E_2(x,k_\perp,\Delta_\perp=0,\bfk\cdot\Dp=0),\\
    f^\perp(x,\bfk^2)&=& F_2^k(x,k_\perp,\Delta_\perp=0,\bfk\cdot\Dp=0),\\
    g^\perp(x,\bfk^2)&=& G^k_2(x,k_\perp,\Delta_\perp=0,\bfk\cdot\Dp=0),\\
    h(x,\bfk^2)&=& H_2^\Delta(x,k_\perp,\Delta_\perp=0,\bfk\cdot\Dp=0),\\
    f_3(x,\bfk^2)&=& F_3(x,k_\perp,\Delta_\perp=0,\bfk\cdot\Dp=0),\\
    h_3^\perp(x,\bfk^2)&=& H_3^k(x,k_\perp,\Delta_\perp=0,\bfk\cdot\Dp=0).
\end{eqnarray}
The unpolarized pion TMDs $f_1$, $e$, $f^\perp$, and $f_3$ are T-even in nature and are found to be non-zero in our case, while the remaining TMDs are T-odd and they comes out to be zero. These TMDs have been plotted with respect to $x$ and $\bfk$ in Fig. \ref{realtmds5}. As expected, the higher-twist TMDs exhibit quark distributions predominantly in the small-$x$ region due to the presence of $1/x$ and $1/x^2$ factors in the GTMDs. The twist-$4$ TMD, $e(x,\bfk^2)$, has the highest peak value compared to the other TMDs and is distributed within the range $0 \leq x \leq 0.5$. Additionally, we observed that these TMDs vanish for $\bfk \geq 0.4$ GeV.
\par The first Mellin moments $\langle x\rangle$ of TMDs $f_1$, $e$, $f^\perp$, and $f_3$ are found to be $0.5$, $0.39$, $0.38$, and $0.30$ respectively. This indicates that the leading-twist TMDs carry higher longitudinal and transverse momentum from the initial pion compared to the higher-twist TMDs. Since the Wilson line is taken to be unity, the tilde terms do not contribute to the total TMDs.
\par There is no available experimental data and very limited lattice simulation data for leading-twist TMDs \cite{LatticeParton:2023xdl}. However, recent studies on pion TMD extraction have been reported in Ref. \cite{Cerutti:2022lmb}. Theoretical studies on higher-twist TMDs of pseudoscalar mesons have also been discussed in Refs. \cite{Puhan:2023ekt,Zhang:2020ecj,Lorce:2016ugb}. These TMDs satisfy the relations
\begin{align}
	x\,e(x,\textbf{k}^2_\perp) & = 
	 \frac{m_{u(\bar d)}}{M_{\pi}}\,f_1(x,\textbf{k}^2_\perp),\label{Eq:eom-e}\\
 	x\,f^{\perp}(x,\textbf{k}^2_\perp) & = 
	 f_1(x,\textbf{k}^2_\perp),\label{Eq:eom-fperp}\\
	x^2f_3(x,\textbf{k}^2_\perp) & = 
	 \frac{\textbf{k}^2_\perp+m^2_{u(\bar d)}}{M^{2}_{\pi}}\;f_1(x,\textbf{k}^2_\perp),
     \label{Eq:eom-f4}
\end{align}
and positivity constraints
\begin{align}
	f_1(x,\textbf{k}^2_\perp) \ge 0, \label{Eq:f1-inequality} &  \\
    	f_3(x,\textbf{k}^2_\perp) \ge 0. \label{Eq:f4-inequality} &
\end{align}

\section{GTMDs to PDFs}
The valence quark PDFs corresponding to various TMDs, GPDs, and GTMDs can be obtained as \cite{Meissner:2008ay}
\begin{align}
	f_1(x) & =\int d^2 \bfk f_1(x,\textbf{k}^2_\perp) & = \int_{GTMD} d^2 \bfk F_1(x,\bfk,0,0) &=F_{1(GPD)}(x,0,0),
   \\
   e(x) & =\int d^2 \bfk e(x,\textbf{k}^2_\perp) &= \int_{GTMD} d^2 \bfk E_2(x,\bfk,0,0) &=E_{2(GPD)}(x,0,0),
   \\
   f_3(x) & =\int d^2 \bfk f_3(x,\textbf{k}^2_\perp) &= \int_{GTMD} d^2 \bfk F_3(x,\bfk,0,0) &=F_{3(GPD)}(x,0,0).
   \label{f3pdf}
 \end{align} 
 Here, $f_1(x)$, $e(x)$, and $f_3(x)$ are twist-$2$, twist-$3$, and twist-$4$ PDFs, respectively. These PDFs have been plotted with respect to the longitudinal momentum fraction $x$ in Fig. \ref{realtmds7}. All the PDFs show positive distributions over the entire range of $x$. The sum rules obeyed by these PDFs are
\begin{align}
  \int d x f_1(x) & = N_q\,, \\
  \sum_q\int d x e(x)& = \frac{\sigma_{\pi}}{m_{q(\bar q)}}\,, \label{leo}\\
  \int d x x e(x) 	& = \frac{m_{u(\bar d)}}{M_{\pi}}\;N_q\,, 
    \end{align}
  \begin{align}
  2\int d x\,f_3(x) 	& = N_q, 	\label{Eq:f4-sum-rule}
\end{align}
with $N_q=1$. The $\sigma_{\pi}$ term on the right-hand side of Eq. \eqref{leo} refers to the scalar form factor at zero-momentum transfer \cite{Efremov:2002qh,Lorce:2016ugb} and it carries the crucial information about the mass of the pion. In our previous work, we also evolved the unpolarized $f_1(x)$ PDF to higher $Q^2$ using next-to-next-to-leading order (NNLO) DGLAP evolution equations \cite{Miyama:1995bd,Hirai:1997gb,Hirai:2011si} and compared the results with available experimental data \cite{Puhan:2024jaw,Puhan:2023ekt}. 

\par Currently, very few experimental results are available for the unpolarized $f_1(x)$ PDF \cite{Aicher:2010cb,Conway:1989fs}. However, the upcoming electron ion collider is going to provide more data through Sullivan process \cite{AbdulKhalek:2021gbh,Accardi:2012qut}.

\section{Conclusion}
In this study, we have investigated the internal structure of the pion through generalized transverse momentum-dependent parton distributions (GTMDs) within the light-cone quark model framework. By solving the quark-quark correlator, we derived expressions for each Dirac matrix structure $\Gamma$ in terms of light-front wave functions (LFWFs). From these, we extracted the twist-$2$, twist-$3$, and twist-$4$ GTMDs, identifying that $12$ out of the $16$ possible GTMDs are non-vanishing. Notably, the twist-$2$ distribution $H_1^k$, the twist-$3$ distributions $\tilde{E}_2$ and $\tilde{H}_2$, and the twist-$4$ distribution $H_3^k$ are found to vanish in our model.

Additionally, we obtained the transverse momentum-dependent parton distributions (TMDs) and generalized parton distributions (GPDs) from the corresponding GTMDs. We also calculated the valence quark electromagnetic form factors (FFs) and parton distribution functions (PDFs) up to twist-4. The predicted value for the pion’s elastic charge radius, 0.558 fm, shows good agreement with results from other theoretical models, supporting the validity of our approach.

Overall, the findings offer deeper insights into the internal dynamics of the pion and underscore the significance of higher-twist effects. Future extensions of this work will involve the inclusion of higher Fock-state components to achieve a more comprehensive description of the pion’s internal structure.
 
 \section{Acknowledgement}
 H. D.  would like to thank  the Science and Engineering Research Board,  Anusandhan-National Research Foundation,  Government of India under the scheme SERB-POWER Fellowship (Ref No.  SPF/2023/000116) for financial support. 
	
	 \bibliography{sample}{}

\begin{thebibliography}{100}

\bibitem{Gross:2022hyw}
F.~Gross et~al.
\newblock {50 Years of Quantum Chromodynamics}.
\newblock {\em Eur. Phys. J. C}, 83:1125, 2023.

\bibitem{Brodsky:1997de}
S.~J. Brodsky, H.~C. Pauli, and S.~S. Pinsky.
\newblock {Quantum chromodynamics and other field theories on the light cone}.
\newblock {\em Phys. Rept.}, 301:299, 1998.

\bibitem{Zhang:1997dd}
Wei-Min Zhang.
\newblock {A Weak coupling treatment of nonperturbative QCD dynamics to heavy
  hadrons}.
\newblock {\em Phys. Rev. D}, 56:1528--1548, 1997.

\bibitem{Magradze:1999um}
B.~A. Magradze.
\newblock {Analytic approach to perturbative QCD}.
\newblock {\em Int. J. Mod. Phys. A}, 15:2715--2734, 2000.

\bibitem{CTEQ:1993hwr}
Raymond Brock et~al.
\newblock {Handbook of perturbative QCD: Version 1.0}.
\newblock {\em Rev. Mod. Phys.}, 67:157--248, 1995.

\bibitem{Donnachie:1993it}
A.~Donnachie and P.~V. Landshoff.
\newblock {Proton structure function at small q**2}.
\newblock {\em Z. Phys. C}, 61:139, 1994.

\bibitem{Markovych:2023tpa}
Matthew Markovych and Asli Tandogan.
\newblock {Analytic Evolution of DGLAP Equations}.
\newblock 4 2023.

\bibitem{Wang:2016sfq}
Rong Wang and Xurong Chen.
\newblock {Dynamical parton distributions from DGLAP equations with nonlinear
  corrections}.
\newblock {\em Chin. Phys. C}, 41(5):053103, 2017.

\bibitem{Aybat:2011ge}
S.~Mert Aybat, John~C. Collins, Jian-Wei Qiu, and Ted~C. Rogers.
\newblock {The QCD Evolution of the Sivers Function}.
\newblock {\em Phys. Rev. D}, 85:034043, 2012.

\bibitem{Aybat:2011ta}
S.~Mert Aybat, Alexei Prokudin, and Ted~C. Rogers.
\newblock {Calculation of TMD Evolution for Transverse Single Spin Asymmetry
  Measurements}.
\newblock {\em Phys. Rev. Lett.}, 108:242003, 2012.

\bibitem{Bautista:2016xnp}
I.~Bautista, A.~Fernandez~Tellez, and Martin Hentschinski.
\newblock {BFKL evolution and the growth with energy of exclusive $J/\psi$ and
  $\Upsilon$ photoproduction cross sections}.
\newblock {\em Phys. Rev. D}, 94(5):054002, 2016.

\bibitem{Mukherjee:2023snp}
Swagato Mukherjee, Vladimir~V. Skokov, Andrey Tarasov, and Shaswat Tiwari.
\newblock {Unified description of DGLAP, CSS, and BFKL evolution: TMD
  factorization bridging large and small x}.
\newblock {\em Phys. Rev. D}, 109(3):034035, 2024.

\bibitem{Collins:1981uw}
J.~C. Collins and D.~E. Soper.
\newblock {Parton Distribution and Decay Functions}.
\newblock {\em Nucl. Phys. B}, 194:445, 1982.

\bibitem{Martin:1998sq}
A.~D. Martin et~al.
\newblock {Parton distributions: A New global analysis}.
\newblock {\em Eur. Phys. J. C}, 4:463, 1998.

\bibitem{Gluck:1994uf}
M.~Gluck, E.~Reya, and A.~Vogt.
\newblock {Dynamical parton distributions of the proton and small x physics}.
\newblock {\em Z. Phys. C}, 67:433, 1995.

\bibitem{Gluck:1998xa}
M.~Gl\"uck, E.~Reya, and A.~Vogt.
\newblock {Dynamical parton distributions revisited}.
\newblock {\em Eur. Phys. J. C}, 5:461, 1998.

\bibitem{Diehl:2003ny}
M.~Diehl.
\newblock {Generalized parton distributions}.
\newblock {\em Phys. Rept.}, 388:41, 2003.

\bibitem{Garcon:2002jb}
M.~Garcon.
\newblock {An Introduction to the generalized parton distributions}.
\newblock {\em Eur. Phys. J. A}, 18:389, 2003.

\bibitem{Belitsky:2005qn}
A.~V. Belitsky and A.~V. Radyushkin.
\newblock {Unraveling hadron structure with generalized parton distributions}.
\newblock {\em Phys. Rept.}, 418:1, 2005.

\bibitem{Sharma:2023ibp}
S.~Sharma and H.~Dahiya.
\newblock {Exploring twist-4 chiral-even GPDs in the light-front quark-diquark
  model}.
\newblock {\em Nucl. Phys. B}, 1001:116522, 2024.

\bibitem{Barone:2001sp}
V.~Barone, A.~Drago, and P.~G. Ratcliffe.
\newblock {Transverse polarisation of quarks in hadrons}.
\newblock {\em Phys. Rept.}, 359:1, 2002.

\bibitem{Diehl:2015uka}
M.~Diehl.
\newblock {Introduction to GPDs and TMDs}.
\newblock {\em Eur. Phys. J. A}, 52(6):149, 2016.

\bibitem{Puhan:2023ekt}
Satyajit Puhan, Shubham Sharma, Navpreet Kaur, Narinder Kumar, and Harleen
  Dahiya.
\newblock {T-even TMDs for the spin-0 pseudo-scalar mesons upto twist-4 using
  light-front formalism}.
\newblock {\em JHEP}, 02:075, 2024.

\bibitem{Boussarie:2023izj}
Renaud Boussarie et~al.
\newblock {TMD Handbook}.
\newblock 4 2023.

\bibitem{Puhan:2023hio}
Satyajit Puhan and Harleen Dahiya.
\newblock {Leading twist T-even TMDs for the spin-1 heavy vector mesons}.
\newblock {\em Phys. Rev. D}, 109(3):034005, 2024.

\bibitem{Sharma:2024lal}
Shubham Sharma, Satyajit Puhan, Narinder Kumar, and Harleen Dahiya.
\newblock {TMD Relations: Insights from a Light-Front Quark\textendash{}Diquark
  Model}.
\newblock {\em PTEP}, 2024(10):103B05, 2024.

\bibitem{Echevarria:2022ztg}
Miguel~G. Echevarria, Patricia~A. Gutierrez~Garcia, and Ignazio Scimemi.
\newblock {GTMDs and the factorization of exclusive double Drell-Yan}.
\newblock {\em Phys. Lett. B}, 840:137881, 2023.

\bibitem{Sharma:2024arf}
Shubham Sharma, Sameer Jain, and Harleen Dahiya.
\newblock {Unraveling subleading twist GTMDs of proton using light-front
  quark-diquark model}.
\newblock {\em Phys. Rev. D}, 110(7):074025, 2024.

\bibitem{Bhattacharya:2017bvs}
Shohini Bhattacharya, Andreas Metz, and Jian Zhou.
\newblock {Generalized TMDs and the exclusive double Drell\textendash{}Yan
  process}.
\newblock {\em Phys. Lett. B}, 771:396--400, 2017.
\newblock [Erratum: Phys.Lett.B 810, 135866 (2020)].

\bibitem{Collins:2004nx}
J.~C. Collins and A.~Metz.
\newblock {Universality of soft and collinear factors in hard-scattering
  factorization}.
\newblock {\em Phys. Rev. Lett.}, 93:252001, 2004.

\bibitem{Polchinski:2002jw}
J.~Polchinski and M.~J. Strassler.
\newblock {Deep inelastic scattering and gauge / string duality}.
\newblock {\em JHEP}, 05:012, 2003.

\bibitem{Cheng:2023kmt}
Peng Cheng, Yang Yu, Hui-Yu Xing, Chen Chen, Zhu-Fang Cui, and Craig~D.
  Roberts.
\newblock {Perspective on polarised parton distribution functions and proton
  spin}.
\newblock {\em Phys. Lett. B}, 844:138074, 2023.

\bibitem{COMPASS:2010hwr}
M.~G. Alekseev et~al.
\newblock {Quark helicity distributions from longitudinal spin asymmetries in
  muon-proton and muon-deuteron scattering}.
\newblock {\em Phys. Lett. B}, 693:227--235, 2010.

\bibitem{Gao:2022uhg}
Xiang Gao, Andrew~D. Hanlon, Jack Holligan, Nikhil Karthik, Swagato Mukherjee,
  Peter Petreczky, Sergey Syritsyn, and Yong Zhao.
\newblock {Unpolarized proton PDF at NNLO from lattice QCD with physical quark
  masses}.
\newblock {\em Phys. Rev. D}, 107(7):074509, 2023.

\bibitem{CMS:2012tdr}
Serguei Chatrchyan et~al.
\newblock {Measurement of the Top-Quark Mass in $t\bar{t}$ Events with Dilepton
  Final States in $pp$ Collisions at $\sqrt{s}=7$ TeV}.
\newblock {\em Eur. Phys. J. C}, 72:2202, 2012.

\bibitem{Aicher:2010cb}
Matthias Aicher, Andreas Schafer, and Werner Vogelsang.
\newblock {Soft-gluon resummation and the valence parton distribution function
  of the pion}.
\newblock {\em Phys. Rev. Lett.}, 105:252003, 2010.

\bibitem{Conway:1989fs}
J.~S. Conway et~al.
\newblock {Experimental Study of Muon Pairs Produced by 252-GeV Pions on
  Tungsten}.
\newblock {\em Phys. Rev. D}, 39:92, 1989.

\bibitem{AbdulKhalek:2021gbh}
R.~Abdul~Khalek et~al.
\newblock {Science Requirements and Detector Concepts for the Electron-Ion
  Collider}: {EIC Yellow Report}.
\newblock {\em Nucl. Phys. A}, 1026:122447, 2022.

\bibitem{Accardi:2012qut}
A.~Accardi et~al.
\newblock {Electron Ion Collider: The Next QCD Frontier}: {Understanding the
  glue that binds us all}.
\newblock {\em Eur. Phys. J. A}, 52(9):268, 2016.

\bibitem{Barry:2021osv}
P.~C. Barry, Chueng-Ryong Ji, N.~Sato, and W.~Melnitchouk.
\newblock {Global QCD Analysis of Pion Parton Distributions with Threshold
  Resummation}.
\newblock {\em Phys. Rev. Lett.}, 127(23):232001, 2021.

\bibitem{Tangerman:1994eh}
R.~D. Tangerman and P.~J. Mulders.
\newblock {Intrinsic transverse momentum and the polarized Drell-Yan process}.
\newblock {\em Phys. Rev. D}, 51:3357, 1995.

\bibitem{Zhou:2009jm}
J.~Zhou, F.~Yuan, and Z.~T. Liang.
\newblock {Transverse momentum dependent quark distributions and polarized
  Drell-Yan processes}.
\newblock {\em Phys. Rev. D}, 81:054008, 2010.

\bibitem{Mulders:1995dh}
P.~J. Mulders and R.~D. Tangerman.
\newblock {The Complete tree level result up to order 1/Q for polarized deep
  inelastic leptoproduction}.
\newblock {\em Nucl. Phys. B}, 461:197, 1996.
\newblock [Erratum: Nucl.Phys.B 484, 538--540 (1997)].

\bibitem{Collins:2002kn}
J.~C. Collins.
\newblock {Leading twist single transverse-spin asymmetries: Drell-Yan and deep
  inelastic scattering}.
\newblock {\em Phys. Lett. B}, 536:43, 2002.

\bibitem{Bacchetta:2017gcc}
A.~Bacchetta et~al.
\newblock {Extraction of partonic transverse momentum distributions from
  semi-inclusive deep-inelastic scattering, Drell-Yan and Z-boson production}.
\newblock {\em JHEP}, 06:081, 2017.
\newblock [Erratum: JHEP 06, 051 (2019)].

\bibitem{Bacchetta:2006tn}
A.~Bacchetta et~al.
\newblock {Semi-inclusive deep inelastic scattering at small transverse
  momentum}.
\newblock {\em JHEP}, 02:093, 2007.

\bibitem{Brodsky:2002cx}
S.~J. Brodsky, D.~S. Hwang, and I.~Schmidt.
\newblock {Final state interactions and single spin asymmetries in
  semiinclusive deep inelastic scattering}.
\newblock {\em Phys. Lett. B}, 530:99, 2002.

\bibitem{Ji:2004wu}
X.d. Ji, J.~p. Ma, and F.~Yuan.
\newblock {QCD factorization for semi-inclusive deep-inelastic scattering at
  low transverse momentum}.
\newblock {\em Phys. Rev. D}, 71:034005, 2005.

\bibitem{Catani:2015vma}
S.~Catani et~al.
\newblock {Vector boson production at hadron colliders: transverse-momentum
  resummation and leptonic decay}.
\newblock {\em JHEP}, 12:047, 2015.

\bibitem{Ji:1996nm}
X.~D. Ji.
\newblock {Deeply virtual Compton scattering}.
\newblock {\em Phys. Rev. D}, 55:7114, 1997.

\bibitem{Favart:2015umi}
L.~Favart et~al.
\newblock {Deeply Virtual Meson Production on the nucleon}.
\newblock {\em Eur. Phys. J. A}, 52(6):158, 2016.

\bibitem{Lorce:2011dv}
C.~Lorce, B.~Pasquini, and M.~Vanderhaeghen.
\newblock {Unified framework for generalized and transverse-momentum dependent
  parton distributions within a 3Q light-cone picture of the nucleon}.
\newblock {\em JHEP}, 05:041, 2011.

\bibitem{Kanazawa:2014nha}
K.~Kanazawa et~al.
\newblock {Twist-2 generalized transverse-momentum dependent parton
  distributions and the spin/orbital structure of the nucleon}.
\newblock {\em Phys. Rev. D}, 90(1):014028, 2014.

\bibitem{Chakrabarti:2017teq}
D.~Chakrabarti et~al.
\newblock {Quark Wigner distributions and spin-spin correlations}.
\newblock {\em Phys. Rev. D}, 95(7):074028, 2017.

\bibitem{Kumar:2017xcm}
N.~Kumar and C.~Mondal.
\newblock {Wigner distributions for an electron}.
\newblock {\em Nucl. Phys. B}, 931:226, 2018.

\bibitem{Kaur:2019kpi}
Navdeep Kaur and Harleen Dahiya.
\newblock {Quark Wigner Distributions and GTMDs of Pion in the Light-Front
  Holographic Model}.
\newblock {\em Eur. Phys. J. A}, 56(6):172, 2020.

\bibitem{Sharma:2023tre}
S.~Sharma and H.~Dahiya.
\newblock {Twist-4 proton GTMDs in the light-front quark\textendash{}diquark
  model}.
\newblock {\em Eur. Phys. J. A}, 59(10):235, 2023.

\bibitem{Sharma:2023qgb}
S.~Sharma and H.~Dahiya.
\newblock {Analysis of the higher twist GTMD $F_{31}$ for proton in the
  light-front quark-diquark model}.
\newblock In {\em {30th International Workshop on Deep-Inelastic Scattering and
  Related Subjects}}, 7 2023.

\bibitem{Bastami:2020rxn}
S.~Bastami et~al.
\newblock {Structure of the nucleon at leading and subleading twist in the
  covariant parton model}.
\newblock {\em Phys. Rev. D}, 103(1):014024, 2021.

\bibitem{Meissner:2008ay}
Stephan Meissner, Andreas Metz, Marc Schlegel, and Klaus Goeke.
\newblock {Generalized parton correlation functions for a spin-0 hadron}.
\newblock {\em JHEP}, 08:038, 2008.

\bibitem{Meissner:2009ww}
S.~Meissner, A.~Metz, and M.~Schlegel.
\newblock {Generalized parton correlation functions for a spin-1/2 hadron}.
\newblock {\em JHEP}, 08:056, 2009.

\bibitem{Anikin:2001ge}
I.~V. Anikin and O.~V. Teryaev.
\newblock {Wandzura-Wilczek approximation from generalized rotational
  invariance}.
\newblock {\em Phys. Lett. B}, 509:95--105, 2001.

\bibitem{Ma:2018ysi}
Zhi-Lei Ma and Zhun Lu.
\newblock {Quark Wigner distribution of the pion meson in light-cone quark
  model}.
\newblock {\em Phys. Rev. D}, 98(5):054024, 2018.

\bibitem{Zhang:2020ecj}
Jin-Li Zhang, Zhu-Fang Cui, Jialun Ping, and Craig~D Roberts.
\newblock {Contact interaction analysis of pion GTMDs}.
\newblock {\em Eur. Phys. J. C}, 81(1):6, 2021.

\bibitem{Luan:2024dvc}
Xiaoyan Luan and Zhun Lu.
\newblock {Higher-twist generalized parton distributions of the pion and kaon
  at zero skewness in the light-cone quark model}.
\newblock {\em Phys. Rev. D}, 110(7):074022, 2024.

\bibitem{Kaur:2018ewq}
Navdeep Kaur, Narinder Kumar, Chandan Mondal, and Harleen Dahiya.
\newblock {Generalized Parton Distributions of Pion for Non-Zero Skewness in
  AdS/QCD}.
\newblock {\em Nucl. Phys. B}, 934:80--95, 2018.

\bibitem{Pasquini:2014ppa}
B.~Pasquini and P.~Schweitzer.
\newblock {Pion transverse momentum dependent parton distributions in a
  light-front constituent approach, and the Boer-Mulders effect in the
  pion-induced Drell-Yan process}.
\newblock {\em Phys. Rev. D}, 90(1):014050, 2014.

\bibitem{Kaur:2019jfa}
Navdeep Kaur and Harleen Dahiya.
\newblock {Transverse momentum dependent parton distributions of pion in the
  light-front holographic model}.
\newblock {\em Int. J. Mod. Phys. A}, 36(08n09):2150052, 2021.

\bibitem{Noguera:2015iia}
Santiago Noguera and Sergio Scopetta.
\newblock {Pion transverse momentum dependent parton distributions in the Nambu
  and Jona-Lasinio model}.
\newblock {\em JHEP}, 11:102, 2015.

\bibitem{Lu:2012hh}
Zhun Lu, Bo-Qiang Ma, and Jiacai Zhu.
\newblock {Boer-Mulders function of the pion in the MIT bag model}.
\newblock {\em Phys. Rev. D}, 86:094023, 2012.

\bibitem{Chen:2019lcm}
Jiunn-Wei Chen, Huey-Wen Lin, and Jian-Hui Zhang.
\newblock {Pion generalized parton distribution from lattice QCD}.
\newblock {\em Nucl. Phys. B}, 952:114940, 2020.

\bibitem{Cerutti:2022lmb}
Matteo Cerutti, Lorenzo Rossi, Simone Venturini, Alessandro Bacchetta, Valerio
  Bertone, Chiara Bissolotti, and Marco Radici.
\newblock {Extraction of pion transverse momentum distributions from Drell-Yan
  data}.
\newblock {\em Phys. Rev. D}, 107(1):014014, 2023.

\bibitem{Lepage:1980fj}
G.~Peter Lepage and Stanley~J. Brodsky.
\newblock {Exclusive Processes in Perturbative Quantum Chromodynamics}.
\newblock {\em Phys. Rev. D}, 22:2157, 1980.

\bibitem{Pasquini:2023aaf}
Barbara Pasquini, Simone Rodini, and Simone Venturini.
\newblock {Valence quark, sea, and gluon content of the pion from the parton
  distribution functions and the electromagnetic form factor}.
\newblock {\em Phys. Rev. D}, 107(11):114023, 2023.

\bibitem{Qian:2008px}
Wen Qian and Bo-Qiang Ma.
\newblock {Vector meson omega-phi mixing and their form factors in light-cone
  quark model}.
\newblock {\em Phys. Rev. D}, 78:074002, 2008.

\bibitem{Brodsky:2000xy}
Stanley~J. Brodsky, Markus Diehl, and Dae~Sung Hwang.
\newblock {Light cone wave function representation of deeply virtual Compton
  scattering}.
\newblock {\em Nucl. Phys. B}, 596:99--124, 2001.

\bibitem{Xiao:2002iv}
Bo-Wen Xiao, Xin Qian, and Bo-Qiang Ma.
\newblock {The Kaon form-factor in the light cone quark model}.
\newblock {\em Eur. Phys. J. A}, 15:523--527, 2002.

\bibitem{Choi:1996mq}
H.~M. Choi and Chueng-Ryong Ji.
\newblock {Light cone quark model predictions for radiative meson decays}.
\newblock {\em Nucl. Phys. A}, 618:291--316, 1997.

\bibitem{Bacchetta:2020vty}
Alessandro Bacchetta, Francesco~Giovanni Celiberto, Marco Radici, and Pieter
  Taels.
\newblock {Transverse-momentum-dependent gluon distribution functions in a
  spectator model}.
\newblock {\em Eur. Phys. J. C}, 80(8):733, 2020.

\bibitem{Kaur:2019jow}
Satvir Kaur and Harleen Dahiya.
\newblock {Study of kaon structure using the light-cone quark model}.
\newblock {\em Phys. Rev. D}, 100(7):074008, 2019.

\bibitem{Zhang:2024adr}
Jin-Li Zhang.
\newblock {Kaon GTMDs in the Dyson-Schwinger equations using contact
  interaction}.
\newblock 9 2024.

\bibitem{Acharyya:2024enp}
Ritwik Acharyya, Satyajit Puhan, and Harleen Dahiya.
\newblock {Quark spin-orbit correlations in spin-0 and spin-1 mesons using the
  light-front quark model}.
\newblock {\em Phys. Rev. D}, 110(3):034020, 2024.

\bibitem{Lorce:2025ayr}
C\'edric Lorc\'e and Qin-Tao Song.
\newblock {Spin-orbit correlation and spatial distributions for spin-0
  hadrons}.
\newblock {\em Phys. Lett. B}, 864:139433, 2025.

\bibitem{Efremov:2002qh}
A.~V. Efremov and P.~Schweitzer.
\newblock {The Chirally odd twist 3 distribution e(a)(x)}.
\newblock {\em JHEP}, 08:006, 2003.

\bibitem{QCDSF:2007ifr}
D.~Br\"ommel et~al.
\newblock {The Spin structure of the pion}.
\newblock {\em Phys. Rev. Lett.}, 101:122001, 2008.

\bibitem{Nematollahi:2024wrj}
H.~Nematollahi and K.~Azizi.
\newblock {Unpolarized valence GPDs and form factors of the pion in the
  modified chiral quark model}.
\newblock {\em Phys. Rev. D}, 111(1):014011, 2025.

\bibitem{Broniowski:2024oyk}
Wojciech Broniowski and Enrique Ruiz~Arriola.
\newblock {Gravitational form factors of the pion and meson dominance}.
\newblock {\em Phys. Lett. B}, 859:139138, 2024.

\bibitem{Zhang:2021shm}
Jin-Li Zhang, Meng-Yun Lai, Hong-Shi Zong, and Jia-Lun Ping.
\newblock {Pion generalized parton distributions and light-front wave functions
  in the Nambu\textendash{}Jona-Lasinio model}.
\newblock {\em Nucl. Phys. B}, 966:115387, 2021.

\bibitem{Arrington:2021biu}
J.~Arrington et~al.
\newblock {Revealing the structure of light pseudoscalar mesons at the
  electron\textendash{}ion collider}.
\newblock {\em J. Phys. G}, 48(7):075106, 2021.

\bibitem{Roberts:2021nhw}
Craig~D. Roberts, David~G. Richards, Tanja Horn, and Lei Chang.
\newblock {Insights into the emergence of mass from studies of pion and kaon
  structure}.
\newblock {\em Prog. Part. Nucl. Phys.}, 120:103883, 2021.

\bibitem{Dorokhov:2011ew}
Alexander~E. Dorokhov, Wojciech Broniowski, and Enrique Ruiz~Arriola.
\newblock {Generalized Quark Transversity Distribution of the Pion in Chiral
  Quark Models}.
\newblock {\em Phys. Rev. D}, 84:074015, 2011.

\bibitem{Amendolia:1984nz}
S.~R. Amendolia et~al.
\newblock {A Measurement of the Pion Charge Radius}.
\newblock {\em Phys. Lett. B}, 146:116--120, 1984.

\bibitem{Choi:2024ptc}
Ho-Meoyng Choi and Chueng-Ryong Ji.
\newblock {Consistency of the pion form factor and unpolarized transverse
  momentum dependent parton distributions beyond leading twist in the
  light-front quark model}.
\newblock {\em Phys. Rev. D}, 110(1):014006, 2024.

\bibitem{Allen:1998hb}
T.~W. Allen and W.~H. Klink.
\newblock {Pion charge form-factor in point form relativistic dynamics}.
\newblock {\em Phys. Rev. C}, 58:3670--3673, 1998.

\bibitem{Alexandrou:2021ztx}
Constantia Alexandrou, Simone Bacchio, Ian Cloet, Martha Constantinou, Joseph
  Delmar, Kyriakos Hadjiyiannakou, Giannis Koutsou, Colin Lauer, and Alejandro
  Vaquero.
\newblock {Scalar, vector, and tensor form factors for the pion and kaon from
  lattice QCD}.
\newblock {\em Phys. Rev. D}, 105(5):054502, 2022.

\bibitem{Ropertz:2018stk}
Stefan Ropertz, Christoph Hanhart, and Bastian Kubis.
\newblock {A new parametrization for the scalar pion form factors}.
\newblock {\em Eur. Phys. J. C}, 78(12):1000, 2018.

\bibitem{Wang:2022mrh}
Xiaobin Wang, Zanbin Xing, Jiayin Kang, Kh\'epani Raya, and Lei Chang.
\newblock {Pion scalar, vector, and tensor form factors from a contact
  interaction}.
\newblock {\em Phys. Rev. D}, 106(5):054016, 2022.

\bibitem{Born:1991dp}
R.~Born, T.~Hurth, K.~Schilcher, and Y.~L. Wu.
\newblock {A QCD calculation of the pion scalar form-factor sigma term}.
\newblock {\em Phys. Lett. B}, 266:463--466, 1991.

\bibitem{Puhan:2024jaw}
Satyajit Puhan, Navpreet Kaur, and Harleen Dahiya.
\newblock {Transverse and spatial structure of light to heavy pseudoscalar
  mesons in light-cone quark model}.
\newblock {\em Phys. Rev. D}, 111(1):014008, 2025.

\bibitem{Kumano:2020ijt}
S.~Kumano and Qin-Tao Song.
\newblock {Transverse-momentum-dependent parton distribution functions up to
  twist 4 for spin-1 hadrons}.
\newblock {\em Phys. Rev. D}, 103(1):014025, 2021.

\bibitem{LatticeParton:2023xdl}
Min-Huan Chu et~al.
\newblock {Transverse-momentum-dependent wave functions of the pion from
  lattice QCD}.
\newblock {\em Phys. Rev. D}, 109(9):L091503, 2024.

\bibitem{Lorce:2016ugb}
C.~Lorc\'e, B.~Pasquini, and P.~Schweitzer.
\newblock {Transverse pion structure beyond leading twist in constituent
  models}.
\newblock {\em Eur. Phys. J. C}, 76(7):415, 2016.

\bibitem{Miyama:1995bd}
M.~Miyama and S.~Kumano.
\newblock {Numerical solution of Q**2 evolution equations in a brute force
  method}.
\newblock {\em Comput. Phys. Commun.}, 94:185--215, 1996.

\bibitem{Hirai:1997gb}
M.~Hirai, S.~Kumano, and M.~Miyama.
\newblock {Numerical solution of Q**2 evolution equations for polarized
  structure functions}.
\newblock {\em Comput. Phys. Commun.}, 108:38, 1998.

\bibitem{Hirai:2011si}
M.~Hirai and S.~Kumano.
\newblock {Numerical solution of $Q^2$ evolution equations for fragmentation
  functions}.
\newblock {\em Comput. Phys. Commun.}, 183:1002--1013, 2012.

\end{thebibliography}
				\bibliographystyle{unsrt}

\end{document}